\pdfoutput=1
\documentclass[11pt]{article}

\usepackage{etex}
\reserveinserts{28}

\usepackage{amsmath,amssymb,amsfonts}
\usepackage[multiple,stable]{footmisc}
\usepackage[amsmath,amsthm,thmmarks]{ntheorem}
\allowdisplaybreaks[4]
\usepackage[noconfig]{refstyle}
\usepackage[dvipsnames]{xcolor}
\usepackage[hyperfootnotes=false, linktocpage=true, colorlinks, citecolor=blue, linkcolor=blue, urlcolor=Maroon]{hyperref}
\usepackage{array,tabularx,booktabs,geometry,subfigure,graphicx,longtable}
\usepackage[bf]{caption}
\usepackage{tikz}
\usetikzlibrary{shapes,arrows}
\tikzstyle{block} = [rectangle, draw, text width=7em, text centered, rounded corners, minimum height=3em]
\usepackage{graphicx,color}
\usetikzlibrary{positioning,arrows}

\usepackage{cite}
\usepackage{arydshln}
\usepackage{tikz}
\usetikzlibrary{positioning,automata,decorations.pathreplacing,shapes}
\usepackage[english]{babel}
\usepackage{microtype}

\usepackage[all,cmtip]{xy}
\usetikzlibrary{matrix, arrows}

\usepackage{hyperref}
\usepackage{extarrows}

\let\eqref=\relax
\newref{eq}{name={eq.~},Name={Eq.~},names={eqs.~},Names={Eqs.~},rngtxt={-},refcmd=(\ref{#1})}
\newref{f}{name={footnote~},Name={Footnote~},names={footnotes~},Names={Footnotes~}}
\newref{app}{name={appendix~},Name={Appendix~},names={appendixes~},Names={Appendixes~}}
\newref{tab}{name={table~},Name={Table~},names={tables~},Names={Tables~}}
\newref{ch}{name={chapter~},Name={Chapter~},names={chapters~},Names={Chapters~}}
\newref{sec}{name={section~},Name={Section~},names={sections~},Names={Sections~}}
\newref{fig}{name={figure~},Name={Figure~},names={figures~},Names={Figures~}}
\numberwithin{equation}{section}
\newcommand{\eref}[1]{(\ref{#1})}

\newcommand{\eeq}{\end{equation}}
\newcommand{\beq}{\begin{equation}}
\newcommand{\ba}{\begin{array}}
\newcommand{\ea}{\end{array}}

\newcommand{\cO}{{\cal O}}

\newcommand{\IP}{\mathbb P}
\newcommand{\IC}{\mathbb C}

\def\a{\alpha}

\def\D{\Delta}

\def\IP{\mathbb{P}}

\def\IZ{\mathbb{Z}}

\def\IC{\mathbb{C}}

\newcommand{\id}{\mathbf{1}}

\def\cN{\mathcal{N}}
\def\cO{\mathcal{O}}

\def\cS{\mathcal{S}}

\def\clap#1{\hbox to 0pt{\hss#1\hss}}

\newcommand{\be}{\begin{equation}}
\newcommand{\ee}{\end{equation}}
\newcommand{\bea}{\begin{equation}\begin{aligned}}	
\newcommand{\eea}{\end{aligned}\end{equation}}		

\newcommand{\iddots}{\mathinner{\mkern2mu\raise1pt\hbox{.}\mkern2mu \raise4pt\hbox{.}\mkern2mu\raise7pt\hbox{.}\mkern1mu}}

\providecommand{\id}{\leavevmode\hbox{\small$\mathrm{1}$\kern-3.8pt\normalsize$\mathrm{1}$}}
\def\fnote#1#2{\begingroup\def\thefootnote{#1}\footnote{#2}
     \addtocounter{footnote}{-1}\endgroup}

\begin{document}

\vspace{1cm}

\title{
       \vskip 40pt
       {\huge  Multiple Fibrations in Calabi-Yau Geometry \\ and String Dualities}
       }
       \date{}
\maketitle

\begin{center}
\author{Lara B. Anderson${}{}$, Xin Gao${}{}$, James Gray${}{}$ and Seung-Joo Lee${}{}$}
\end{center}

\begin{center} {\small ${}${\it Physics Department, Robeson Hall, Virginia Tech, Blacksburg, VA 24061, USA}}\\
\fnote{}{lara.anderson@vt.edu,  xingao@vt.edu, jamesgray@vt.edu, seungsm@vt.edu}
\end{center}

\begin{abstract}
\noindent In this work we explore the physics associated to Calabi-Yau (CY) $n$-folds that can be described as a fibration in more than one way. Beginning with F-theory vacua in various dimensions, we consider limits/dualities with M-theory, type IIA, and heterotic string theories. Our results include many  M-/F-theory correspondences in which distinct F-theory vacua -- associated to different elliptic fibrations of the same CY $n$-fold -- give rise to the same M-theory limit in one dimension lower. Examples include $5$-dimensional correspondences between $6$-dimensional theories with Abelian, non-Abelian and superconformal structure, as well as examples of higher rank Mordell-Weil geometries. In addition, in the context of heterotic/F-theory duality, we investigate the role played by multiple $K3$- and elliptic fibrations in known and novel string dualities in $8$-, $6$- and $4$-dimensional theories. Here we systematically summarize nested fibration structures and comment on the roles they play in T-duality, mirror symmetry, and $4$-dimensional compactifications of F-theory with G-flux. This investigation of duality structures is made possible by geometric tools developed in a companion paper \cite{us_to_appear}.
\end{abstract}

\thispagestyle{empty}
\setcounter{page}{0}
\newpage

\tableofcontents

\section{Introduction: Multiple Fibrations and String Dualities}\label{sec:Intro}
F-theory has proven to be a flexible and extensive framework for studying the possible effective field theories arising from string compactifications in various dimensions.  Because F-theory itself is defined via ``geometrizing" the axio-dilaton of Type IIB string theory \cite{Vafa:1996xn}, any systematic study of F-theory vacua must necessarily be linked to a study of the geometry of elliptically (or more generally genus one) fibered Calabi-Yau (CY) manifolds. Significantly, the set of all genus one fibered CY $3$-folds is known to be finite \cite{gross_finite} and recent progress \cite{2016arXiv160802997D} has given evidence of finiteness for genus one fibered CY $4$- and $5$-folds. A central motivation of the classification of fibered CY $3$-folds in \cite{grassi,gross_finite} was that these results may be an important step towards establishing the finiteness of the set of \emph{all CY $3$-folds}.

Within this framework, the forms that elliptically fibered CY manifolds can take have been of interest in both mathematics and physics. A key result due to Nakayama \cite{nakayama} guarantees that any elliptically fibered manifold is birational to a so-called Weierstrass model. As a result, Weierstrass models have played a significant role not only in classifying CY geometries, but also in defining the physics associated to F-theory in various dimensions. From the point of view of F-theory, Weierstrass models correspond to a minimal, irreducible form of the torus fiber that can be directly linked to the axio-dilaton. In addition, in the study of F-theory effective field theories, Weierstrass models are useful in that they can correspond to generically singular geometries (``non-Higgsable" effective theories \cite{Morrison:2012np,Morrison:2014lca}) and as their moduli are tuned, give rise to many different singularity types and hence, CY resolutions. Taking this point of view, a Weierstrass model at a suitable singular point in its moduli space can often be resolved to produce a smooth CY $n$-fold. This provides a method of constructing elliptically fibered CY manifolds with distinct Hodge numbers, Chern classes, etc. which must be counted in the current ``zoo" of known CY manifolds. 

With these observations in mind, here we will take the reverse viewpoint and begin with F-theory on a given elliptically/genus one fibered CY manifold of a more general form. We will work towards Weierstrass models and F-theory EFTs by starting from resolved geometries and more generally with the known datasets of CY manifolds. \emph{The goal will be to observe whether or not there are global features of the CY total space that are ``hidden" from the Weierstrass description?}

\begin{figure}[t!]
\centering
\includegraphics[width=14cm] {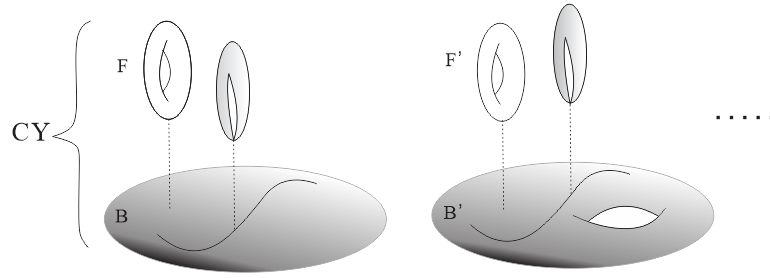}
\caption{An illustration of multiple genus-one fibrations in a single Calabi-Yau $n$-fold: $\pi: X \to B$ and $\pi': X \to B'$}.
\label{multy_fib}
\end{figure}

We will note that there is at least one significant feature which is difficult to observe from the Weierstrass-focused approach described above. This is the possibility of \emph{multiple fibrations} within a single CY total space. We will refer to a CY $n$-fold as \emph{multiply elliptically fibered} (or genus one fibered in the case without section) when it admits multiple descriptions of the form $\pi_i: X_{n} \longrightarrow B^{(i)}_{n-1}$ with elliptic fiber $\mathbb{E}_{(i)b}=\pi^{-1}(b\in B^{(i)}_{n-1})$ (denoted succinctly by $\pi_i: X_{n} \stackrel{\mathbb{E}_{(i)}}{\longrightarrow} B^{(i)}_{n-1}$).  That is,
\beq\label{manyfibs}
\xymatrix{
& X_{n} \ar[ld]_{\mathbb{E}_{(1)}} \ar[d]^{\mathbb{E}_{(2)}}  \ar[rd]^{\mathbb{E}_{(i)}} &\\
B^{(1)}_{n-1} & B^{(2)}_{n-1} \ldots & B^{(i)}_{n-1}}
\eeq
For each fibration, the structure of the singular fibers, discriminant locus, fibral divisors and Mordell-Weil group can be different, as can the topology of the base manifolds $B^{(i)}_{n-1}$. This is illustrated in Figure \ref{multy_fib}. According to Nakayama's theorem \cite{nakayama}, a Weierstrass model can be formed for \emph{each} of the fibrations above\footnote{More specifically, a Weierstrass model can be found for each fibration which admits a section and in the case of genus one fibrations without section, the Jacobian can be found \cite{artin}.} and the resulting F-theory vacua explicitly determined.

The goal of this work is to study such multiple fibrations in explicit CY geometries and to enumerate the F-theory vacua they lead to, as well their relationship to string dualities. In this work we focus primarily on correspondences between F-theory and M-theory vacua in $5$- and $6$-dimensions and heterotic/F-theory duality in $8$-, $6$- and $4$-dimensions. Related questions about multiple fibrations and weakly coupled Type IIB orientifold limits \cite{Sen:1996vd,Sen:1997gv,Aluffi:2007sx,Collinucci:2008zs,Collinucci:2008pf} or more general weakly coupled limits \cite{Clingher:2012rg,Heckman:2013sfa}, we leave to future work.

In order to have explicit examples of multiply fibered geometries, the tools developed here and in a companion paper \cite{us_to_appear} are illustrated using the dataset of CY manifolds constructed as complete intersections in products of projective spaces (CICYs) \cite{Candelas:1987kf,Gray:2013mja,Gray:2014kda,Gray:2014fla,Constantin:2016xlj}. However many of the tools and observations could equally well be applied to complete intersections in toric varieties \cite{Kreuzer:2000xy,Rohsiepe:2005qg,Braun:2011ux} or the recently constructed gCICY manifolds \cite{Anderson:2015iia,Anderson:2015yzz,Berglund:2016yqo}. A complete survey of multiple fibrations in the context of CICY $3$- and $4$-folds is currently underway \cite{usscanning}. Here we will focus on highlighting the types of multiple fibrations that can arise and relate these geometric results to known and novel string dualites.

\vspace{.2cm}

Central results of the present work include the following:
\begin{itemize}
\item We relate distinct F-theory vacua associated to different fibers in \eref{manyfibs} via their shared M-theory limits. Because they all arise from a single CY $n$-fold (and its associated complex structure moduli space), we find networks of distinct theories that all inherit their infinitesimal deformations from the same manifold. They are all limits of one moduli space of theories.
\item The possible collection of linked fibrations in \eref{manyfibs} can frequently include \emph{non-flat fibers} \cite{Morrison:1996na,Morrison:1996pp,Candelas:2000nc} and hence the effective physics includes superconformal theories (SCFTs) coupled to the usual ${\cal N}=1$ supergravity + gauge theory in $6$-dimensions.
\item In the case of non-flat fibers, we find Weierstrass models and verify the presence of superconformal loci in the associated discriminant locus arising at higher co-dimension in the base geometry. Systematic studies of such superconformal loci in the literature to date have focused on blowing up the base geometry to obtain smooth CY resolutions. The geometries with non-flat fibrations studied here provide alternate CY resolutions and thus may have interesting consequences for systematic constructions of fibered CY geometries (see e.g. \cite{Johnson:2016qar}).
\item In the context of heterotic/F-theory duality (and heterotic/Type IIA duality) we observe the important role played by \emph{nested fibrations}. To study F-theory, we require the existence of an elliptic fibration and in order to have a heterotic dual one must further demand a $K3$ fibration (usually also with section). The compatibility of such fibrations --  i.e. the $K3$ fiber itself admitting an elliptic fibration -- plays a key role in the duality map \cite{Friedman:1997yq}. In the presence of \emph{multiple $K3$ fibrations} the role of nested fibrations becomes important. We explore the consequences in heterotic/F-theory duality for multiple $K3$ fibrations (of a single CY geometry $Y_{n+1}$) and study three important classes of fibration:
\begin{enumerate}
\item \emph{Case 1}: $Y_{n+1}$ contains multiple $K3$ fibrations -- all sharing an elliptic fibration. In this case there is a single F-theory vacuum dual to multiple heterotic backgrounds.
\item \emph{Case 2}: $Y_{n+1}$ contains multiple $K3$ fibrations with distinct elliptic fibrations. In this case we find distinct F-theory vacua, each with a heterotic dual all leading to the same effective theory upon dimensional reduction on $S^1$.
\item \emph{Case 3}: $Y_{n+1}$ contains a single $K3$ fibration with multiple elliptic fibrations. In this case there are multiple F-theory/heterotic backgrounds possible, all associated to the \emph{same} heterotic manifold $X_{n}$.
\end{enumerate}
\item Most of the heterotic/F-theory dualities we study involve multiple fibrations appearing in the F-theory geometry (as above). However, we also find novel dualities by considering the role of multiple elliptic fibrations in $4$-dimensional, ${\cal N}=1$ compactifications of the heterotic string. In this case the heterotic effective theory is independent of the choice of an elliptic fibration -- leading to a collection of possibly distinct dual F-theory geometries, $Y_{4}$ and G-flux, all of which share the same EFT.
\end{itemize}

The structure of this paper is as follows. In Section \ref{sec:geometry} we lay out a brief geometric survey of elliptic and $K3$-fibrations in known datasets of CY $3$-folds and $4$-folds and observe that the presence of \emph{multiple fibrations} appears to be ubiquitous in all known systematic constructions of CY $n$-folds. In Section \ref{m_f_duals} we begin our study of multiple fibrations by exploring the consequences of such geometries for M-/F-theory correspondences in $5$/$6$-dimensions. F-theory vacua associated to distinct elliptic fibrations of a single CY $3$-fold are systematically analyzed. As expected, dimensionally reducing these theories on $S^1$ and moving to the Coulomb branch leads to the same $5$-dimensional theory in all cases. We find networks of theories linking F-theory vacua with Abelian, non-Abelian gauge symmetries and even theories coupled to SCFTs. Examples of CY $3$-folds with higher rank Mordell-Weil group (i.e. ${\rm rk}(MW)=4$) are presented. In Section \ref{section:hetf} we turn to heterotic/F-theory duality (and heterotic/type IIA duality in $4$-dimensions) and study possible structures for multiple $K3$ fibrations. We review how many known string dualities -- including $E_8 \times E_8$/$SO(32)$ heterotic duality \cite{Candelas:1997pq} in $8$-dimensions and the well-known $6$-dimensional duality of Duff, Minasian and Witten \cite{Duff:1996rs} can be realized geometrically via multiple fibrations. Moreover, we find examples of new dualities in $6$- and $4$-dimensions by considering the structure of nested elliptic and $K3$ fibrations in geometries serving as backgrounds for both the heterotic theory and F-theory. In Section \ref{conc} we summarize our results and outline a host of other areas in which multiple fibrations in CY manifolds may play a role in new physics. The Appendices provide a collection of useful results on genus-one fibered CY manifolds and their discriminant loci.

We turn now to explicit studies of multiple elliptic and $K3$-fibrations in known datasets of smooth CY geometries.

\section{The Geometry of Multiple Fibrations}\label{sec:geometry}

It appears that the vast majority of all known Calabi-Yau manifolds are genus-one fibered \cite{Rohsiepe:2005qg,Johnson:2014xpa,Gray:2014fla,Johnson:2016qar,Candelas:2012uu}. It is also suspected that they are multiply fibered, that is that they can be written in more than one way as a genus-one fibration. Indeed this has been shown to be true in the case of CICY three and four-folds \cite{Gray:2014fla,usscanning}. 

As an example consider the following configuration matrix.
 \begin{eqnarray} \label{thisone}
 \quad X =\def\arraystretch{1.2}\left[\ba{c|cc} 
\IP^1 &  1 & 1 \\
\IP^2 &  1 & 2 \\ 	
\IP^1 &  1 & 1 \\
\IP^1 & 1 & 1 \\ 
\ea\right]
 \end{eqnarray}
This matrix defines a family of Calabi-Yau manifolds in the ambient space $\IP^1 \times \IP^2 \times \IP^1 \times \IP^1$. Each column of integers specifies a defining relation of the manifold by giving its polynomial multi-degree in the homogeneous coordinates of the ambient projective spaces. One can see the multiple fibrations in the configuration matrix (\ref{thisone}) as follows. Consider splitting the configuration matrix up into two pieces, one describing the base and the other the fiber. In this example we can achieve this in two ways.
\beq \label{thelad}
\def\arraystretch{1.2}\left[\ba{c|cc} 
\IP^1 &  1 & 1 \\
\IP^2 &  1 & 2 \\ \hdashline
\IP^1 &  1 & 1 \\
\IP^1 & 1 & 1 \\ 
\ea\right]   \;\;\ , \;\;\;\;\;\;\; 
\def\arraystretch{1.2}\left[\ba{c|cc} 
\IP^1 &  1 & 1 \\
\IP^1 & 1 & 1 \\ 
\IP^1 &  1 & 1 \\ \hdashline
\IP^2 &  1 & 2 \\
\ea\right]  \ .
\eeq
In the above the first rewriting of the configuration matrix (\ref{thisone}) is as a genus one fibration over the base $\IP^1\times \IP^1$. To see this, consider picking any point on the direct product of the last two $\IP^1$'s in the ambient space. Substituting the coordinates of this point into the two defining relations we would obtain a specific complex structure for two defining relations depending only upon the coordinates in the first two projective space factors. The degrees of the equations in the remaining variables are described by
\begin{eqnarray}
\def\arraystretch{1.2}\left[\ba{c|cc} 
\IP^1 &  1 & 1 \\
\IP^2 &  1 & 2 \\ \ea\right]  \ .
\end{eqnarray}
This is the configuration matrix of a Calabi-Yau one-fold -- a torus. As we change the point we choose in $\IP^1 \times \IP^1$ the complex structure describing the associated torus fiber will change, and so we have a non-trivial fibration of a genus-one curve over that base.

Similarly, the second configuration matrix in (\ref{thelad}) describes the configuration as a genus-one fibration over $\IP^2$. Clearly these two fibrations are inequivalent (they do not even have the same base) and thus even this very simple configuration matrix admits multiple genus-one fibrations.

\begin{figure}[!tbp]
  \centering
  \begin{minipage}[b]{0.45\textwidth}
    \includegraphics[scale=0.55]{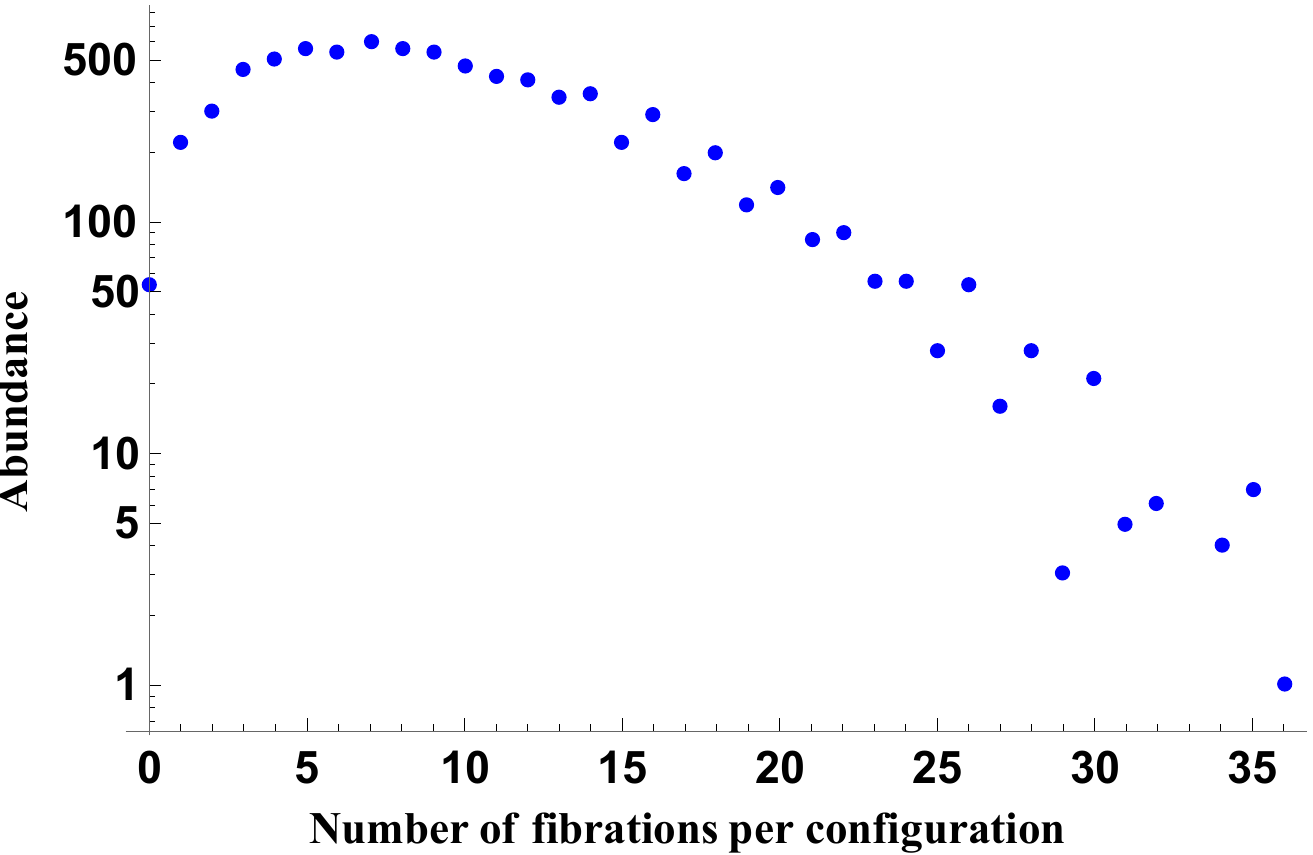}
\caption{\footnotesize  The abundance of CICY threefold configuration matrices, in the standard list, exhibiting a given number of genus one fibrations which are visible directly in the configuration matrix \cite{usscanning}.}\label{f1}
  \end{minipage}
  \hspace{0.1cm}
  \begin{minipage}[b]{0.45\textwidth}
   \includegraphics[scale=0.3]{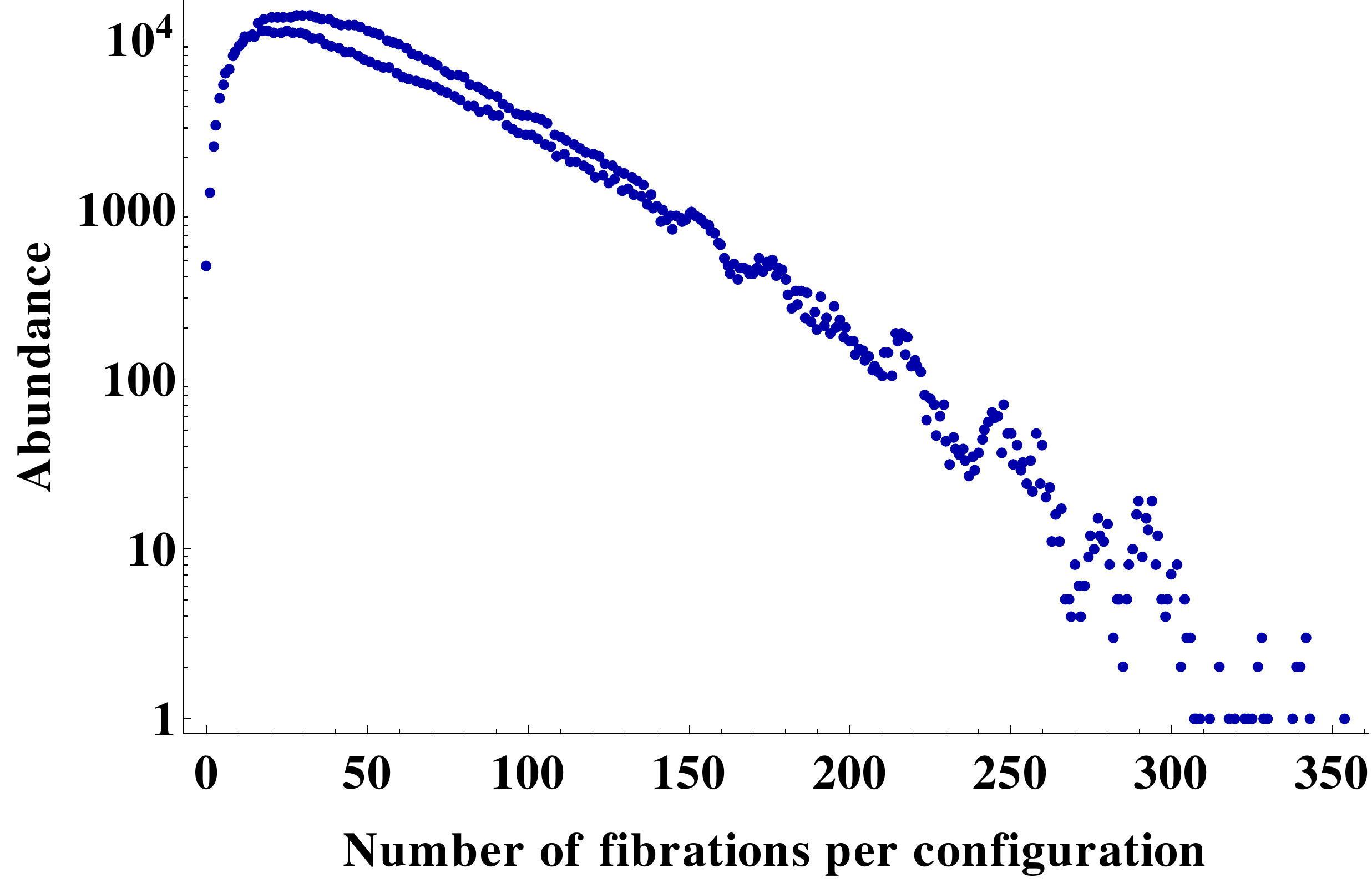}
\caption{\footnotesize  The abundance of CICY fourfold configuration matrices, in the standard list, exhibiting a given number of genus one fibrations which are visible directly in the configuration matrix \cite{Gray:2014fla}.}\label{f2}
  \end{minipage}
\end{figure}

Fibrations of the type we are describing here are referred to as Obvious Genus-One Fibrations (OGFs) as they are manifest in the configuration matrix. As was briefly mentioned above, almost all CICYs admit multiple fibrations of this kind. Of the 7,890 CICY three-fold configuration matrices 7,837 admit at least one such fibration, with the average number of inequivalent fibrations per manifold being 9.85. For the CICY four-folds 921,420 out of 921,497 cases admit such a fibration with the average manifold being OGF'd in 54.6 different ways \cite{Gray:2014fla}. Extreme cases also exist -- there is one CICY four-fold that admits 354 different OGFs \cite{Gray:2014fla}. This rich structure of multiple fibrations in CICYs is illustrated in Figures \ref{f1} and \ref{f2}.

Fibrations in CICYs are not restricted to genus-one curves. The vast majority of known CICYs of dimension $n$ are fibered by CICYs of dimension $n-1$, which are themselves fibered by CICYs of dimension $n-2$, and so forth \cite{usscanning}. For example, 7,768 out of 7,890 CICY three-folds have obvious K3 fibrations, with an average of 3.1 different K3 fibrations per configuration. Almost all of these K3 fibers are also genus-one fibered and there are 103,513 different pairs of nested genus-one fibrations within K3 fibrations in the CICY data set \cite{usscanning}. Note that this number is larger than the total number of elliptic fibrations of the CICY three-folds since a given elliptic fiber may appear in multiple different $K3$ fibrations of a given manifold. For more details, classifications and complex examples of such structure see \cite{Gray:2014fla,usscanning,us_to_appear}.

\vspace{0.2cm}
 
Clearly, the above kind of fibration analysis generalizes to complete intersections in other types of ambient spaces, and indeed some completely general statements can be made. The existence of a genus-one fibration in a Calabi-Yau $n$-fold has been conjectured by Kollar \cite{Kollar:2012pv} to be determined by the following criteria:

\vspace{0.3cm}

\noindent \emph{Conjecture \cite{Kollar:2012pv}: Let $X$ be a Calabi-Yau $n$-fold. Then $X$ is genus-one fibered iff there exists a $(1,1)$-class $D$ in $H^2(X, \mathbb{Q})$ such that $D \cdot C \geq 0$ for every algebraic curve $C \subset X$, $D^{{\rm dim}(X)}=0$ and $D^{{\rm dim}(X)-1} \neq 0$.}

\vspace{0.2cm}

\noindent In the case that $X$ is a Calabi-Yau threefold this conjecture has been proven subject to the additional constraints that $D$ is effective or $D \cdot c_2(X) \neq 0$ \cite{oguiso,wilson}. Phrased simply these criteria are characterizing the existence of a fibration by \emph{characterizing the existence of the base manifold of that fibration}. In particular, the role of the divisor $D$ above is that of one pulled back from the base, $B$, where the fibration of $X$ is written $\pi: X \to B$. The existence of $D=\pi^*(D_{base})$ makes it possible to define the form dual to points on the base (i.e. $D^{{\rm dim}(X)-1}$) which in turn determines the class of the genus-one fiber itself. This allows us to cleanly explain in general cases what we mean both by ``choosing a fibration" of $X$ and ``exchanging a given pair of fibrations" within $X$\footnote{It should be noted that the existence of a fibration structure within a smooth Calabi-Yau $n$-fold with $n>2$ is a deformation invariant quantity ({i.e.}, given a fibered manifold, every small deformation is also fibered)\cite{Kollar:2012pv,wilson,wilson2}. Indeed this must clearly be the case if the above conjecture is to make sense.}.  

\vspace{0.1cm}

As a simple example, consider the second of the two genus-one fibrations in (\ref{thelad}). This manifold has Hodge numbers $h^{1,1}=4,h^{2,1}=50$. By inspection of the K\"ahler cone of $X$ (spanned by the restriction of the hyperplanes, $D_1$, $D_2$, $D_3$, $D_4$, from each factor of the ambient space $\IP^1 \times \mathbb{P}^1 \times \mathbb{P}^1 \times \IP^2$) and from the triple intersection numbers, $d_{rst}$, of $X$, it can be readily verified that one divisor, $D$ in $X$ satisfying the criteria given above is given simply by $D_4$. For the first fibration in (\ref{thelad}) a relevant divisor is $D_3+D_4$ (where in this fibration the two divisors, $D_3$ and $D_4$, are associated to the base $\IP^1 \times \IP^1$ factors). 

\vspace{0.2cm}

In the above we have started with a description of the total space of the Calabi-Yau manifold being considered, being agnostic about what should be chosen as fiber and base. We have then identified the multiple genus-one fibrations explicitly given this starting point. Such an approach is somewhat different to that pursued in much of the F-theory literature. There, it is often the case that one simply picks a base manifold and writes down a Weierstrass model over that choice. Such a construction has the disadvantage that the multiple fibration structure might not be as easy to see as in the examples above. However, the physics of the associated F-theory model is easier to obtain in such an approach as much of the technology that has been developed in this regard revolves around a Weierstrass description. It is important, therefore, if we are studying fibrations such as those in (\ref{thelad}), that we know how to put them in Weierstrass form.

In order to put an elliptically fibered Calabi-Yau manifold in Weierstrass form, one must first obtain an explicit section to that fibration. Given a section one can then use a procedure due to Nakayama~\cite{deligne,nakayama} to blow down those components of the fibers which do not generically intersect the zero section and obtain a Weierstrass description. Importantly for the program being pursued here, the technology to perform such computations explicitly has been developed in a companion paper to this one \cite{us_to_appear} (see also \cite{cicy-package}). We refer the reader to that discussion for the details of how such analysis is performed, along with a review of how the Jacobian of a fibration is obtained and other details.

\vspace{0.3cm}

In the following Sections, we will explore many examples of geometries with multiple fibration structures and examine ways that the choice of a fibration ({i.e.}, an ``orientation" of the total space into fiber/base) can yield insights into effective theories and string dualities in various dimensions.


\section{Multiple Fibrations and the M-theory/F-theory Correspondence}\label{m_f_duals}

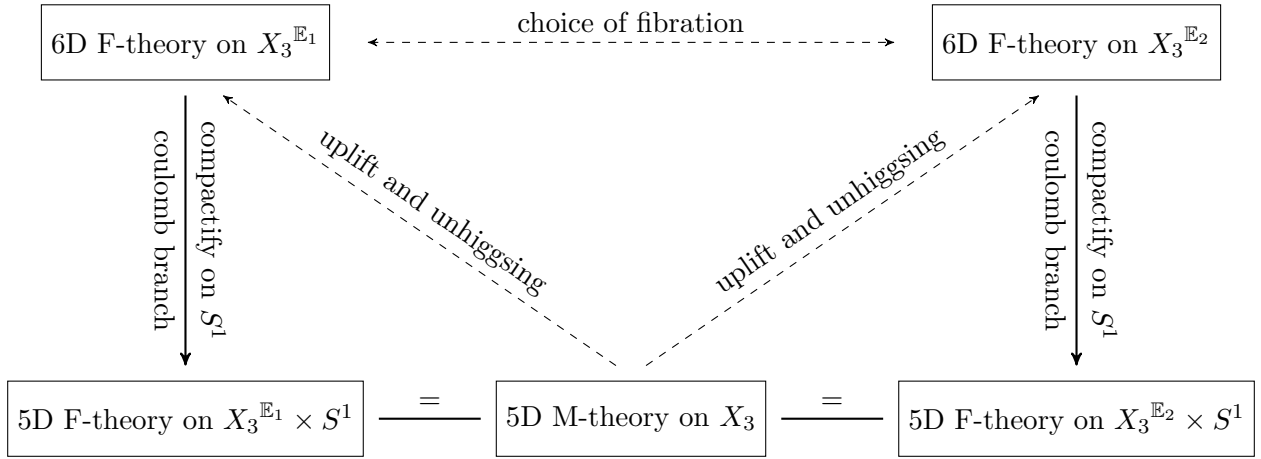
\begin{figure}[t!]
\centering
\begin{tikzpicture}[>=stealth',
box/.style={rectangle,draw,minimum width=3cm,  minimum height=1cm},
line/.style={->,shorten >=3mm, shorten <=3mm},
mylabel/.style={text width=7em, text centered} ]

\node[box] (source) { 6D F-theory  on $ {X_3}^{\mathbb{E}_1}$};
\node[box, right=8cm of source] (dest) { 6D F-theory on ${X_3}^{\mathbb{E}_2}$};
\node[box, below=4.5cm of source, anchor=center] (relay1) {5D F-theory on $ {X_3}^{\mathbb{E}_1} \times S^1$ };
\node[box, below=4.5cm  of dest, anchor=center] (relay2) {5D F-theory on  $ {X_3}^{\mathbb{E}_2} \times S^{1}$};
\node[box, below right=4.5cm and 4cm of source, anchor=center] (relay) {5D M-theory on $X_3$};

\draw  [->, thick] ([yshift=-2mm]source.south) -- node[below,sloped] {coulomb branch} node[above,sloped]{compactify on $S^1$} ([yshift=2mm]relay1.north); 

\draw [<->, dashed] ([xshift=5mm]source.east) -- node[above,sloped] {choice of fibration}  ([xshift=-5mm]dest.west); 

\draw[->, thick] ([yshift=-2mm] dest.south) -- node[below,sloped] {coulomb branch} node[above,sloped]{compactify on $S^{1}$} ([yshift=2mm]relay2.north); 

\draw [=,thick] ([xshift=2mm]relay1.east) -- node[above,sloped] {=} node[below,sloped] {}([xshift=-2mm]relay.west); 
\draw [=,thick] ([xshift=2mm]relay.east) -- node[above,sloped] {=}node[below,sloped] {}  ([xshift=-2mm]relay2.west); 

\draw[->, dashed] ([yshift=2mm,xshift=-2mm] relay.north) -- node[below,sloped] {} node[above,sloped]{uplift and unhiggsing} ([yshift=-2mm, xshift=5mm]source.south); 

\draw[->, dashed] ([yshift=2mm,xshift=2mm] relay.north) -- node[below,sloped] {} node[above,sloped]{uplift and unhiggsing} ([yshift=-2mm, xshift=-5mm]dest.south);

\end{tikzpicture}
\caption{ \it Correspondences between different compactifications of M-theory and F-theory associated to geometries with the same total space.}
\label{f:duality}
\end{figure}

In this Section we will consider the consequences of multiple elliptic fibrations in a single smooth Calabi-Yau manifold for the correspondence between F-theory and M-theory compactifications (for recent work on this correspondence see \cite{Bonetti:2011mw}). More precisely, we will consider F-theory compactified on a multiply elliptically fibered, Calabi-Yau threefold $X_3$. For each choice of elliptic fibration, we can obtain a Weierstrass model associated to the original geometry by the procedures discussed in \cite{us_to_appear}. The associated F-theory compactification then gives rise to a particular ${\cal N}=(1,0)$ effective theory in $6$-dimensions.  While the details of the effective theory obtained in this manner depends upon the choice of elliptic fibration and thus the F-theory torus, these $6$-dimensional theories all share the same M-theory limit. That is, if we compactify further on an $S^1$ and go to the Coulomb branch of the resulting $5$-dimensional ${\cal N}=2$ theory\footnote{With $8$ real supercharges.}, then all of the seemingly disparate theories corresponding to different fibrations result in the same physics. The resulting $5$-dimensional gauge theory is, in fact, the same as that obtained by compactifying M-theory on the original Calabi-Yau threefold $X_3$. This relationship between different F-theory and M-theory compactifications is depicted schematically in Figure \ref{f:duality}.

\vspace{0.1cm}

In studying examples of this phenomenon, we will need to extract the physics of the various effective theories from the associated geometries. We will begin by briefly reviewing how this is achieved, before proceeding to give explicit examples of families of theories descending from multiply elliptically fibered CICY three-folds.
	
\subsection{Matter in $6$-dimensional F-theory compactifications}

We will begin by computing and comparing the massless spectrum of the lower dimensional theories that we consider. For  a review of the the well-studied effective physics of F-/M-theory in $6$/$5$-dimensions see \cite{Bonetti:2011mw,Grimm:2012yq}. Briefly, there are a variety of multiplets that arise in the effective theories obtained in six dimensions by compactifying F-theory on an elliptic fibration described by a Weierstrass model. These include
\begin{itemize}
\item A single gravitational multiplet.
\item Tensor multiplets. In the simple cases that we will consider, the multiplicity of tensor multiplets are determined by the topology of the base of the Weierstrass model.
\begin{eqnarray}
n_T = h^{1,1}(B) -1
\end{eqnarray}
\item The number of vectors is determined by the gauge group associated to the low energy effective theory. This can be decomposed into two components.
\begin{eqnarray}
G= \prod_{A} G_A \times U(1)^{r}
\end{eqnarray}
Here $\prod_{A} G_{A}$ is the non-Abelian factors in the gauge group which can be found by an analysis of the codimension one behavior of the discriminant locus of the elliptically fibered threefold \cite{kodaira, tate}. Each such factor $G_A$ is associated to an irreducible component of the discriminant, which we denote by ${\cal S}_A$. The Abelian factors, $U(1)^r$, are associated with the Mordell-Weil group of rational sections of the fibration being considered. Indeed $r={\rm rk}(MW(X))$ is the rank of this group. We will defer a more detailed computation of these geometrical quantities to the examples sections and will content ourselves here by noting that the number of vectors is given by 
\begin{eqnarray}
n_V = r + \sum_A \textnormal{dim} (  \textnormal{\bf adj}\; G_{A}) \;.
\end{eqnarray}
\item Finally the number of hypermultiplets, $n_H$, in the $6$-dimensional theory is given by (see e.g. \cite{Cvetic:2015txa})
\begin{eqnarray} \label{nh}
n_H= n_H^{(\textnormal{codim 2})} + \sum_A g_A (\textnormal{dim} (  \textnormal{\bf adj}\; G_{A}) - \textnormal{rk}(G_A)) +h^{2,1}(X_3)+1
\end{eqnarray}
Here, $g_A$ is the genus of the divisor ${\cal S}_A$ in the base, which can be determined by the following formula,
\bea
g_A= 1+ \frac{1}{2} {\cal S}_{A} \cdot ({\cal S}_{A}+ K_{B_2}),
\eea
where $K_{B_2}$ is the canonical divisor of the base.

The contribution $n_H^{(\textnormal{codim 2})}$ in \eref{nh} arises from fibers over codimension two points in the base. For example, at specific codimension-two loci in the base, where self-intersections of the discriminant locus occur, the rank of the non-Abelian singularity of the fiber will enhance. Such enhancements generically lead to matter states and their representations can be obtained from the branching rules from the adjoint representation of the enhanced gauge algebra into the low energy gauge group of the theory \cite{Katz:1996xe}. Counting $U(1)$ charged matter that is neutral under the non-Abelian gauge group factors is somewhat more involved. Such degrees of freedom are associated with $I_2$ fibers appearing over nodes of the self-intersection of the $I_1$ locus of the discriminant \cite{Braun:2014oya}. The computation of this $U(1)$-charged matter is discussed in detail in Appendix \ref{sec:plane}.
\end{itemize}

\subsubsection{Anomaly cancelation}

The spectrum that is found from a consistent F-theory compactification is always anomaly free. Given this, freedom from anomalies is an excellent check of the calculations we will present. In this paper we will concentrate on anomalies involving only gravitational and non-Abelian degrees of freedom. The relevant constraints on the spectrum are then as follows. Firstly, in terms of pure gravitational anomalies we have \cite{Morrison:1996na,Morrison:1996pp}
\begin{eqnarray}
\label{eq:6dAnomalies}
n_H-n_V+ 29n_T=273\,, \\\label{eq:6dAnomalies2}
9-n_T=a\cdot a \,.
\end{eqnarray}
Here the vector $a^\alpha$ is defined in terms of an expansion of the canonical divisor of the base $K_{B_2}$ with respect to a fixed basis of divisors in that space, $\D_\alpha^{\rm b}$ where $\alpha=1, \dots, h^{1,1}(B_2)$.
\bea
K_{B_2}&=& a^\alpha D^{\rm b}_{\alpha} 
\eea

Cancellation of the mixed non-Abelian/gravitational anomalies leads to the following constraints \cite{Green:1984bx,Sagnotti:1992qw,Kumar:2010ru},
 \bea \label{naganom}
 \textstyle{ \frac{1}{6}}\left( A_{{\rm \bf adj}_A}-\sum_{{R}} x_{{R}} A_{{ R}}\right)=a \cdot \left( \frac{b_{A}}{\lambda_A}\right) \;,
 \eea
 and cancelation of the pure non-Abelian anomalies leads to,
 \bea \label{naanom}
 B_{{\rm \bf adj}_A} - \sum_{{R}} x_{{R}} B_{{R}} = 0  \;\;\;\;\;\;\;\;\;\;\, \\  
\textstyle{\frac{1}{3}} \left( \sum_{{R}} x_{{R}} C_{{R}}-  C_{{\rm \bf adj}_A}  \right) = \left( \frac{b_A}{\lambda_A}\right)^2  \;.
\eea
In both \eref{naganom} and \eref{naanom}, $x_R$ represents the number of matter fields in the $R$ representation of gauge group $G_A$. The factors $\lambda_A$ are, for the cases we will need in this paper, $\lambda_{SU(N)}=1$. The vector $b_A$ is defined by expanding the divisor associated to the gauge group factor $G_A$ in terms of a basis of divisors:
\begin{eqnarray} \label{bdef}
\mathcal{S}_{A}^{{\rm b}} &=& b^\alpha_A  D^{\rm b}_\alpha \;.
\end{eqnarray}
Finally, the coefficients $A_R,B_R$ and $C_R$ are group-theoretic in nature, being defined by the following formulae,
\bea
  {\rm tr}_R\, F^2 = A_R\,{\rm tr}\,F^2, \quad\quad  {\rm tr}_R\, F^4 = B_R\,{\rm tr}\,F^4 + C_R\,({\rm tr}\,F^2)^2 ,
  \eea
  for any representation $R$, where the unlabeled trace is with respect to the fundamental representation. For ${SU}(N)$ with $N>3$, these coefficients take the following values:
\beq
\text{
\begin{tabular}{|c|c|c|c|c|} \hline
 Representation & Dimension & $A_{{R}}$ & $B_{{R}}$ & $C_{{R}}$   \\ \hline
 Fundamental & $N$ &1 & 1& 0 \\ \hline
 Adjoint & $N^2-1$ &$2N$ & $2N$  & 6 
 \\ \hline
 Antisymmetric& $N(N-1)/2$ & $N-2$ & $N-8$ & 3 
 \rule{0pt}{1Em}\\ \hline
\end{tabular}
}
\eeq
For $SU(2)$ and $SU(3)$, the $A_{{R}}$ coefficients in the above table are still correct, while the $B_{{R}}$ vanish. The coefficient $C_{{R}}$, for $SU(2)$ and $SU(3)$, can be computed as the combination $C_{{R}}+\tfrac12 B_{{R}}$ using the values for $B_{{R}}$ and  $C_{{R}}$ in the above table.

\subsection{Matter of $5$-dimensional M-theory compactifications and $6D \to 5D$ reductions}	\label{mred}

The matter content of the $5$-dimensional theory obtained when M-theory is compactified on a smooth Calabi-Yau threefold $X_3$ is rather straightforward in comparison to the structure seen in the previous subsection. We have the following (see e.g. \cite{Cadavid:1995bk}):
\begin{itemize}
\item A single gravitational multiplet.
\item Vector multiplets with multiplicity given by
\begin{eqnarray} \label{5dV}
n_V^{(5D)}=h^{1,1}(X_3)-1\;.
\end{eqnarray}
\item  Hypermultiplets with multiplicity given by
\begin{eqnarray} \label{5dH}
n_H^{(5D)}=  h^{2,1}(X_3)+1\;.
\end{eqnarray}
\end{itemize}
	
If we start with one of the $6$-dimensional theories described in the previous subsection, compactify further on an additional $S^1$, and go to the Coulomb branch, we obtain an ${\cal N}=2$, $5$-dimensional supergravity theory. The resulting multiplet content is determined in terms of the spectrum seen in $6$-dimensions. In particular, we will find a theory with
\begin{eqnarray} \label{5dV2}
n_V^{(5D)}&=& n_V+ n_T+1 \\ \label{neut}
n_H^{(5D)}&=& n_H^{\textnormal{neutral}}
\end{eqnarray}
where in a slight abuse of notation we use the same notation for the vector and hyper multiplets in this theory as we did in that obtained by dimensional reduction of M-theory. 

In fact, given what we have learned about the spectra of the $6$- and $5$-dimensional theories, it is easy to see that the F-theory/M-theory correspondence always works at the level of multiplet content, independently of which fibration we pick to be the F-theory torus. For the hypermultiplets, in going to the Coulomb branch it is clear that all of the complex structure of $X_3$ will become neutral. These, in addition to the universal hypermultiplet, then give $n_H^{\textnormal{neutral}}=h^{2,1}(X_3)+1$. Comparing this to \eref{neut} and \eref{5dH} then makes it obvious that we obtain the correct number of $5$-dimensional hypermultiplets to agree with the associated M-theory compactification.
					
In going to the Coulomb branch the gauge group in the $6$-dimensional theory is broken down to its maximal Abelian subgroup. After undergoing such a process we are left with $n_V = \sum_A \textnormal{rk}(G_A) +r$ vector multiplets. According to \eref{5dV2}, we should then see 
\begin{eqnarray}
n_V^{(5D)} = \sum_A \textnormal{rk} (G_A) +r + n_T+1 = \sum_A \textnormal{rk} (G_A) +r + h^{1,1}(B)
\end{eqnarray}
vector multiplets in the $5$-dimensional compactification of this theory on a circle. According to the formula of Shioda-Tate-Wazir \cite{shioda,shioda2,COM:213767}, applied to this example we have 
\begin{eqnarray} \label{stw}
h^{1,1}(X_3)= h^{1,1}(B) + \sum_A \textnormal{rk}(G_A)+r+1 \;.
\end{eqnarray}
Finally then, we arrive at the expression,
\begin{eqnarray}
n_V^{(5D)} = h^{1,1}(X_3)-1\;,
\end{eqnarray}
which agrees precisely with the result of the M-theory reduction as given in \eref{5dV}.
		
\vspace{0.2cm}

Having described the basics of the M-theory/F-theory correspondence in generality, we now move on to construct some explicit examples of this structure in multiply fibered Calabi-Yau three-folds.		

\subsection{Example of a (non-)Abelian/Abelian correspondence}

As our first example of the M-theory/F-theory correspondence let us consider the family of Calabi-Yau three-folds described by the following configuration matrix.
\begin{eqnarray} \label{case1}
X_3=\left[\begin{array}{c|cccc}
\IP^{1} & 1 & 1 & 0 & 0\\
\IP^{2} & 1 & 0 & 2 & 0  \\
\IP^2 &   0 & 1  &  1 & 1 \\
\IP^2 & 1 & 0 & 1 & 1 
\end{array}\right] 
\end{eqnarray}
As we shall see, this configuration admits 4 different OGFs which we will consider in turn in what follows. The Hodge numbers of \eref{case1} are $h^{1,1}(X_3)=4$ and $h^{2,1}(X_3)=47$. Given this, and \eref{5dV} and (\ref{5dH}), we see that M-theory compactified on such a manifold leads to a $5$-dimensional theory containing $n_V^{(5D)}=3$ vector multiplets and $n_H^{(5D)}=48$ hyper multiplets. 

\subsubsection{A fibration $\mathbb{E}_1$ with $G= SU(2) \times U(1)$}
The first OGF we will consider of the configuration \eref{case1} is as follows.
\beq\label{e.g.trivial}
X_3^{\mathbb{E}_1}=\left[\begin{array}{c|cccc}
\IP^{1} & 1 & 1 & 0 & 0\\
\IP^{2} & 1 & 0 & 2 & 0  \\
\IP^2 &   0 & 1  &  1 & 1 \\\hdashline
\IP^2 & 1 & 0 & 1 & 1 
\end{array}\right] \ 
\eeq 
Here the base of the fibration is $\IP^2$ and thus $h^{1,1}(B)=1$. We wish to obtain the Weierstrass form associated to this fibration, and from this obtain the gauge group and matter content of the associated F-theory model. To achieve this, we will make use of the techniques developed in the companion paper to this one \cite{us_to_appear}.

\subsubsection*{Obtaining Weierstrass form and the gauge group}

To put (\ref{e.g.trivial}) in Weierstrass form, we first need to isolate a section of the fibration. Following Section 2.2 of \cite{us_to_appear}, the first step in doing so is to obtain a set of possible (or ``putative") sections. These are divisor classes that intersect a generic fiber once and which satisfy an additional topological constraint that follows from the requirement that the divisor be birational to the base (in this case $\mathbb{P}^2$). We then follow the discussion of Section 2.3 of that same paper in order to isolate a divisor $S_0$ such that $h^*(\cO(S_0))=(1,0,0,0)$ before obtaining an explicit description of that divisor class which is proven by direct computation to describe a section to the fibration. 

In the case at hand, all of this analysis leads us to conclude that 
\beq
{\cal O}(S_0) = {\cal O}_{X_3^{\mathbb{E}_1}} (-1,1,0,1)
\eeq
is a line bundle corresponding to a suitable divisor which describes the image of a rational section of the fibration (\ref{e.g.trivial}). Since $h^{1,1}(X_3)=4$ and the manifold is ``favorable" \cite{Anderson:2008uw} in that a basis of divisors restricts from the ambient space, the degree of the line bundle is expanded in a basis of restricted hyperplanes from the ambient $\mathbb{P}^n$ factors.

Once a section of the fibration $S_0$ has been isolated, the next step is to follow the Deligne procedure \cite{deligne,nakayama} to put (\ref{e.g.trivial}) in Weierstrass form. This procedure is described for such examples in detail in Section 3.1 of \cite{us_to_appear}. In the case at hand we obtain a specific Weierstrass from, 
\begin{eqnarray} \label{wsf}
y^2= x^3 + f x^2 z^4 + g z^6,
\end{eqnarray}
the details of which are lengthy and dependent upon the initial choice of complex structure made for the defining relations of (\ref{case1}). In particular, the functions of the base coordinates $f\in H^0(B, K_B^{-4})$ and $g\in H^0(B, K_B^{-6})$ are rather complex and obtained numerically in our analysis. Due to their length, we refrain from including them explicitly here.

Given a Weierstrass form such as \eref{wsf}, the discriminant locus of the fibration can be determined simply as 
\begin{eqnarray}
\Delta_W = 27 g^2 + 4 f^3
\end{eqnarray}
Computing $\Delta_W$ in our case, we find that this discriminant locus factorizes as follows.
 \bea \label{loci}
 \Delta_W= (\Delta_{I_2})( \Delta_{I_1})
 \eea
That is the discriminant is made up of an $I_2$ and an $I_1$ locus according to Kodaira's classification \cite{kodaira}. We thus see that the F-theory compactification associated to this Weierstrass model will have an $SU(2)$ factor. 

Finally, we can use the theorem by Shioda-Tate-Wazir \cite{shioda,shioda2,COM:213767}, together with the above analysis and the fact that the fibration (\ref{e.g.trivial}) is flat, in order to determine the Abelian component of the gauge group in this example. From Shioda-Tate-Wazir, applied to this example we know that
\begin{eqnarray}
h^{1,1}(X_3)= h^{1,1}(B) + \sum_A \textnormal{rk} (G_A)+r+1 \\ 
\Rightarrow r = h^{1,1}(X_3) -h^{1,1}(B) - \sum_A \textnormal{rk} ( G_A)-1 = 4-1-1-1 =1 \;.
\end{eqnarray}
 Thus we see that we have a MW group of rank one and thus a single abelian factor in the gauge group. We conclude that the gauge group of the theory obtained by compactifying F-theory on this Weierstrass model is $SU(2) \times U(1)$.
 
It should be noted that the loci in \eref{loci} are obtained numerically in the above analysis and so it is useful to confirm this result using another technique for calculating the discriminant locus, to which we now turn.
 
 \vspace{0.2cm}

The discriminant of the Jacobian of the fibration (\ref{e.g.trivial}) is identical as a locus in the base $\mathbb{P}^2$ to that obtained from the Weierstrass model described above. Indeed, as described in Section 3.2 of \cite{us_to_appear}, the process of ``contractions" of the CICY fiber \cite{Candelas:1987kf} in (\ref{e.g.trivial}), followed by taking the Jacobian ($J(X)$), also turns out to give rise to the same discriminant locus and also benefits from being easier to compute. We follow such a procedure below in several different ways, in order to get codimension 1 or 2 fiber whose Jacobian may be taken easily using results in the literature \cite{Braun:2014qka}. We will consider two different contractions of the configuration matrix (\ref{e.g.trivial}). The first of these involves contracting the $\IP^2$ in the third row of the configuration matrix. The second will involve contracting the first $\IP^1$ and then the $\IP^2$ in the third row.

We first consider contracting the $\IP^2$ in the third row of the fiber. Under such a blow down, the configuration matrix becomes:
\beq \label{ladprime}
{X_{3}'}^{\mathbb{E}_1}=\left[\begin{array}{c|cc}
\IP^1 & 1 & 1 \\
\IP^2 & 1 & 2 \\\hdashline
\IP^2 & 1 & 2
\end{array}\right] \ 
\eeq 
This codimension two fiber is labeled by PALP ID $(4,0)$ \cite{Kreuzer:2002uu} and we can directly obtain the associated Jacobian by using the results of \cite{Braun:2014qka}. One can check explicitly that that the two discriminant loci $\Delta_W$ and $\Delta_{bl_1}$ associated to the Jacobian of (\ref{ladprime}) agree exactly. Moreover, it can be explicitly verified that $\Delta_{bl_1}$ factorizes as an $I_2$ locus times an $I_1$ locus.
\bea \label{lineloc}
\Delta_{bl_1}  \sim  p_1^2    \,p_{34} = (\Delta_{I_2})(\Delta_{I_1}).
\eea
Here, $p_1$ and $p_{34}$ are polynomials of degree $1$ and $34$ respectively in the homogeneous coordinates of the base $\mathbb{P}^2$. We also find that $f$ and $g$ vanish linearly on the locus where $p_1=0$ and thus we do indeed obtain an $I_2$ locus.

As a second method for blowing down the configuration (\ref{e.g.trivial}) we can contract the first $\IP^1$ and the $\IP^2$ in the third row. This leads to a configuration matrix,
\beq
{X_3''}^{\mathbb{E}_1}=\left[\begin{array}{c|c}
\IP^2 &   3 \\\hdashline
\IP^2 & 3
\end{array}\right]  \;.
\eeq 
We follow the procedure described in \cite{artin,Braun:2014qka} to get the discriminant $\Delta_{bl_2}$ of the associated Jacobian. Once again, we find it factorized as:
\bea
\Delta_{bl_2} \sim  p_1^2   \,p_{34} = (\Delta_{I_2})(\Delta_{I_1}).
\eea
Once again $\Delta_W$ of the original manifold exactly matches $\Delta_{bl_2}$, and vice versa and we conclude that all three of the discriminants we have obtained are identical as expected \cite{us_to_appear}.
\bea
\Delta_{W} \sim \Delta_{bl_1} \sim \Delta_{bl_2}  
\eea

\vspace{0.2cm}

In short, via numerous complementary analyses we have been able to show that the gauge group of this compactification is $SU(2) \times U(1)$ and we have obtained an explicit expression for the discriminant locus of the Weierstrass model. We now turn to an identification of the matter spectrum seen in $6$-dimensions.

\subsubsection*{Matter content and anomaly cancelation in $6$-dimensions}

From \eref{nh} we see that there are several contributions to the number of hypermultiplets in a compactification such as the one we are considering here. In the case at hand, it can be seen from the analysis of the previous section that the genus of the $I_2$ locus $g_1=0$. This is a straightforward consequence of the fact that this locus is described as a linear constraint $p_1=0$ inside $\mathbb{P}^2$ and is thus simply a $\mathbb{P}^1$. As a result, the second term in (\ref{nh}) does not contribute to the number of hypermultiplets. We also know that $h^{2,1}(X_3)=47$ in our case and the Weierstrass model is birational to the original configuration (\ref{e.g.trivial}). Thus we need only compute the remaining contribution $n_H^{(\textnormal{codim } 2)}$ which is associated with points (co-dimension two loci) in the base $\mathbb{P}^2$.

One contribution to the number of hypermultiplets comes from charged matter arising at the intersections of the  $I_2$ and $I_1$ loci of the discriminant. In total these two loci intersect $34$ times at $28$ distinct points, $22$ with intersection multiplicity $1$ and $6$ points with intersection multiplicity $2$.  The $22$ points associated with single intersection multiplicities are associated with $I_3$ degenerations of the fiber while the $6$ points of double intersections correspond to type III singularities of the fiber. 

The $22$ type $I_3$ fibers correspond to transverse intersections of the $I_1$ and $I_2$ loci.	Examining the decomposition of the adjoint of $SU(3)$ to $SU(2)$, given by the branching rule ${\bf 8}={\bf 1}+{\bf 2} + { \bf \bar 2} + {\bf 3}$, we see that we expect each such intersection to contribute a hypermultiplet associated to the ${\bf 2}$ representation of $SU(2)$. Thus, in total, we expect such charge matter to contribute $22 \times 2  =44$ hypermultiplets to the low energy spectrum. The type III fibers are located at intersection points where the $I_2$ and $I_1$ loci are tangent, and we do not expect such points to contribute further charged matter.

\vspace{0.2cm}

Next, we must count the matter that is uncharged under $SU(2)$ but which is charged with respect to the abelian factor in the gauge group. These are counted by the number of nodes of the $I_1$ locus. In this example, using the techniques described in Appendix \ref{sec:plane}, we find that there are $185$ such nodes, each of which corresponds to a single $U(1)$ charge hypermultiplet. The important question of determining the $U(1)$ charges of this matter, we leave to future investigation.

\vspace{0.2cm}

Now that we have determined the matter content of the $6$-dimensional theory associated to the Weierstrass model of the fibration (\ref{e.g.trivial}), it can be verified that the result is consistent with anomaly cancelation. We begin by considering the gravitational anomaly cancelation condition (\ref{eq:6dAnomalies}).

In the case at hand the base manifold is $\IP^2$ and thus $h^{1,1}(B)=1$, which indicates that $n_T=0$ in the $6$-dimensional $\cN=(1,0)$ effective theory. Given that the gauge group is $SU(2) \times U(1)$ we have $n_V=4$. Finally, combining all of the contributions listed above to the total number of hypermultiplets, we have from (\ref{nh}) that $n_H=44+185+0+47+1=277$. Combining these three values we find that $n_H-n_V+29n_T = 277-4+0=273$ as desired.

The cancelation of the non-Abelian gauge anomalies and mixed non-Abelian/gravitational anomalies can also be seen to hold correctly. When the base is $\IP^2 $ we have that $a= -3$ and \eref{eq:6dAnomalies2} is satisfied. Furthermore, $b=1$ given that the $I_2$ locus is a hyperplane in $\mathbb{P}^2$. The anomaly cancelation conditions (\ref{naganom}) and (\ref{naanom}) then simplify to the following.
  \bea
  \label{eq:anomaly}
  18 = \sum_{i} A_{R_i} - A_{adj}, \quad\quad 0= \sum_{i} B_{R_i}-B_{adj}, \quad\quad 3=\sum_{i} C_{R_i}-C_{adj}
  \eea
Recall that in our case, in terms of $SU(2)$ charged matter, we have $22$ multiplets in the doublet representation. The relevant coefficients appearing in the above equations are then:
 \bea
 A_{\bf 2} =1, \quad A_{adj} =4, \quad  B_{\bf 2} =0,  \quad B_{adj} =0, \quad C_{\bf 2} =\frac{1}{2},  \quad C_{adj} =8.
  \eea 
These numbers satisfy the anomaly cancelation condition (\ref{eq:anomaly}). Thus we see that the matter content we have obtained is consistent with $6$-dimensional anomaly cancelation, providing a highly non-trivial check of the tools we have used to analyze the complete intersection manifold above.
  
\vspace{0.2cm}  
  
The final comment that we should make in this subsection is that the $6$-dimensional $SU(2) \times U(1)$ theory that we have obtained consistently reduces to the expected $5$-dimensional supergravity, derived from the compactification of M-theory on (\ref{case1}) if we go to the Coulomb branch and dimensionally reduce on a circle. We have $h^{2,1}(X_3)+1$ neutral hypermultiplets in $6$-dimensions which give rise to the same number of hypermultiplets in the $5$-dimensional supergravity. In addition, the 4 vector multiplets that we have in $6$-dimensions reduce to just two massless degrees of freedom in going to the Coulomb branch of the theory. Dimensionally reducing to $5$-dimensions, and remembering that we have no tensor multiplets in $6$-dimensions, (\ref{5dV2}) tells us that this will lead to $3$ vector multiplets in that theory. This is indeed $h^{1,1}(X_3)-1$ and thus agrees with the M-theory compactification.
    
\subsubsection{Other fibrations with purely Abelian gauge groups}
\subsubsection*{ A fibration $\mathbb{E}_2$ with $G=U(1)$}
The configuration matrix (\ref{case1}) admits another OGF with $h^{1,1} (B)=2$.
\beq\label{mr2}
X_3^{\mathbb{E}_2}=\left[\begin{array}{c|c:ccc}
\IP^{2} & 0 & 1 & 2 & 0\\
\IP^{2} & 0 & 1 & 1 & 1  \\\hdashline
\IP^1 &   1 & 1  &  0 & 0 \\
\IP^2 & 1 & 0 & 1 & 1 
\end{array}\right], \ \quad
B=\left[\begin{array}{c|c}
\IP^1 &  1  \\
\IP^2 & 1  
\end{array}\right] =\mathbb{F}_1
\eeq  
Following the same procedure described in previous subsection,  for this fibration it is possible to identify a rational section as 
\beq
\cO(S_0)=\cO_{X_3^{\mathbb{E}_2}} (1,-1,1,1)
\eeq
with  $h^\bullet(X, \cO(S_0)) = (1,0,0,0)$.  Starting with such a section, we can compute the Weierstrass Model as before and study its discriminant locus. In this case, the discriminant is simply comprised of a $I_1$ locus: $\Delta_W=\Delta_{I_1}$. 

Our result for the discriminant of the Weierstrass model can once again be checked by computing the discriminant of Jacobians of contractions of the original configuration (\ref{mr2}). We can blow down one $\IP^2$ to obtain:
\beq\label{}
{X_{3}'}^{\mathbb{E}_2}=\left[\begin{array}{c|c:c}
\IP^2 & 0 & 3 \\\hdashline
\IP^1 & 1 & 1 \\
\IP^2 & 1 & 2
\end{array}\right] \;.
\eeq 
Using the results of \cite{artin,Braun:2014qka} we again find that the discriminant $\Delta_{bl}$ only contains an $I_1$ locus.

Given the above results, and the theorem of Shioda-Tate-Wazir (\ref{stw}), we see that we must have a single $U(1)$ gauge group factor associated to this compactification. That is, the Mordell-Weil group is of rank one (see \cite{us_to_appear} for details on constructing the Mordell-Weil lattice explicitly in configurations such as this). Since the base has $h^{1,1}(B)=2$ we then have that $n_V=1$ and $n_T=1$. Finally, we need to count the number of $U(1)$ charged hypermultiplets. Counting the number of nodes in the $I_1$ locus as before we find that there are $197$ such charged matter fields. Combining this result with (\ref{nh}) we see that we have $n_H=197+h^{2,1}(X_3)+1=245$ in this example.

The results we have obtained here can once again be checked to be consistent with anomaly cancelation. In particular, we see that
\begin{eqnarray}
n_H-n_V+29 n_T = 245 - 1+29 =273 \;,
\end{eqnarray}
as required by the cancelation of the gravitational anomaly. It is also trivial to check from (\ref{5dV2}) and (\ref{neut}) that the theory has the correct M-theory limit.

\subsubsection*{A fibration $\mathbb{E}_3$ with $G=U(1)^2$}
A third OGF admitted by the configuration matrix (\ref{case1}), once more with a $\mathbb{P}^2$ base, is as follows.
\beq\label{}
X_3^{\mathbb{E}_3}=\left[\begin{array}{c|cccc}
\IP^{1} & 1 & 1 & 0 & 0\\
\IP^{2} & 0 & 1 & 1 & 1  \\
\IP^2 &   1 & 0  &  1 & 1 \\\hdashline
\IP^2 & 1 & 0 & 2 & 0
\end{array}\right], \ \quad
B=\IP^2
\eeq  
We identify rational sections as before and find that one appropriate choice for a zero section is given by 
\beq
\cO(S_0)=\cO_{X_3^{\mathbb{E}_3}} (-1, 0, 1, 1)
\eeq
with  $h^\bullet(X, \cO(S_0)) = (1,0,0,0)$.  The Weierstrass Model constructed using this section results in a discriminant which is not factorizable. In other words, $\Delta_W=\Delta_{I_1}$ and there are no non-Abelian factors in the gauge group.

As in previous cases, the structure of the discriminant can be checked by comparison to that of the Jacobian of a contraction of the original configuration matrix. Contracting one of the $\IP^2$'s we arrive at:
\beq\label{}
{X_{3}'}^{\mathbb{E}_3}=\left[\begin{array}{c|cc}
\IP^1 & 1 & 1 \\
\IP^2 & 2 & 1 \\\hdashline
\IP^2 & 3 & 0
\end{array}\right] \;.
\eeq 
This fiber is of the same form as that obtained in (\ref{ladprime}) and its Jacobian can be obtained in the same way. The resulting discriminant $\Delta_{bl}$ agrees with that of the Weierstrass model discussed above.

Given that the gauge group has no non-Abelian component, we find from (\ref{stw}) that the rank of the Mordell-Weil group must be two in this example. Thus we have a $U(1)^2$ symmetry in $6$-dimensions. This tells us that $n_V=2$, and given that the base is $\mathbb{P}^2$ we know that $n_T=0$. All that is left for us to compute is the $U(1)$ charged matter.

Analyzing the  $I_1$ locus, we find that it contains  $227$ nodes, thus giving rise to the same number of charged hypermultiplets. The formula (\ref{nh}) then tells us that $n_H=227+47+1=275$. As in previous examples these results are consistent with anomaly cancelation,
\begin{eqnarray}
n_H-n_V+29 n_T =275-2+0 = 273 \;,
\end{eqnarray}
and reduce to the correct result in the M-theory limit described in Section \ref{mred}.

\subsubsection*{A non-flat fibration $\mathbb{E}_4$}
The geometry admits a fourth OGF with $h^{1,1} (B)=1$.
\beq\label{}
X_3^{\mathbb{E}_4}=\left[\begin{array}{c|cccc}
\IP^{1} & 1 & 1 & 0 & 0\\
\IP^{2} & 1 & 0 & 2 & 0  \\
\IP^2 &   1 & 0  &  1 & 1 \\\hdashline
\IP^2 & 0 & 1 & 1 & 1 
\end{array}\right], \ \quad
B=\IP^2
\eeq  
This fibration, however, is not flat and as such we do not analyze it further here. Studying the physics associated to such fibrations \cite{Braun:2013nqa,Morrison:1996na,Seiberg:1996vs,Morrison:1996pp,Candelas:2000nc}  is beyond the scope of this paper, although we will make a few comments in this direction at the end of this Section. As has been noted in other contexts \cite{Braun:2013nqa}, many geometries in the CICY list exhibit at least one such non-flat OGF and that such structures seem to be fairly common.

\vspace{0.3cm}

\begin{figure}[t!]
\centering
\begin{tikzpicture}
[
>=stealth',
  punktchain/.style={
    rectangle, 
    draw=black,
    text width=12em, 
    minimum height=3em, 
    text centered},
    ar/.style={->,thick,shorten <=8pt,shorten >=8pt,>=stealth}
    ]
    
\node[punktchain](source){ $ \mathbb{E}_1: SU(2) \times U(1)$ \\ $n_V=4, n_T=0, n_H=277$};
\node[punktchain, below left=1.cm and 1cm of source](relay){ $ \mathbb{E}_2:  U(1)$ \\ $n_V=2, n_T=1, n_H=245$};
\node[punktchain, below right=1.cm and 1cm of source](relay2){ $ \mathbb{E}_3: U(1)^2$ \\ $n_V=2, n_T=0, n_H=275$};
\node[punktchain, below right= 1cm and 1cm of relay](relay3){$ \mathbb{E}_4:$ non-flat fiber   };

\draw  [<->, thick] ([xshift=-2mm]source.west) -- node[below,sloped] {}  ([yshift=2mm]relay.north);
\draw  [<->, dashed] ([yshift=-2mm]source.south) -- node[below,sloped] {}  ([yshift=2mm]relay3.north);
\draw  [<->, thick] ([xshift=2mm]source.east) -- node[below,sloped] {}  ([yshift=2mm]relay2.north);
\draw  [<->, double, dashed] ([xshift=2mm]relay.east) -- node[below,sloped] {}  ([xshift=-2mm]relay2.west);
\draw  [<->, dashed] ([yshift=-2mm]relay.south) -- node[below,sloped] {}  ([xshift=-2mm]relay3.west);
\draw  [<->, dashed] ([xshift=2mm]relay3.east) -- node[below,sloped] {}  ([yshift=-2mm]relay2.south);

\end{tikzpicture}
\caption{\it $6D$ F-theory compactifications that share the same 5D M-theory limit with $n_V^{(5D)}=3, n_H^{(5D)}=48$. The double dashed line denotes a link between two Abelian theories while the thick line denotes a correspondence between non-Abelian and Abelian theories.}
\label{f:duality2}
\end{figure}
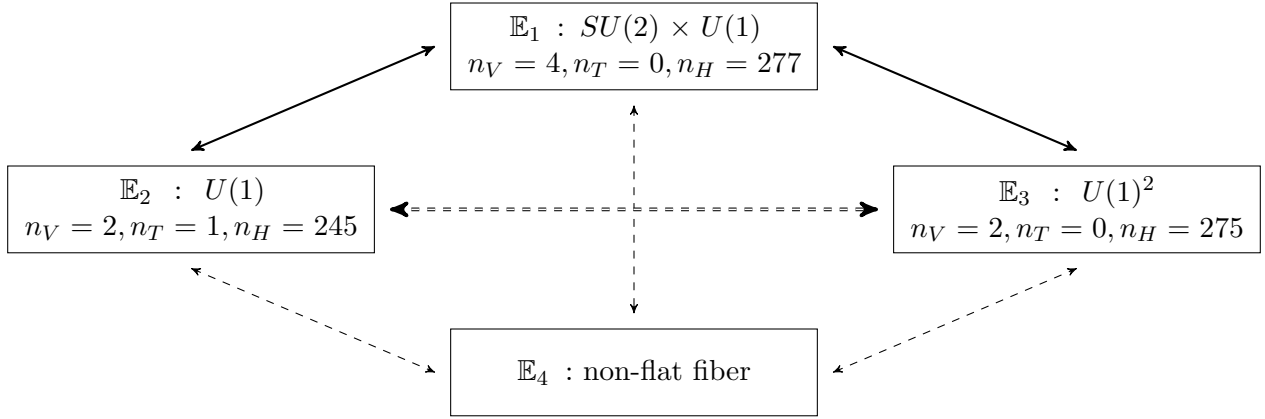

At the end of this lengthy analysis, we have identified four different $6$-dimensional compactifications of F-theory $\mathbb{E}_1, \mathbb{E}_2, \mathbb{E}_3$ and $\mathbb{E}_4$ which share an M-theory limit. The four theories which are linked in this way are illustrated in Figure \ref{f:duality2}.  For three out of four of the $6$-dimensional theories involved in this correspondence, we have been able to describe the multiplet structure that arises in detail and check our computations explicitly by considering both anomaly cancelation and the M-theory limit. It is interesting that such a wide variety of different theories, including non-Abelian models, purely Abelian theories, and even the more exotic physics associated to non-flat fibrations, can be related in this way by a common dimensional reduction of their Coulomb branch. The advantage of the construction we have described here is that because all of these theories were derived from the same initial configuration matrix (\ref{case1}), it was guaranteed from the start that they would share an M-theory limit in this manner.
 
\subsection{Example of a non-Abelian/non-Abelian correspondence}
\label{sec:non-abelian}

By starting with different initial configuration matrices we can generate a huge variety of families of F-theory compactifications with the same M-theory limit. To give another example, let us consider a geometry which gives rise to a correspondence between different non-Abelian $6$-dimensional theories. The three-fold we will consider
\begin{eqnarray}\label{case2}
X_{3}=\left[\begin{array}{c|ccccc}
\IP^{1} & 1 & 1 & 0 & 0 & 0\\
\IP^{1} & 1 & 0 & 1 & 0 & 0  \\
\IP^2 &   0 & 1  &  0 & 2 & 0  \\
\IP^2 & 0 & 0 & 1 & 1 & 1 \\
\IP^2 & 1 & 0 & 0 & 0 & 2  
\end{array}\right]
\end{eqnarray}
has Hodge numbers $h^{1,1}(X_3)=5, \, h^{2,1}(X_3)=25$ and exhibits six different OGF structures.  Here we will focus on three of those OGF's which admit rational sections.

\subsection*{A fibration $\mathbb{E}_1$ with $G=SU(2)\times SU(2)' \times U(1)$}

The first OGF  of (\ref{case2}) that we will consider is as follows:
\beq\label{thisspud}
X_{3}^{\mathbb{E}_1}=\left[\begin{array}{c|ccccc}
\IP^{1} & 1 & 1 & 0 & 0 & 0\\
\IP^{1} & 1 & 0 & 1 & 0 & 0  \\
\IP^2 &   0 & 1  &  0 & 2 & 0  \\
\IP^2 & 0 & 0 & 1 & 1 & 1 \\\hdashline
\IP^2 & 1 & 0 & 0 & 0 & 2  
\end{array}\right], \ \quad
B=\IP^2.
\eeq  
Using the same techniques as in previous subsections we identify $\cO(S_0)=\cO_{X_3^{\mathbb{E}_1}}(-1,0,1,0,0)$ as a rational section. We can then obtain the associated Weierstrass model and compute its discriminant locus $\Delta_W$. The same locus in the base can be obtained by following contracting the two $\IP^1$'s and the second $\IP^2$ in (\ref{thisspud}) to obtain a fiber described as a a simple cubic in $\IP^2$, and then taking the Jacobian. This Jacobian has a discriminant $\Delta_{bl} =\Delta_W$ and both computations agree to give the following result. 
\begin{eqnarray}
\Delta_{bl}=( \Delta_{I_2})(\Delta_{I_2'})(\Delta_{I_1})
\end{eqnarray}
We see that in this example we obtain a $6$-dimensional theory with two $SU(2)$ factors in the low energy gauge group. Use of the Shioda-Tate-Wazir formula (\ref{stw}) then tells us that $rk(MW)=1$ in this example. Thus we conclude that the resulting $6$-dimensional compactification of F-theory exhibits a gauge group $SU(2) \times SU(2)'  \times U(1)$.

Next we must compute the number of charged hypermultiplets in the effective $6$-dimensional theory. The genus of each of the two $I_2$ components in this example is zero, and as such the second term in (\ref{nh}) vanishes. We need then only consider the intersection between the $I_2$, $I_2'$ and $I_1$ loci and the nodes of the $I_1$ locus itself: 
\begin{itemize}
\item $\Delta_{I_2}  \cap \Delta_{I_1}$: 32 points in the base corresponding to $I_3$ singular fibers and 12 points corresponding to type III singular fibers. These contribute 32 doublets under $SU(2)$: $32 \times ({\bf 2},{\bf1})$.
\item $\Delta_{I_2'}  \cap \Delta_{I_1}$: 32 points in the base corresponding to $I_3$ singular fibers and 12 points corresponding to type III singular fibres. These contribute 32 doublets under $SU(2)'$: $32 \times ({\bf 1},{\bf 2})$.
\item $\Delta_{I_2}  \cap \Delta_{I_2'}$: 4 points in the base corresponding to $I_4$ singular fibers. These contribute 4 bi-doublets under $SU(2) \times SU(2)'$: $ 4 \times ({\bf 2}, {\bf 2'}) $.
\item $\Delta_{I_1}$ self-intersection: The $I_1$ locus exhibits $110$ nodes which contribute $110$  $SU(2)\times SU(2)'$ singlets charged under the $U(1)$ gauge group factor.
\end{itemize}

Combining all of these results together, and including the $h^{2,1}(X_3)+1$ neutral hypermultiplets that arise in this configuration, we find that we have $n_H=32\times 2 + 32 \times 2 + 4 \times 4 +110 + 25 +1= 280$. Using this information, together with the fact that $n_V=7$ and $n_T=0$ we can now proceed to consider anomaly cancelation. The gravitational anomaly cancelation condition is clearly satisfied with $n_H-n_V+29n_T= 280 -7+0 =273$ as desired. A slightly lengthier computation also shows that the anomalies associated with the two non-Abelian factors in the gauge group also cancel.

\subsection*{A fibration $\mathbb{E}_2$ with $G=SU(2) \times U(1)$}
The second OGF of (\ref{case2}) that we consider has $h^{1,1} (B)=2$ and is described as follows:
\beq\label{chappymcchapperson}
X_3^{\mathbb{E}_2}=\left[\begin{array}{c|c:cccc}
\IP^{1} & 0 & 1 & 1 & 0 & 0\\
\IP^{2} & 0 & 0 & 1 & 1 & 1  \\
\IP^2 &   0 & 1  &  0 & 0 & 2  \\\hdashline
\IP^2 & 1 & 0 & 0 & 2 & 0 \\
\IP^1 & 1 & 1 & 0 & 0 & 0  
\end{array}\right], \ \quad
B=\left[\begin{array}{c|c}
\IP^{2} & 1 \\
\IP^{1} & 1 
\end{array}\right]=\mathbb{F}_1.
\eeq  
For this elliptic fibration, we identify $\cO(S_0)=\cO_{X_3^{\mathbb{E}_2}}(-1,0,1,0,1)$ as a rational section. Following either the procedure for obtaining the associated Weierstrass model or the procedure involving contracting the configuration matrix and then taking the Jacobian we obtain the same description of the discriminant locus in the base $\mathbb{F}_1$. In particular, we find that $\Delta_W=\Delta_{bl}= (\Delta_{I_2})(\Delta_{I_1})$ and thus we have a single $SU(2)$ non-Abelian factor in the gauge group. Using the Shioda-Tate-Wazir theorem as in previous examples we can then show that the Mordell-Weil group is of rank one leading to a total low energy gauge group of $SU(2)\times U(1)$.

The charged matter in this fibration comes from both the intersection of the $I_2$ and $I_1$ loci and the self-intersection of the $I_1$ locus. Using the techniques we have described, one can show that the $I_2$ locus intersects the $I_1$ locus at $52$ points. Of these points, $40$ are associated with $I_3$ fibers, and $12$ with type III fibers. In total, then, this leads to $40$ hypermultiplets which are doublets of $SU(2)$. To compute the charged $U(1)$ matter we examine the $I_1$ locus and show that it exhibits $142$ nodes. In total then, using (\ref{nh}) and the fact that the genus of the $I_2$ component of the discriminant is zero, we see that we have $n_H=40 \times 2+142 + 25+1=248$ hypermultiplets in the $6$-dimensional theory associated to the fibration (\ref{chappymcchapperson}).

Combining the information we have obtained about the $6$-dimensional theory associated to the fibration  (\ref{chappymcchapperson}), we can easily see that gravitational anomaly cancelation holds. Indeed we have that $n_H-n_V+29 n_T= 248-4+29=273$. A slightly more involved calculation, following along exactly the same lines as those described in earlier subsections, also shows that the non-Abelian anomalies cancel too.

 \subsection*{A non-flat fibration $\mathbb{E}_3$}
As was the case for the configuration (\ref{case1}), the geometry (\ref{case2}) also admits a non-flat fibration with a section:
\beq\label{}
X_{3}^{\mathbb{E}_3}=\left[\begin{array}{c|ccccc}
\IP^{1} & 1 & 1 & 0 & 0 & 0\\
\IP^{1} & 1 & 0 & 1 & 0 & 0  \\
\IP^2 &   0 & 0  &  1 & 1 & 1  \\
\IP^2 & 1 & 0 & 0 & 0 & 2 \\\hdashline
\IP^2 & 0 & 1 & 0 & 2 & 0  
\end{array}\right], \ \quad
B=\IP^2.
\eeq 
Once again, we will save a discussion of non-flat fibers for the end of this Section.

\vspace{0.3cm} 
  
\begin{figure}[t!]
\centering
\begin{tikzpicture}
[
>=stealth',
  punktchain/.style={
    rectangle, 
    draw=black,
    text width=12em, 
    minimum height=3em, 
    text centered},
    ar/.style={->,thick,shorten <=8pt,shorten >=8pt,>=stealth}
    ]
    
\node[punktchain](source){ $ \mathbb{E}_1: SU(2)^2 \times U(1)$ \\ $n_V=7, n_T=0, n_H=280$};
\node[punktchain, below left=1.67cm and 0.5cm of source](relay){ $ \mathbb{E}_2: SU(2) \times U(1)$ \\ $n_V=4, n_T=1, n_H=248$};
\node[punktchain, below right=1.67cm and 0.5cm of source](relay2){ $ \mathbb{E}_3:$ non-flat fiber};

\draw  [<->, thick] ([xshift=-2mm]source.west) -- node[below,sloped] {}  ([yshift=2mm]relay.north);
\draw  [<->, dashed] ([xshift=2mm]source.east) -- node[below,sloped] {}  ([yshift=2mm]relay2.north);
\draw  [<->, dashed] ([xshift=2mm]relay.east) -- node[below,sloped] {}  ([xshift=-2mm]relay2.west);

\end{tikzpicture}
\caption{\it $6D$ compactifications of F-theory that share a M-theory limit. In this example the $6D$ theories have different non-Abelian gauge groups but nevertheless give rise to the same theory in $5D$ with $n_V^{(5D)}=4$ and $n_H^{(5D)}=26$.}
\label{f:duality3}
\end{figure}
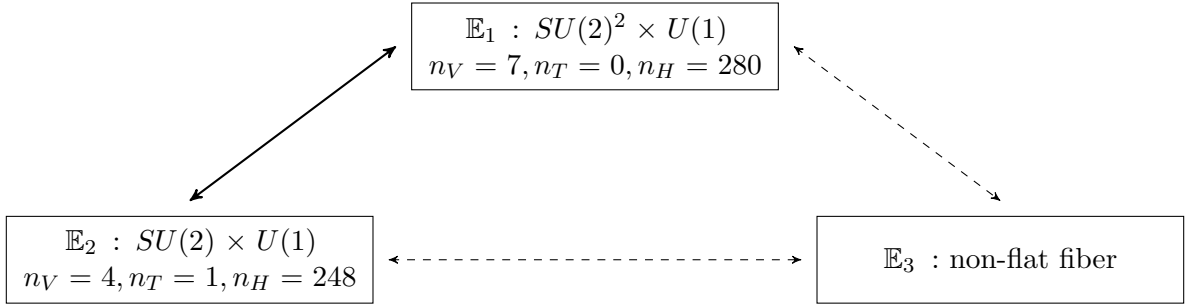

After completing our analysis of the configuration matrix (\ref{case2}) we conclude that it exhibits $6$ OGFs. Of these three admit a section and one of those is a non-flat fibrations. We have computed the spectrum associated to the Weierstrass model of the two flat fibrations with section and have shown that we achieve a consistent anomaly free theory in both cases. Both $6$-dimensional theories are non-Abelian, with different gauge groups, and have the same $5$-dimensional M-theory limit with $n_V^{(5D)}=4$ and $n_H^{(5D)}=26$. The structure we have elucidated here is depicted in Figure \ref{f:duality3}.

\subsection{Higher rank Mordell-Weil group} \label{9waysup}

There is currently no known bound on the rank of the Mordell-Weil group that can be achieved in an elliptically fibered Calabi-Yau three-fold, and as a result, it is interesting to consider whether the novel fiber types explored here can naturally give rise to examples with high rank? For recent systematic studies of fiber-types with non-trivial Mordell-Weil group see \cite{Morrison:2012ei,Borchmann:2013jwa,Cvetic:2013nia,Borchmann:2013hta,Cvetic:2013qsa,Cvetic:2015ioa} (and for genus one fibrations \cite{Braun:2014oya,Morrison:2014era,Anderson:2014yva}). Here we note that the favorable CICYs exhibit a rich structure of cases with higher rank Mordell-Weil groups that can easily be studied with the techniques we have been discussing. As an example, we present an example of a geometry which exhibits nine different fibration structures and where the total space has $h^{1,1}(X_3)=7, h^{2,1}(X_3)=26$. Among the nine fibrations, two of them, $\mathbb{E}_1$ and $\mathbb{E}_8$, are of Mordell-Weil rank $r=4$.  

Let us begin our discussion with the fibration,
\bea
\label{eq:hmw1}
X_3^{\mathbb{E}_1}=\left[
\begin{array}{c|c:cccccccc}
\IP^1 &  0 & 0 & 1 & 1 & 0 & 0 & 0 & 0 & 0 \\
\IP^2 &  0 & 0 & 1 & 0 & 1 & 1 & 0 & 0 & 0 \\
\IP^2 &  0 & 0 & 0 & 1 & 0 & 0 & 0 & 1 & 1 \\
\IP^2 &  0 & 1 & 0 & 0 & 1 & 0 & 0 & 1 & 0 \\
\IP^2 &  0 & 0 & 0 & 0 & 0 & 1 & 1 & 0 & 1 \\\hdashline
\IP^1 &  1 & 1 & 0 & 0 & 0 & 0 & 0 & 0 & 0 \\
\IP^2 &  1 & 0 & 0 & 1 & 0 & 0 & 1 & 0 & 0 \\
\end{array}
\right]\;.
\eea
In this case, we can identify $\cO(S_0)=\cO_{X_3^{\mathbb{E}_1}}(1,-1,0,0,1,0,0)$ as a good choice of zero section. Computing the discriminant locus of the fibration in the same manner as for previous examples, we find that it only contains an $I_1$ locus. Since $h^{1,1}(B)=2$, the Shioda-Tate-Wazir decomposition of the Picard lattice tells us that the Mordell-Weil group is of rank 4. Thus the gauge group of the associated $6$-dimensional compactification of F-theory is simply $U(1)^4$. 

It is possible to study this rank $4$ Mordell-Weil group of sections explicitly. Following the discussion in \cite{us_to_appear}, we can find ``putative sections" that obey certain necessary topological constraints that must hold for any true section. Parameterizing a general line bundle associated to a putative section as $\cO(S)=\cO_{X_3^{\mathbb{E}_1}}(b_1, b_2, b_3, b_4, b_5, b_6, b_7)$, the relevant objects form a four-parameter family of possibilities:
\bea
b_1&=-1 - 3 k_2 - k_3 - k_4 \ , \\
b_2&=k_1 \ , \\
b_3&=1 - k_1 + 2 k_2 \ , \\
b_4&=k_3 \ , \\
b_5&=k_4 \ , \\
b_6&= -k_1 + k_1^2 + 3 k_2 - 2 k_1 k_2 + 4 k_2^2 + 2 k_2 k_3 + k_3^2 + k_4 +  3 k_2 k_4 + k_3 k_4 + k_4^2 \ , \\
b_7&=1 -  3 k_1 + 2 k_1^2 + 11 k_2 -  7 k_1 k_2 + 14 k_2^2 + 3 k_3 -  k_1 k_3 +  8 k_2 k_3 + 2 k_3^2 + 2 k_4 \\
&\quad -  k_1 k_4 +  7 k_2 k_4 + 2 k_3 k_4 + 2 k_4^2 \ ,
\eea
where $k_1, \cdots, k_4 \in \IZ$.
If we further require that the zeroth cohomology of the line bundle associated to the putative sections should be equal to one, we can narrow down our search for rational sections and, in fact, find explicit generators for the Mordell-Weil lattice. For the case at hand, a suitable set of generators is ${\cal O}_{X_3^{\mathbb{E}_1}}(1,-1,0,1,0,0,0)$, ${\cal O}_{X_3^{\mathbb{E}_1}}(1,0,-1,0,1,0,1)$, ${\cal O}_{X_3^{\mathbb{E}_1}}(0,0,1,-1,0,1,0)$ and ${\cal O}_{X_3^{\mathbb{E}_1}}(-1,0,1,0,0,0,1)$. More details of how such computations are performed, as well as a detailed discussion of the addition law and Shioda map in geometries like this can be found in \cite{us_to_appear}.

We can again calculate the matter content of the F-theory compactification associated to the fibration $\mathbb{E}_1$ and check that gravitational and non-Abelian gauge anomaly cancellation holds. The case at hand has $n_T=1$, $n_V=4$ and $n_H=248$ and indeed all such consistency checks are passed.

Let us analyze the other eight OGFs associated to the configuration matrix (\ref{eq:hmw1}).
\bea
X_3^{\mathbb{E}_2}=\left[
\begin{array}{c|c:cccccccc}
\IP^1 & 0 & 1 & 1 & 0 & 0 & 0 & 0 & 0 & 0 \\
\IP^2 &  0 & 1 & 0 & 1 & 0 & 0 & 1 & 0 & 0 \\
\IP^2 & 0 & 0 & 0 & 1 & 0 & 0 & 0 & 1 & 1 \\
\IP^2 &  0 & 0 & 1 & 0 & 1 & 0 & 0 & 1 & 0 \\
\IP^2 &  0 & 0 & 0 & 0 & 0 & 1 & 1 & 0 & 1 \\\hdashline
\IP^1 &  1 & 0 & 0 & 1 & 0 & 0 & 0 & 0 & 0 \\
\IP^2 &  1 & 0 & 0 & 0 & 1 & 1 & 0 & 0 & 0 \\
\end{array}
\right];
\quad
X_3^{\mathbb{E}_3}=\left[
\begin{array}{c|c:cccccccc}
\IP^1 & 0 & 0 & 1 & 1 & 0 & 0 & 0 & 0 & 0 \\
\IP^2 & 0 & 0 & 1 & 0 & 1 & 1 & 0 & 0 & 0 \\
\IP^2 & 0 & 1 & 0 & 1 & 0 & 0 & 1 & 0 & 0 \\
\IP^2 & 0 & 0 & 0 & 1 & 0 & 0 & 0 & 1 & 1 \\
\IP^2 & 0 & 0 & 0 & 0 & 0 & 1 & 1 & 0 & 1 \\\hdashline
\IP^1 & 1 & 1 & 0 & 0 & 0 & 0 & 0 & 0 & 0 \\
\IP^2 & 1 & 0 & 0 & 0 & 1 & 0 & 0 & 1 & 0 \\
\end{array}
\right] ; \nonumber\\
X_3^{\mathbb{E}_4}=\left[
\begin{array}{c|ccccccccc}
\IP^1 & 1 & 1 & 0 & 0 & 0 & 0 & 0 & 0 & 0 \\
\IP^1 &  0 & 0 & 1 & 1 & 0 & 0 & 0 & 0 & 0 \\
\IP^2 &  0 & 0 & 1 & 0 & 1 & 1 & 0 & 0 & 0 \\
\IP^2 &  0 & 0 & 0 & 1 & 0 & 0 & 0 & 1 & 1 \\
\IP^2 &  0 & 1 & 0 & 0 & 1 & 0 & 0 & 1 & 0 \\
\IP^2 &  0 & 0 & 0 & 0 & 0 & 1 & 1 & 0 & 1 \\\hdashline
\IP^2 &  1 & 0 & 0 & 1 & 0 & 0 & 1 & 0 & 0 \\
\end{array}
\right];\quad
X_3^{\mathbb{E}_5}=\left[
\begin{array}{c|ccccccccc}
\IP^2 &   0 & 0 & 1 & 0 & 1 & 1 & 0 & 0 & 0 \\
\IP^2 &   1 & 0 & 0 & 1 & 0 & 0 & 1 & 0 & 0 \\
\IP^2 &   0 & 0 & 0 & 1 & 0 & 0 & 0 & 1 & 1 \\
\IP^2 &   0 & 1 & 0 & 0 & 1 & 0 & 0 & 1 & 0 \\
\IP^2 &   0 & 0 & 0 & 0 & 0 & 1 & 1 & 0 & 1 \\\hdashline
 \IP^1 &  1 & 1 & 0 & 0 & 0 & 0 & 0 & 0 & 0 \\
\IP^1 &   0 & 0 & 1 & 1 & 0 & 0 & 0 & 0 & 0 \\
\end{array}
\right]; \nonumber\\
X_3^{\mathbb{E}_6}=\left[
\begin{array}{c|ccccccccc}
\IP^1 & 1 & 1 & 0 & 0 & 0 & 0 & 0 & 0 & 0 \\
\IP^1 & 0 & 0 & 1 & 1 & 0 & 0 & 0 & 0 & 0 \\
\IP^2 & 0 & 0 & 1 & 0 & 1 & 1 & 0 & 0 & 0 \\
\IP^2 & 1 & 0 & 0 & 1 & 0 & 0 & 1 & 0 & 0 \\
\IP^2 & 0 & 0 & 0 & 1 & 0 & 0 & 0 & 1 & 1 \\
\IP^2 & 0 & 1 & 0 & 0 & 1 & 0 & 0 & 1 & 0 \\\hdashline
\IP^2 & 0 & 0 & 0 & 0 & 0 & 1 & 1 & 0 & 1 \\
\end{array}
\right];\quad
X_3^{\mathbb{E}_7}=\left[
\begin{array}{c|ccccccccc}
\IP^1 & 1 & 1 & 0 & 0 & 0 & 0 & 0 & 0 & 0 \\
\IP^1 & 0 & 0 & 1 & 1 & 0 & 0 & 0 & 0 & 0 \\
\IP^2 & 0 & 0 & 1 & 0 & 1 & 1 & 0 & 0 & 0 \\
\IP^2 & 1 & 0 & 0 & 1 & 0 & 0 & 1 & 0 & 0 \\
\IP^2 & 0 & 0 & 0 & 1 & 0 & 0 & 0 & 1 & 1 \\
\IP^2 & 0 & 0 & 0 & 0 & 0 & 1 & 1 & 0 & 1 \\\hdashline
\IP^2 & 0 & 1 & 0 & 0 & 1 & 0 & 0 & 1 & 0 \\
\end{array}
\right]; \nonumber\\
X_3^{\mathbb{E}_8}=\left[
\begin{array}{c|c:cccccccc}
\IP^1 &  1 & 1 & 0 & 0 & 0 & 0 & 0 & 0 & 0 \\
\IP^1 &  0 & 0 & 1 & 1 & 0 & 0 & 0 & 0 & 0 \\
\IP^2 &  0 & 0 & 1 & 0 & 1 & 1 & 0 & 0 & 0 \\
\IP^2 &  1 & 0 & 0 & 1 & 0 & 0 & 1 & 0 & 0 \\
\IP^2 &  0 & 1 & 0 & 0 & 1 & 0 & 0 & 1 & 0 \\
\IP^2 &  0 & 0 & 0 & 0 & 0 & 1 & 1 & 0 & 1 \\\hdashline
\IP^2 &  0 & 0 & 0 & 1 & 0 & 0 & 0 & 1 & 1 \\
\end{array}
\right];\quad
X_3^{\mathbb{E}_9}=\left[
\begin{array}{c|ccccccccc}
\IP^1 &  1 & 1 & 0 & 0 & 0 & 0 & 0 & 0 & 0 \\
\IP^1 &  0 & 0 & 1 & 1 & 0 & 0 & 0 & 0 & 0 \\
\IP^2 &  1 & 0 & 0 & 1 & 0 & 0 & 1 & 0 & 0 \\
\IP^2 &  0 & 0 & 0 & 1 & 0 & 0 & 0 & 1 & 1 \\
\IP^2 &  0 & 1 & 0 & 0 & 1 & 0 & 0 & 1 & 0 \\
\IP^2 &  0 & 0 & 0 & 0 & 0 & 1 & 1 & 0 & 1 \\\hdashline
\IP^2 &  0 & 0 & 1 & 0 & 1 & 1 & 0 & 0 & 0 \\
\end{array}
\right].
\eea

 Direct computation shows that the genus of the components of the discriminant associated to non-Abelian gauge group factors are all zero in these examples.

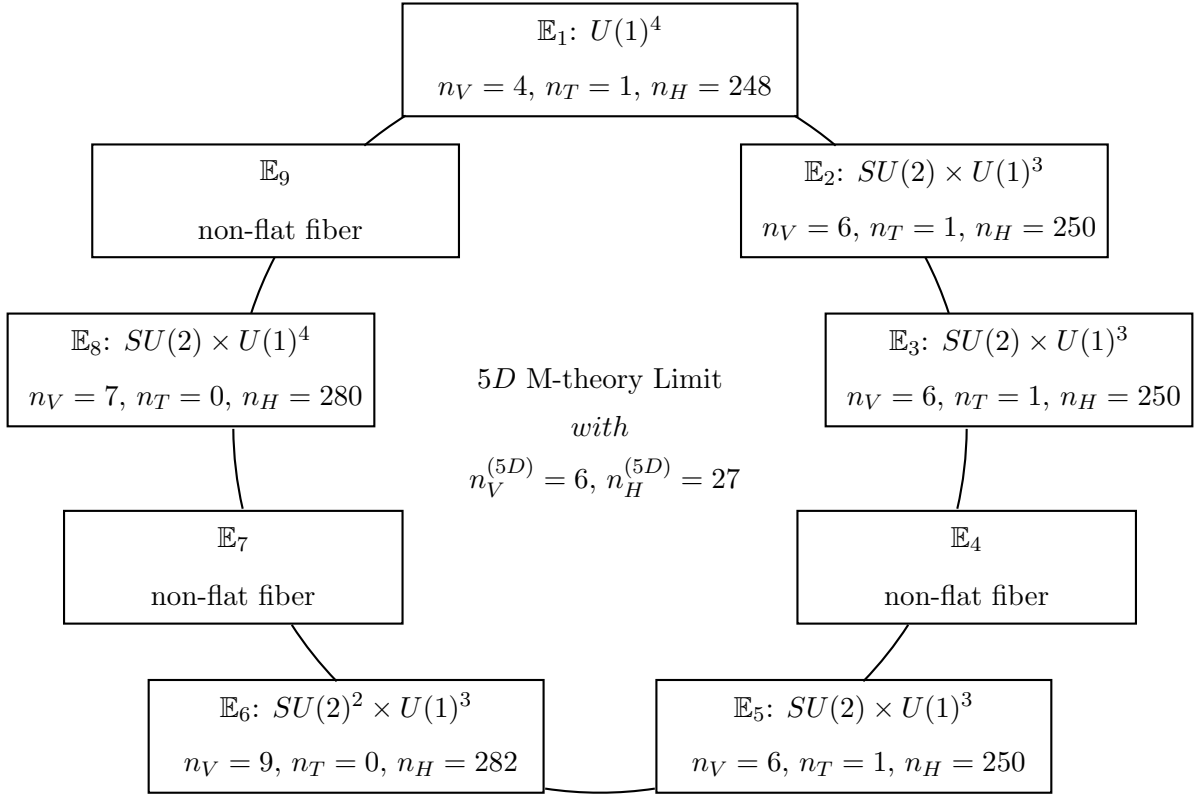
\begin{figure}[t!]
\centering
\begin{tikzpicture}[thick, scale=0.75]

\draw  (-3.5,5.5) rectangle (3.5,7.5);      
\draw (0,7) node {$\mathbb{E}_1$: $U(1)^4$};
\draw (0,6) node {{ $n_V=4$, $n_T=1$, $n_H=248$}};  

\draw  (-2.5,3) rectangle (-9,5);                       
\draw (-5.75,4.5) node {$\mathbb{E}_9$};
\draw (-5.75,3.5) node {{ non-flat fiber}}; 

\draw  (-4,0) rectangle (-10.5,2);                
\draw (-7.25,1.5) node {$\mathbb{E}_8$: $SU(2) \times U(1)^4$};
\draw (-7.25,0.5) node {{ $n_V=7$, $n_T=0$, $n_H=280$}}; 

\draw  (-3.5,-1.5) rectangle (-9.5,-3.5);       
\draw (-6.5,-2) node {$\mathbb{E}_7$ };
\draw (-6.5,-3) node {{non-flat fiber}}; 

\draw  (-1,-4.5) rectangle (-8,-6.5);                
\draw (-4.5,-5) node {$\mathbb{E}_6$: $SU(2)^2 \times U(1)^3$};
\draw (-4.5,-6) node {{ $n_V=9$, $n_T=0$, $n_H=282$}};

\draw  (2.5,5) rectangle (9,3);              
\draw (5.75,4.5) node {$\mathbb{E}_2$: $SU(2) \times U(1)^3$};
\draw (5.75,3.5) node {{ $n_V=6$, $n_T=1$, $n_H=250$}};

\draw  (4,2) rectangle (10.5,0);               
\draw (7.25,1.5) node {$\mathbb{E}_3$: $SU(2)  \times U(1)^3$};
\draw (7.25,0.5) node {{ $n_V=6$, $n_T=1$, $n_H=250$}};

\draw  (3.5,-1.5) rectangle (9.5,-3.5);            
\draw (6.5,-2) node {$\mathbb{E}_4$};
\draw (6.5,-3) node {{non-flat fiber}};

\draw  (1,-6.5) rectangle (8,-4.5);           
\draw (4.5,-5) node {$\mathbb{E}_5$: $SU(2)  \times U(1)^3$};
\draw (4.5,-6) node {{ $n_V=6$, $n_T=1$, $n_H=250$}};

\draw (0,0.8) node {$5D$ M-theory Limit};   
\draw (0,0) node {$with$}; 
\draw (0,-0.9) node {{ $n_V^{(5D)}=6$, $n_H^{(5D)}=27$}}; 

\draw[thick] ([shift=(50:6.5cm)]0,0) arc (50:58.:6.5cm); *
\draw[thick] ([shift=(18:6.5cm)]0,0) arc (18:27.5:6.5cm); *
\draw[thick] ([shift=(-0.4:6.5cm)]0,0) arc (-0.4:-13:6.5cm); *
\draw[thick] ([shift=(-32.8:6.5cm)]0,0) arc (-32.8:-44.1:6.5cm); *
\draw[thick] ([shift=(-81.4:6.5cm)]0,0) arc (-81.4:-98.6:6.5cm); *
\draw[thick] ([shift=(-147.2:6.5cm)]0,0) arc (-147.2:-135.9:6.5cm);*
\draw[thick] ([shift=(-179.6:6.5cm)]0,0) arc (-179.6:-167:6.5cm); *
\draw[thick] ([shift=(162:6.5cm)]0,0) arc (162:152.5:6.5cm);
\draw[thick] ([shift=(130:6.5cm)]0,0) arc (130:122.0:6.5cm);

\end{tikzpicture}
 \caption{\it  F-theory models in 6D with the same 5D M-theory limit where $n_V^{(5D)}=6$ and $n_H^{(5D)}=27$. }
  \label{fig:9-dual}
\end{figure}

\begin{itemize}
\item $\mathbb{E}_2$: We can identify $\cO(S_0)=\cO_{X_3^{\mathbb{E}_2}}(1,-1,0,0,1,0,0)$ as a good choice of zero section. The gauge group is $G=SU(2) \times U(1)^3$, containing $n_T=1$ tensor multiplets, $n_V=6$ vector multiplets and $n_H=250$ hyper multiplets. Among these hyper multiplets, there are $22$ $SU(2)$ doublets and $179$ $SU(2)$ singlets charged under the $U(1)$s. The generators of the rank 3 Mordell-Weil group can be taken to be ${\cal O}_{X_3^{\mathbb{E}_2}}(0,0,-1,1,1,0,0)$, ${\cal O}_{X_3^{\mathbb{E}_2}}(-1,1,0,0,0,0,0)$ and ${\cal O}_{X_3^{\mathbb{E}_2}}(0,1,0,0-1,0,1)$.
\item $\mathbb{E}_3$:  We can identify $\cO(S_0)=\cO_{X_3^{\mathbb{E}_3}}(0,-1,0,0,1,0,1)$ as a good choice of zero section. The gauge and matter content is the same as $\mathbb{E}_2$. The generators of the Mordell-Weil group can be taken to be ${\cal O}_{X_3^{\mathbb{E}_3}}(0,0,-1,0,1,1,0)$, ${\cal O}_{X_3^{\mathbb{E}_3}}(0,0,0,-1,1,0,1)$ and ${\cal O}_{X_3^{\mathbb{E}_3}}(0,1,1,0,-1,0,0)$.
\item $\mathbb{E}_4$:  Is a non-flat fibration.
\item $\mathbb{E}_5$:  We can identify $\cO(S_0)=\cO_{X_3^{\mathbb{E}_5}}(0,-1,0,0,1,1,0)$ as a good choice of zero section. The gauge group is $G=SU(2) \times U(1)^3$, containing $n_T=1$ tensor multiplets, $n_V=6$ vector multiplets and $n_H=250$ hyper multiplets. Among these hyper multiplets, there are $28$ $SU(2)$ doublets and $167$ $SU(2)$ singlets charged under the $U(1)$s. The generators of the Mordell-Weil group can be taken to be ${\cal O}_{X_3^{\mathbb{E}_5}}(-1,0,0,0,1,0,1)$, ${\cal O}_{X_3^{\mathbb{E}_5}}(0,0,1,-1,0,1,0)$ and ${\cal O}_{X_3^{\mathbb{E}_5}}(1,1,0,0,-1,0,0)$.
\item $\mathbb{E}_6$:  We can identify $\cO(S_0)=\cO_{X_3^{\mathbb{E}_6}}(-1,0,0,0,0,1,0)$ as a good choice of zero section. The gauge group is $G=SU(2)^2 \times U(1)^3$, containing $n_T=0$ tensor multiplets, $n_V=9$ vector multiplets and $n_H=282$ hyper multiplets. Among these hyper multiplets, there is $1$ bi-doublets, $40$ doublets and $171$ $SU(2)$ singlets that are charged under the $U(1)$ factors. The generators of the Mordell-Weil group can be taken to be ${\cal O}_{X_3^{\mathbb{E}_6}}(0,0,-1,0,0,1,1)$, ${\cal O}_{X_3^{\mathbb{E}_6}}(0,0,0,0,-1,1,1)$ and ${\cal O}_{X_3^{\mathbb{E}_6}}(1,0,0,0,1,-1,0)$.
\item $\mathbb{E}_7$:  Is a non-flat fibration.
\item $\mathbb{E}_8$:  We can identify $\cO(S_0)=\cO_{X_3^{\mathbb{E}_8}}(0,-1,0,1,0,0,1)$ as a good choice of zero section. The gauge group is $G=SU(2) \times U(1)^4$, containing $n_T=0$ tensor multiplets, $n_V=7$ vector multiplets and $n_H=280$ hyper multiplets.   Among these hyper multiplets, there are $22$ $SU(2)$ doublets and $209$ $SU(2)$ singlets that are charged under the $U(1)$ factors. The generators of the rank 4 Mordell-Weil group can be taken to be ${\cal O}_{X_3^{\mathbb{E}_8}}(0,-1,1,0,0,0,0)$, ${\cal O}_{X_3^{\mathbb{E}_8}}(-1,0,0,1,0,0,0)$, ${\cal O}_{X_3^{\mathbb{E}_8}}(1,0,0,-1,0,1,0)$ and ${\cal O}_{X_3^{\mathbb{E}_8}}(0,0,1,0,-1,0,1)$.
\item $\mathbb{E}_9$:   Is a non-flat fibration.
\end{itemize} 
The gravitational and non-Abelian anomaly cancellation conditions hold for all the six effective theories (arising from flat fibrations). The nine $6$-dimensional F-theory models we have described in this subsection share the same M-theory limit in $5$-dimensions with $n_V^{(5D)}=6$, $n_H^{(5D)}=27$. These results are summarized in Figure \ref{fig:9-dual}.

\vspace{0.3cm}

In the discussions so far, we have not gone into details in the cases of fibrations which are not flat. For the rest of this Section we will give several comments on the geometry of non-flat fibrations, based on one of the explicit examples above:
\beq
X_3^{\mathbb{E}_4}=\left[
\begin{array}{c|ccccccccc}
\IP^1 & 1 & 1 & 0 & 0 & 0 & 0 & 0 & 0 & 0 \\
\IP^1 &  0 & 0 & 1 & 1 & 0 & 0 & 0 & 0 & 0 \\
\IP^2 &  0 & 0 & 1 & 0 & 1 & 1 & 0 & 0 & 0 \\
\IP^2 &  0 & 0 & 0 & 1 & 0 & 0 & 0 & 1 & 1 \\
\IP^2 &  0 & 1 & 0 & 0 & 1 & 0 & 0 & 1 & 0 \\
\IP^2 &  0 & 0 & 0 & 0 & 0 & 1 & 1 & 0 & 1 \\\hdashline
\IP^2 &  1 & 0 & 0 & 1 & 0 & 0 & 1 & 0 & 0 \\
\end{array}
\right] \ . 
\eeq
To begin with, from the first column of the configuration matrix, one can immediately see that there exists a point in the $\IP^2$ base, the fiber over which is two-dimensional. Explicitly, defining the polynomial associated to the first column as $P_1 = x_{1,0} p_1(\bold x_7) + x_{1,1} p_2(\bold x_7)$, the fiber is non-flat above the unique point given by the intersection of the two linear functions $p_1=p_2=0$ in the base $\mathbb{P}^2$. 

Such a non-flat fiber is, in particular, itself a divisor, and is defined by vanishing of the global holomorphic section of $\cO_{X_3^{\mathbb{E}_4}}(-1,0,0,0,0,0,1)$ (see Appendix~{A} of~\cite{us_to_appear} for a detailed analysis). Even in the presence of a non-flat fiber, however, one may still proceed to find sections to the genus-one fibration and can confirm that the following line bundles, 
\bea
\cO_{X_3^{\mathbb{E}_4}}(S_0)\,&=\,\cO_{X_3^{\mathbb{E}_4}}(1, 0, 1, 0, -1, 0, 0) \ , \\
\cO_{X_3^{\mathbb{E}_4}}(S_1)\,&=\,\cO_{X_3^{\mathbb{E}_4}}(0, -1, 1, 0, 0, 0, 0) \ , \\
\cO_{X_3^{\mathbb{E}_4}}(S_2)\,&=\,\cO_{X_3^{\mathbb{E}_4}}(0, -1, 0, 1, 0, 0, 1) \ , \\
\cO_{X_3^{\mathbb{E}_4}}(S_3)\,&=\,\cO_{X_3^{\mathbb{E}_4}}(0, 1, -1, 0, 1, 0, 0) \ , \\
\cO_{X_3^{\mathbb{E}_4}}(S_4)\,&=\,\cO_{X_3^{\mathbb{E}_4}}(0, 1, -1, 0, 0, 1, 0) \ , \\
\cO_{X_3^{\mathbb{E}_4}}(S_5)\,&=\,\cO_{X_3^{\mathbb{E}_4}}(0, 0, 0, -1, 1, 1, 0) \ , 
\eea
correspond to smooth, rational sections, which we denote as $\sigma_0, \sigma_1, \cdots, \sigma_5$, respectively. 
When naively applied with respect to the zero section $\sigma_0$, the usual section arithmetic \cite{Morrison:2012ei,Grimm:2015wda} leads to a general integer combination of the form, 
\beq
{\rm Div}(\sum_{a=1}^5 l_a \sigma_a) \sim \sum_{i=1}^7 \beta_i J_i  \ , 
\eeq
where the coefficients $\beta_i$ are given in terms of the $l_a$ as
\bea\label{sec-arith}
\beta_1&=1 - l_1 - l_2 - l_3 - l_4 - l_5 \ , \\
\beta_2&=-l_1 - l_2 + l_3 + l_4 \ , \\
\beta_3&=1 - l_2 - 2 l_3 - 2 l_4 - l_5 \ , \\
\beta_4&=l_2 - l_5 \ , \\
\beta_5&=-1 + l_1 + l_2 + 2 l_3 + l_4 + 2 l_5 \ , \\
\beta_6&=l_4 + l_5 \ , \\
\beta_7&=-3 l_1 + 3 l_1^2 - 6 l_2 +  7 l_1 l_2 + 7 l_2^2 - 8 l_3 + 4 l_1 l_3 + 9 l_2 l_3 + 8 l_3^2 - 5 l_4 + l_1 l_4 \\
& \quad+ 6 l_2 l_4 + 10 l_3 l_4  + 5 l_4^2 - 6 l_5 + 6 l_1 l_5 + 6 l_2 l_5 + 9 l_3 l_5 + 6 l_4 l_5 + 6 l_5^2 \ . 
\eea
The following two observations teach us that the section arithmetic must be modified in the presence of a non-flat fiber. Firstly, non-effective putative sections can be generated by~\eref{sec-arith}: For instance, with $\left(l_a\right)_{a=1}^5 = (0,0,1,-1,0)$, one obtains the associated line bundle, $\cO_{X_3^{\mathbb{E}_4}}(1, 0, 1, 0, 0, -1, 0)$, whose bundle valued cohomology is computed as
\beq
h^\bullet(X_3, \cO_{X_3^{\mathbb{E}_4}}(1, 0, 1, 0, 0, -1, 0))=(0,0,0,0) \ . 
\eeq
Secondly, via~\eref{sec-arith} one can generate putative sections that are a true sections with a vertical-divisor piece attached. For instance, with $\left(l_a\right)_{a=1}^5 = (2,-1,0,1,-1)$, one obtains the putative section class, $\cO_{X_3^{\mathbb{E}_4}}(0, 0, 1, 0, -1, 0, 1)$, which proves, using the techniques developed in~\cite{us_to_appear}, to be the zero section ($\sim \cO_{X_3^{\mathbb{E}_4}}(1, 0, 1, 0, -1, 0, 0)$) attached to the non-flat divisor ($\sim \cO_{X_3^{\mathbb{E}_4}}(-1,0,0,0,0,0,1)$).  

With these observations, one must first recall that the arithmetic being used here was proven in~\cite{Grimm:2015wda} under the assumption that the blow-up divisors do not contribute. Such an assumption was motivated by the consistency of F-theory effective theories. In the presence of a non-flat fiber, on the other hand, we have just learnt that a modification to this arithmetic is necessary. In particular, it is interesting to note that in the second case above the naive application of the usual rules has lead to a true section, up to a shift by the non-flat fiber class.
 
Independently of discussions on section arithmetic, it is worth mentioning that in all of the non-flat examples we have analyzed, a naive application of the Picard lattice decomposition of Shioda-Tate-Wazir gives rise to a six-dimensional spectrum that is completely free of (non-Abelian) gauge and gravitational anomalies. To be specific, we note that the anomalies are apparently satisfied when the number of Abelian vector multiplets is identified as the (potential) rank of the Mordell-Weil group obtained by naively applying the Tate-Shioda-Wazir theorem to the non-flat fibration. It should also be emphasized that the $n_H^{({\rm codim} 2)}$ in \eref{nh} appears to require contributions not only from nodes but also from ``tacnodes'' of the $I_1$ locus (see Appendix~\ref{sec:plane} for the description of tacnodes).

Section arithmetic, as well as the Shioda-Tate-Wazir decomposition structure, in the presence of a non-flat fiber provides an interesting topic for further investigation. We leave a careful study of the geometry and the physics of such non-flat fibrations to future work.

\section{Multiple Fibrations and Heterotic/F-theory Duality}\label{section:hetf}
In the following sections we will briefly outline some of the dualities and relationships between theories that can be understood through heterotic/F-theory duality in the presence of multiple fibrations. To begin, compactifications of the heterotic string and F-theory are believed to be dual whenever the underlying geometries of the two theories take the form
\beq\label{theduals}
\text{Heterotic on}~~\pi_{h}: X_n \stackrel{\mathbb{E}}{\longrightarrow} B_{n-1}~~\Leftrightarrow ~~\text{F-theory on}~~ \tau_{f}: Y_{n+1} \stackrel{K3}{\longrightarrow} B_{n-1} 
\eeq
where $X_n$ is an elliptically fibered (with section) CY $n$-fold and $Y_{n+1}$ is a compatibly $K3$ and elliptically fibered (both with section) CY $n+1$-fold. This correspondence is an adiabatic extension of an $8$-dimensional duality \cite{Vafa:1996xn} ({i.e.} heterotic theory compactified on $T^2$ related to F-theory on a $K3$ surface), fibered over a shared base manifold $B_{n-1}$ to obtain lower-dimensional dualities. 

As has been recently noted \cite{Donagi:2012ts}, heterotic/F-theory duality is one form of a weakly coupled limit of F-theory (similar to so-called Sen limits \cite{Sen:1997gv,Aluffi:2007sx,Collinucci:2008pf,Collinucci:2008zs,Clingher:2012rg} which connect F-theory to weakly coupled Type IIB orientifold theories and even more general weakly coupled limits of the theory \cite{Donagi:2012ts,Heckman:2013sfa}).
The universal framework for discussing such a duality is the notion of a semi-stable degeneration \cite{mumford,Donagi:2012ts}, in which the CY manifold degenerates from a smooth manifold into a fiber product of two \emph{log Calabi-Yau} varieties, $M_1,M_2$, glued along a common effective divisor, $D \subset M_i$, $i=1,2$:
\beq\label{semi_stab_degen}
Y_{n+1} \to M_1 \cup_D M_2 \ . 
\eeq
A log Calabi-Yau variety is defined to be a pair $(M, D)$ where $M$ is a variety and $D \subset M$ an effective divisor with $K_{(M,D)}\equiv K_M +D$ trivial ({i.e.} vanishing log canonical class). Such a variety admits a unique $(n+1,0)$-form which is holomorphic on $M\backslash D$ and has at most logarithmic poles along $D$, whose residue is the holomorphic $n$-form on $D$. In general the fiber product in \eref{semi_stab_degen} is a singular variety which can be deformed back into a CY manifold by the smoothing theorem of \cite{kawamata}.

The canonical example of such a degeneration is the well-known \emph{stable degeneration limit} \cite{Morrison:1996na,Morrison:1996pp,Friedman:1997yq} of heterotic/F-theory duality in which the F-theory geometry $Y_{n+1}$ in \eref{theduals} degenerates as in \eref{semi_stab_degen}. In this limit the log-CY ``halves" of $Y_{n+1}$ take the form of fibered $(n+1)$-folds, $\pi_i: M_i \to B_{n-1}$ with fibers given by rationally elliptically fibered surfaces (commonly referred to as a $dP_9$ surface in the physics literature). Here the $(n+1)$-folds $M_i$ are not themselves Calabi-Yau manifolds, but the divisor upon which they are glued (as in \eref{semi_stab_degen}) is Calabi-Yau. In fact, here $D$ is simply the CY $n$-fold, $X_n$ -- the heterotic compactification geometry in \eref{theduals}. Moreover, in the stable degeneration limit, the geometric moduli of $M_1, M_2$ correspond to the moduli of the pair of slope-stable, holomorphic bundles (with structure group embedded into $E_8 \times E_8$) appearing in the heterotic compactification (see \cite{Friedman:1997yq,Aspinwall:1998bw,Anderson:2014gla} for reviews). To understand the results of the following sections, it will be useful to briefly review here some of aspects of this standard correspondence to remind the reader of several key geometric features of this duality.

First, the fibration structures in \eref{theduals} can be written even more explicitly. The paired heterotic/F-theory geometries given in \eref{theduals} involves both elliptic and $K3$ fibered manifolds. In particular, the F-theory geometry, $Y_{n+1}$ must be compatibly $K3$ and elliptically fibered. The requirement of these two fibration structures implies further that $Y_{n+1}$ be elliptically fibered over a complex $n$-dimensional base, ${\cal B}_n$ which is in turn rationally fibered:
\begin{equation}
\begin{array}{lllll}
&~~~~Y_{n+1}&\xrightarrow{~~\mathbb{E}~~}&{\cal B}_n&\\
 &K3~\Big\downarrow&&~\Big\downarrow~\mathbb{P}^1& \\
&~~~~B_{n-1} &\xleftrightarrow{~~=~~} &B_{n-1}&
\end{array}
\label{nested_fib}
\end{equation}
The existence of a section in any two of the non-trivial fibrations above is enough to guarantee the existence of a section in the third fibration ({i.e.} if ${\cal B}_n \stackrel{\mathbb{P}^1}{\longrightarrow} B_{n-1}$ and $Y_{n+1} \stackrel{\mathbb{E}}{\longrightarrow} {\cal B}_n$ both admit sections then so does the fibration $Y_{n+1} \stackrel{K3}{\longrightarrow} B_{n-1}$). 

For heterotic/F-theory duality then, there are three possibilities that are immediately relevant in the context of multiple fibrations in the F-theory CY geometry, $Y_{n+1}$:
\begin{enumerate}
\item \emph{Case 1}: $Y_{n+1}$ admits multiple $K3$ fibrations which \emph{share an elliptic fibration}. In order for this to occur the nested fibration structure in \eref{nested_fib} implies that ${\cal B}_n$, must admit multiple rational ({i.e.} $\mathbb{P}^1$) fibrations:
\beq\label{multp1_first}
 \xymatrix{
& {\cal B}_n \ar[ld]^{\rho}_{\mathbb{P}^1} \ar[rd]^{\mathbb{P}^1}_{\rho'} &\\
B_{n-1}& & B'_{n-1}}
\eeq
with $\rho: {\cal B}_n \to B_{n-1}$ and $\rho': {\cal B}_n \to B'_{n-1}$. Since the effective physics of F-theory depends only on the structure of the elliptic fibration $\pi_f: Y_{n+1} \to {\cal B}_n$ it is clear that the theory does not change depending on ``which way up" the base ${\cal B}_n$ is oriented in \eref{multp1_first}. However, clearly the construction of the dual heterotic theory is markedly different ({i.e.}, different base manifolds $B_{n-1}$ in \eref{theduals} and different semi-stable degenerations as in \eref{semi_stab_degen}). In this case we see that in order for heterotic/F-theory duality to hold there must exist a further heterotic/heterotic correspondence between compactifications on two different CY geometries (with vector bundles over them),
\beq\label{two_hets}
\pi_{h}: X_n \to B_{n-1}~~~~~ \text{and}~~~~\pi'_{h}: X'_n \to B'_{n-1} \ , 
\eeq
which \emph{yield the same effective theory}. In this case we see that the multiple fibration structure of $Y_{n+1}$ yields a true \emph{string duality} in the usual sense.

\item \emph{Case 2}: $Y_{n+1}$ admits multiple $K3$ fibrations with \emph{distinct elliptic fibrations}. In this case we are once again led to multiple heterotic geometries as in \eref{two_hets}, but if the condition of a shared elliptic fiber to the two $K3$ surfaces is dropped, the effective physics for each choice of fibration can be different. This is similar to the examples encountered in F-/M-theory correspondence in Section \ref{m_f_duals}. As in those examples, the expectation is that the related heterotic geometries (as in \eref{two_hets}) and the F-theory vacua (corresponding to the two choices of elliptic fibration) will all lead to the same effective physics upon a circle reduction and Higgsing in one dimension lower. In this case, we see not a duality of theories, but a \emph{correspondence leading to a shared Coulomb branch in a lower dimensional theory}. This correspondence includes a shared region of moduli space and the structure of the common lower dimensional theory can yield important insights into the distinct higher dimensional theories. See \cite{Bonetti:2011mw,Bonetti:2013cza,Grimm:2013oga,Grimm:2015mua} for useful tools in such uplifts of M-/F-theory in $5(6)$- and $3(4)$-dimensions.

At the level of the effective theory, the dimensional reduction and shared Coulomb branch of these theories is the same as that analyzed in Section \ref{m_f_duals}. As a result, we will not further explore such correspondences in this Section.

\item \emph{Case 3}: $Y_{n+1}$ admits only one $K3$ fibration with \emph{multiple elliptic fibrations}. In this case the geometry takes the form 
\beq\label{same_k3}
\xymatrix{
{}& *+[r]{Y_{n+1}}
\ar[dl]^{\pi}_{\mathbb{E}}\ar[dr]^{\mathbb{E}}_{\pi'}  & {}\\
{\cal B}_n\ar[rd]^{\mathbb{P}^1}_{\rho} &{}&*+[r]{{\cal B'}_{n}}\ar[dl]^{\rho'}_{\mathbb{P}^1}\\
{}&B_{n-1}&{}
}
\eeq
The dual heterotic geometry must be a Calabi-Yau $n$-fold which is elliptically fibered over $B_{n-1}$ but in general the two ``orientations" of the elliptic fibration (and hence the two different $\mathbb{P}^1$ bases to that elliptic fibration) correspond to distinct weakly coupled limits of the F-theory geometry. As a result, the dual heterotic theories will be different, but clearly connected in a broader moduli space. In addition, it is important to note that these different heterotic theories will be defined over \emph{the same compactification CY geometry}, $\pi_h: X \to B_{n-1}$.
\end{enumerate}
See Table~\ref{tb:three_cases} for a summary of the fibration structures in the above three cases.

\begin{table}
\begin{center}
\begin{tabular}{|c ||c | c|}
\hline
Case & $K3$ Fibration & Elliptic Fibration\\ \hline
1& distinct & shared \\
2& distinct & distinct \\
3 & shared & distinct \\ \hline
\end{tabular}\end{center}
\caption{Three different Cases for the multiple nested-fibration structures of the $K3$ and elliptic fibrations that can be found within a Calabi-Yau manifold. Such multiple fibrations lead to novel links between various effective theories via either heterotic/F-theory duality or F-/M-theory correspondence.} 
\label{tb:three_cases}
\end{table}

In the following Subsections we will explore each of the possibilities above in various dimensions. In each it is important to realize that in determining the heterotic/F-theory dual theories, the notion of a weak coupling limit implicit in a semi-stable degeneration is characterized by two limits in the dual geometries. In the heterotic theory, we require the limit of \emph{large volume of $X_n$ and weak coupling} and in the F-theory geometry this corresponds to a limit in which semi-stable degeneration occurs as in \eref{semi_stab_degen} and the volume of the $\mathbb{P}^1$ fiber in $\rho: {\cal B}_n \to B_{n-1}$ (which is related to the heterotic dilaton under the duality \cite{Morrison:1996na,Morrison:1996pp}) is also taken to the appropriate limit. In the situation where the CY geometry admits multiple fibrations as described above it is clear that the semi-stable degenerations and the role of the heterotic dilaton will, in general, differ for distinct choices of fibration. Thus, it should be expected that these correspondences should generically involve relationships between perturbative and non-perturbative degrees of freedom in the theory and different values of the heterotic coupling. These possibilities will be concretely illustrated in the following Subsections.

As a final observation, it should be noted in the context of this work that heterotic/F-theory duality has proven to be a rich framework in which to explore the possible vacua of heterotic and F-theory compactifications. Far from being a rare occurrence within known constructions of CY manifolds, the fibration structures in \eref{theduals} seem in fact to be generic (see Section \ref{sec:geometry} and \cite{us_to_appear} for a discussion). Moreover, since it is known (in $8$-, $6$- and $4$-dimensions) that the number of distinct classes of heterotic compactifications over elliptically fibered manifolds is finite \cite{Anderson:2014gla}, heterotic/F-theory duality presents a well-constrained framework with which to explore the effective theories. This has been employed in \cite{Rajesh:1998ik,Berglund:1999qk,Anderson:2014gla} to characterize and bound degrees of freedom in both heterotic and F-theory compactifications. We will further explore such constraints in Section \ref{eta_bounds}. In addition, heterotic/F-theory duality has recently shed further light on $U(1)$'s and discrete symmetries in F-theory \cite{Cvetic:2015uwu,Anderson:2015cqy,Cvetic:2016ner}.

To explore the effects of multiple fibrations in heterotic/F-theory duality we will consider $8$-, $6$- and $4$-dimensional compactifications in turn and see that multiple fibrations can play a role in a number of correspondences -- some well-known string dualities and some novel relationships.

\subsection{Heterotic/F-theory duality in $8$-dimensions and T-duality}
The duality between the $E_8 \times E_8$ heterotic string and F-theory in $8$-dimensions played a fundamental role in initially defining F-theory itself \cite{Vafa:1996xn}. As we will see below, the dual $8$-dimensional theories provide one of the simplest contexts in which the role of multiple fibrations and string dualities can be completely described. 

In the $8$-dimensional theory we consider the heterotic string (either $E_8 \times E_8$ or $SO(32)$) compactified on a two torus, $T^2$, and F-theory as defined by an elliptically fibered $K3$ manifold:
\beq\label{8d_dual}
\text{Heterotic on}~~T^2~~\Leftrightarrow ~~\text{F-theory on}~~ \pi_{f}: K3 \stackrel{\mathbb{E}}{\longrightarrow} \mathbb{P}^1
\eeq
Phrased in the language of log-semi-stable degenerations given above, the possible dualities available to F-theory in $8$-dimensions are remarkably constrained. As shown in \cite{Clingher:2003ui}, a $K3$ surface admits \emph{only two} classes of the log semi-stable degenerations. The first is the usual ``stable degeneration limit" \cite{Friedman:1997yq} to the weakly coupled $E_8 \times E_8$ heterotic theory. This takes the form of the limit described above in which the $K3$ surface degenerates as a fiber product of two rationally elliptically fibered surfaces
\beq\label{stab_degenk3}
K3 \to dP_9 \cup_{T^2} dP_9
\eeq
glued along a torus (the heterotic compactification geometry). In $8$-dimensions, the only other log semi-stable degeneration limit is of the form
\beq\label{so32lim}
K3 \to M_1 \cup_{T^2} M_2
\eeq
where $M_1,M_2$ are \emph{rationally fibered surfaces} (the generic fiber to the degenerate geometry in \eref{so32lim} is two smooth rational curves meeting in two points \cite{Clingher:2003ui}). As shown in \cite{Clingher:2003ui} this limit corresponds to the $SO(32)$ heterotic string and is essentially the \emph{only other distinct log semi-stable degeneration of a $K3$ surface}  (see \cite{Donagi:2012ts} for a discussion).

In the context of this work, an important observation was made by Candelas and Skarke in \cite{Candelas:1997pq} where it was noted that these two limits can be thought of as \emph{two elliptic fibrations of the same $K3$ surface}. Note that this is an example of ``Case 3" fibration structure in Table~\ref{tb:three_cases}, as depicted by \eref{same_k3}, where in this simple geometry the base $B_{n-1}$ is a point. A simple toric realization was provided in \cite{Candelas:1997pq}.
 
In terms of the toric data,\footnote{Readers are kindly referred to \cite{fulton} for an introduction to toric geometry.} the dual polyhedron $\Delta^*$ for the geometry in question is determined by the convex hull of the following lattice points in $N\simeq \IZ^3$,
 \beq
 (1,0,0)\ ,~~~(0,1,0)\ ,~~~(-2,-3,6)\ ,~~~(-2,-3,-6) \ . 
 \eeq
The corresponding $K3$ surface has a Picard lattice of rank $18$, with two moduli appearing as embedding ({i.e.}, complex structure) moduli for the given toric realization. The fiber associated to the $E_8 \times E_8$ theory is the standard ``Weierstrass" $\mathbb{P}_{1,2,3}[6]$ fiber, realized by the convex hull of the points,
\beq\label{e8_triangle}
 (1,0,0)\ ,~~~(0,1,0)\ ,~~~(-2,-3,0) \ . 
 \eeq
In this case the two ($dP_9$) ``halves" of the $K3$ and the appropriate stable degeneration limit of \eref{stab_degenk3} can be identified from the halves of the polyhedron divided by the triangle in \eref{e8_triangle}. Each half is an $E_8$ ``top" in the notation of \cite{Candelas:1996su,Candelas:1997pq,Candelas:2012uu} and corresponds to an extended Dynkin Diagram of $E_8$ ({i.e.}, the resolution of an $E_8\times E_8$ singularity).

\begin{figure}
\centering \includegraphics[width=0.3\textwidth]{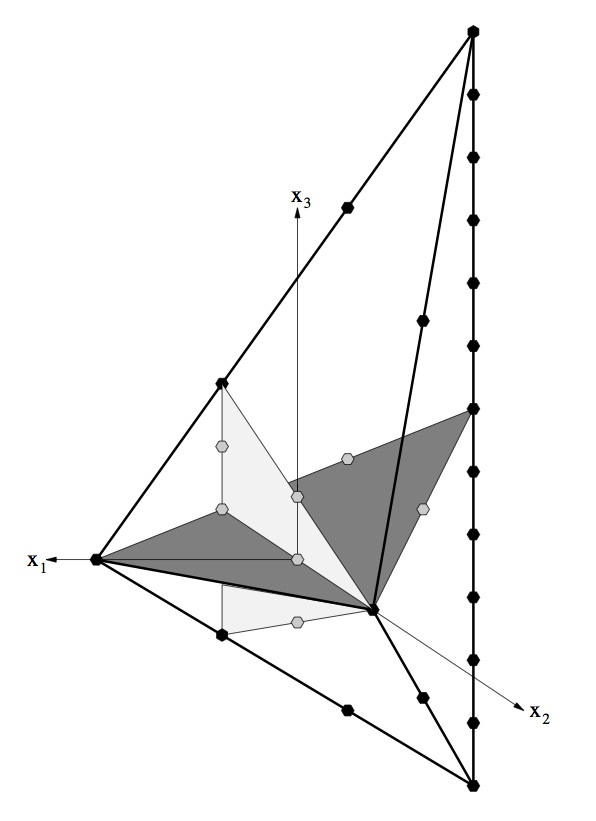} \caption{A toric polyhedron from~\cite{Candelas:1997pq} that can be used to define a $K3$ manifold with ${\rm rk}(Pic(X))=18$ as a toric hypersurface. As noted in \cite{Candelas:1997pq} this polyhedron admits two distinct elliptic fibrations. The two fibrations of the manifold can be viewed as the resolutions of singular varieties with gauge group $E_8 \times E_8$ and $SO(32)$, respectively. The highlighted triangles represent the two elliptic fibers.} \label{fig:k3}
\end{figure}

 The second elliptic fibration is described by the convex hull of the points,
 \beq\label{so32_triangle}
  (0,1,0)\ ,~~~(0,-1,-2)\ ,~~~(0,-1,2) \ . 
 \eeq
 This triangle is dual to the Newton polyhedron of $\mathbb{P}_{1,1,2}[4]$. This fiber type is compatible with the existence of \emph{two sections} to this elliptic fibration as required for the $SO(32)$ theory \cite{Aspinwall:1996nk} (note that in addition to the zero section, the second section in this case is a torsion element of the Mordell-Weil group). Here the decomposition in \eref{so32lim} is asymmetric and this can be seen by the way that the triangle in \eref{so32_triangle} divides the divisors: one side consists of a single point, while the other side contains 17 divisors filling out the extended Dynkin diagram of $SO(32)$ (see \cite{Candelas:1997pq} for further details). The polyhedron with the two elliptic fiber ``triangles" highlighted is shown in Figure \ref{fig:k3}.
 
In this simple case the fact the two fibers are respresentative of $E_8 \times E_8$ and $SO(32)$ heterotic theories provides a simple link between \emph{interchanging two elliptic fibrations and string dualities}. It is well known that the $E_8 \times E_8$ and $SO(32)$ heterotic theories can be related to each other by \emph{T-duality} whenever the compactification geometry includes a circle factor. Thus, by dimensionally reducing both sides of these heterotic/F-theory dual pairs on $S^1$ (and choosing appropriate Wilson lines), it is clear in this case that fiber orientation and T-duality must correspond. 

Indeed, one can go further and note that in this case since the heterotic geometry of $T^2$ is topologically $S^1 \times S^1$, it should be possible to understand the two limits of the theory even in $8$-dimensions. It is helpful to recall the steps leading to a correspondence between the two heterotic theories in the case of compactification on $S^1$: 1) First, beginning at a general point in the moduli space of the $SO(32)$ theory, Wilson lines can be chosen to break the symmetry group to $SO(16) \times SO(16)$. 2) Considering $SO(32)$ theory with large radius for the $S^1$, one can let $r\to 1/r$. This leads in the T-dual description to an $E_8 \times E_8$ theory with $SO(16)\times SO(16)$ symmetry. 3) In the $E_8 \times E_8$ theory, one can continuously deform the Wilson lines to return to a generic point in the moduli space of the $E_8 \times E_8$ theory. It is really the second step which is crucial to understand the duality and this argument can be applied to the heterotic theory on an $n$-torus, $T^n$~\cite{Berkooz:1996iz}.

In the context of the concrete geometry given above, these two, T-dual limits can be understood using the criteria of Kollar \cite{Kollar:2012pv} discussed in Section \ref{sec:geometry}. An application of the criteria, as outlined in Conjecture $1$ in Section \ref{sec:geometry}, shows that there are two fibrations and hence, two divisors $D_1,D_2$ capable of describing the $\mathbb{P}^1$ bases of the two elliptic fibration\footnote{Note that for $K3$ surfaces with high rank Picard lattices, there can be many fibrations. See for example, \cite{Braun:2013yya} for tools to systematically find all such fibrations and \cite{McOrist:2010jw} for more general (U-duality) correspondences possible in $8$- (and lower) dimensions.}. In the F-theory geometry then, the T-duality of the heterotic theory can be manifested as an interchange of these base $\mathbb{P}^1$s and \emph{and hence, the elliptic fibrations}. That is,
\beq
\text{Heterotic:}~~SO(32) \stackrel{r \leftrightarrow 1/r}{\longleftrightarrow} E_8 \times E_8~~~~ \Leftrightarrow~~\text{F-theory:} ~~~D_1 \stackrel{\mathbb{E}_1 \leftrightarrow \mathbb{E}_2}\longleftrightarrow D_2
\eeq
The fact that there are manifestly two elliptic fibrations in this $K3$ surface is a confirmation of the known/simple structure of the $8$-dimensional effective theories theory (and the log-semi-stable degenerations mentioned in the previous Subsection). In general, it is possible for CY $n$-folds to admit many more than two fibrations and the corresponding string dualities in consideration must necessarily be more complex. We turn now to these lower dimensional theories.

\subsection{Multiple fibrations and $6$-dimensional theories}
In $6$-dimensions multiple fibrations can begin to play an even more non-trivial role. Unlike $K3$ surfaces, for CY $3$-folds, the existence of a fibration structure ({i.e.}, $\mathbb{E}$- or $K3$-fibration) is a \emph{topological invariant} \cite{Kollar:2012pv} and cannot change under infinitesimal deformations of the geometry. We will briefly review here a few of the canonical $6$-dimensional dualities and their relationship to fibration structures. It should be noted that many of the observations about heterotic/F-theory duality can be extended/linked to heterotic/Type IIA duality in $4$-dimensions. For the sake of brevity we focus only on the F-theory description here and leave the logical extensions of these results to Type IIA dualities implicit.

\subsubsection{Heterotic $E_8 \times E_8$ and $SO(32)$ duality}
To begin simply, it is clear that having established the $8$-dimensional duality described in the previous Subsection, this can be fibered over a shared base manifold $\mathbb{P}^1$ to provide a $6$-dimensional example of the ``Case 3" correspondence of fibrations in Table~\ref{tb:three_cases}. It is well known that an elliptically fibered CY 3-fold in Weierstrass form over the base $\mathbb{F}_4$:
\beq\label{f4_geom}
\pi_f: Y_3 \to \mathbb{F}_4
\eeq
is generically singular with a non-Higgsable $SO(8)$ symmetry \cite{Morrison:1996pp}. This geometry can be dual to \emph{either} the 
$SO(32)$ (with instanton number $24$) or the $E_8 \times E_8$ theory (with instanton embedding $(8,16)$). The Weierstrass model in \eref{f4_geom} can be tuned to become still more singular, ranging over possible subgroups of each theory. In the case of resolution of such a geometry (including the generic $SO(8)$ singularity) we expect that the $K3$ fiber of $Y_3$ will admit more than one elliptic fibration as described above.  With this expected adiabatic realization of the familiar dual theories in hand, we turn next to a correspondence that cannot be realized in the $8$-dimensional theory.

\subsubsection{Heterotic/Heterotic duality and the dual F-theory geometry}\label{dmw_recap}
One of the most well established and understood examples of a $6$-dimensional duality is the detailed heterotic/heterotic correspondence discovered by Duff, Minasian, and Witten in \cite{Duff:1996rs}. There it was observed that the $E_8 \times E_8$ heterotic theory compactified on a $K3$ surface with $G_i$-bundles, $V_i \to K3$ ($i=1,2$) chosen such that $G_i \subset E_8$ and $c_2(V_1)=c_2(V_2)=12$ (the so-called ``symmetric embedding") admits a self-duality. This duality not only inverts the gauge coupling and dualizes the antisymmetric tensor, but also acts non-trivially on hypermultiplets and relates perturbative and non-perturbative gauge sectors of the heterotic theory \cite{Duff:1996rs}.

The relationship of this remarkable duality to F-theory was realized immediately \cite{Morrison:1996na} and in particular, it was observed in \cite{Morrison:1996na,Aspinwall:1996mw} that the existence of the two dual heterotic theories on $K3$ implied that each heterotic theory must give rise to an F-theory dual -- and that, in fact, this could be understood as a single dual CY 3-fold geometry with multiple fibrations. The requirement of a symmetric embedding of equal instantons into each $E_8$ factor indicated that (under the usual heterotic/F-theory dictionary in $6$-dimensions \cite{Morrison:1996na,Morrison:1996pp,Bershadsky:1996nh}) the F-theory dual geometry should be 
\beq\label{symm_weier}
\pi_f: Y_3 \to \mathbb{F}_0=\mathbb{P}^1 \times \mathbb{P}^1 \ , 
\eeq
and the heterotic/heterotic duality of \cite{Duff:1996rs} could clearly be realized as \emph{two $K3$ fibrations of the same CY $3$-fold} simply by choosing which $\mathbb{P}^1$ in $\mathbb{F}_0$ to consider as part of the $K3$ fiber. These two $K3$ fibrations clearly share an elliptic fibration and are a simple example of the ``Case 1" fibration structure in Table~\ref{tb:three_cases}.

To study multiple fibrations more generally, it will be useful to review several observations about the duality given above, first made in \cite{Morrison:1996na}. The dual theories must involve an interchange of perturbative/non-perturbative origins for gauge fields in the dual heterotic/F-theory geometries.  This can be realized readily even in the simplest form of the doubly-fibered theory. Consider a Weierstrass model
\beq\label{stand_weier}
y^2=x^3 +f_{(8,8)}(\bold u, \bold v)x + g_{(12,12)}(\bold u, \bold v)
\eeq
where $\bold u=(u_0:u_1)$ and $\bold v = (v_0:v_1)$ are homogeneous coordinates on $\mathbb{P}^1_{\bold u} \times\mathbb{P}^1_{\bold v}$, $f \in H^0(\mathbb{P}^1 \times\mathbb{P}^1, K_{B_2}^{-4})$ and $g \in H^0(\mathbb{P}^1 \times\mathbb{P}^1, K_{B_2}^{-6})$ and $K_{B_2}=\cO(-2,-2)$. For general choices of the complex structure moduli this Weierstrass model is smooth -- corresponding to a completely broken $E_8 \times E_8$ symmetry in the dual heterotic theory. By tuning the complex structure it is possible to consider larger unbroken symmetry groups in $6$-dimensions. Let us consider here the effect of such a tuning on \emph{both} $K3$-fibrations: $\rho_{\bold u}: Y_3 \to \mathbb{P}^1_{\bold u}$ and $\rho_{\bold v}: Y_3 \to \mathbb{P}^1_{\bold v}$, where the two $\mathbb{P}^1$ bases are defined by choosing a rational fiber/base in \eref{symm_weier} and \eref{stand_weier}.

To choose a heterotic dual, a base $\mathbb{P}^1$ to the $K3$-fibration is selected and the coefficients of $f_8$ and $g_{12}$ expanded with respect to such a basis. Let $u_0,u_1$ be coordinates on the fiber $\mathbb{P}^1_f$ and $v_0,v_1$ be coordinates on the base $\mathbb{P}^1_b$ of $B_2=\mathbb{F}_0$. Following the notation of \cite{Morrison:1996na,Morrison:1996pp,Bershadsky:1996nh} consider the patch in $\mathbb{F}_0$ where $u_0=v_0=1$ and expand $f,g$ in \eref{stand_weier} as
\begin{align}\label{p1p1_weier}
&f \sim \sum_{i=0}^8 u_1^if_8^{(i)}(v_1) \\
&g \sim \sum_{j=0}^{12}u_1^jg_{12}^{(j)}(v_1)
\end{align}
For this orientation of the $\mathbb{P}^1$ fiber of $B_2$, it is possible to tune a non-abelian singularity while keeping the dual heterotic $K3$ surface smooth. The argument that will follow can be done for any symmetry $G \subset E_8$, but here we will illustrate it for the simple example of $E_7$ symmetry. An $E_7$ singularity in the fiber of $Y_3$ requires $(f,g)$ vanish to degrees $(3,5)$ on a divsior within $B_2$ \cite{kodaira}. This means that $f,g$ take the form
\begin{align}\label{e7_eg}
&f \sim  u_1^3 f_8^{(3)}(v_1)+u_1^4f_8^{(4)}(v_1) + \ldots \\
&g \sim u_1^5 g_{12}^{(5)}(v_1)+ u_1^6 g_{12}^{(6)}(v_1)+ \ldots \nonumber
\end{align}
({i.e.}, all coefficients $f_8^{(i)}$ and $g_{12}^{(j)}$ with $i<3$ and $j<5$ are set to zero). For this choice, an $E_7$ singularity is associated to a stack of 7-branes wrapping the $\mathbb{P}^1_{b}$ divisor, located at the point $u_1=0$ in the $\mathbb{P}^1_{f}$. 

To read off the heterotic dual theory, a semi-stable degeneration limit must be identified as in \eref{semi_stab_degen}. In this case, the heterotic dual $K3$ surface ($D$ in \eref{semi_stab_degen}) is determined by the order $4,6$ coefficients in $f,g$ respectively \cite{Morrison:1996na,Morrison:1996pp,Friedman:1997yq}. The $K3$ Weierstrass model takes the form,
\beq\label{k3gen}
Y^2=X^3 + f_8^{(4)}(v_1)X + g_{12}^{(6)}(v_1) \ , 
\eeq
with $f_8^{(4)}$ and $g_{12}^{(6)}$ \emph{general polynomials} over the base $\mathbb{P}^1$ of the elliptically fibered $K3$ surface $\pi_h: K3 \to \mathbb{P}^1$. In this case, the heterotic dual theory would consist of a smooth $K3$ surface with a perturbatively realized $E_7$ symmetry obtained from an $SU(2)$ bundle on a smooth $K3$.

The tuning above is standard in heterotic/F-theory duality in $6$-dimensions. However, as noted in \cite{Morrison:1996na} a more interesting observation is possible if the roles of fiber and base $\mathbb{P}^1$ in $B_2$ are switched above while \emph{holding the complex structure fixed}. Under this exchange, the \emph{elliptic} fibration of $Y_3$ is unchanged, and thus, the $6$-dimensional F-theory effective theory is still manifestly the same ({i.e.}, non-Abelian $E_7$ symmetry). However, exchanging the two $\mathbb{P}^1$s in \eref{e7_eg} clearly changes which divisor $D$ serves as the dual $K3$ surface in the stable degeneration limit of \eref{semi_stab_degen}. Upon interchanging of the roles of the $\mathbb{P}^1$s, we have a new expansion (again, without changing the complex structure in \eref{stand_weier}):
\begin{align}
&f \sim \sum_{i=0}^8 v_1^i\tilde{f}_8^{(i)}(u_1) \ ,  \\
&g \sim \sum_{j=0}^{12}v_1^j\tilde{g}_{12}^{(j)}(u_1) \ .
\end{align}
With the tuning given above, it is possible to once again read off the coefficients, $\tilde{f}_8^{(4)}$ and $\tilde{g}_{12}^{(6)}$, which will form the coefficients of the new dual $K3$ surface. With the complex structure fixed as described, the heterotic geometry takes the form,
\beq
Y^2=X^3 +(a_3 u_1^3+ a_4u_1^4 + \ldots)X + (b_5u_1^5+b_6 u_1^6+ \ldots)  \ , 
\eeq
with $a_i, b_j$ constants. Rather than the general $K3$ Weierstrass model of \eref{k3gen}, here we see that choice of the tuned complex structure for $Y_3$ in \eref{e7_eg} has forced the dual $K3$ to be singular for this choice of fibration in $Y_3$. Moreover, the form of the singularity is exactly an $E_7$ type singularity, as expected. Thus, \emph{the interchanging of fibrations in $Y_3$ results in changing the gauge symmetry from one with a perturbative origin to a non-perturbative origin in the dual heterotic theory}. This was illustrated above for $E_7$ symmetry but an analagous calculation can be obtained for any tuning of a $G$-type singularity in the Weierstrass model \eref{p1p1_weier} over $\mathbb{P}^1 \times \mathbb{P}^1$. As a final note, it was demonstrated in \cite{Berkooz:1996iz} that the heterotic/heterotic duality of \cite{Duff:1996rs} was once more related to T-duality and that the non-perturbative physics described by the singular $K3$ surface described above could be understood in the context of $SO(32)$ small instantons on the singular $K3$. Unlike the example of the previous Subsection, in this case the bundle with structure group in $SO(32)$ satisfies $c_2(V)=12$ (and the symmetric, $(12,12)$ embedding in the $E_8 \times E_8$ theory).

Another confirmation \cite{Morrison:1996na} that the duality of \cite{Duff:1996rs} is realized by interchanging the $K3$ fibrations in $Y_3$ can be seen by inspection of the heterotic coupling and its image under heterotic/F-theory duality. As explained in \cite{Morrison:1996na,Morrison:1996pp}, given F-theory on $\pi_f: Y_3 \to \mathbb{F}_n$, the heterotic string coupling in the $6$-dimensional theory is mapped to the following ratio of the volumes of the $\mathbb{P}^1$ base and fiber of $\mathbb{F}_n$:
\beq\label{coupling_match}
e^{-2 \phi}= \frac{k_b}{k_f}
\eeq
where $k_b$ and $k_f$ denote the volumes of the two $\mathbb{P}^1$'s. Thus, it is clear that the interchange of the two $K3$ fibrations in $\pi_f: Y_3 \to \mathbb{P}^1 \times \mathbb{P}^1$ ({i.e.} $k_b \leftrightarrow k_f$) corresponds to an inversion of the heterotic coupling, exactly as expected. In summary, the heterotic/heterotic duality in $6$-dimensions of \cite{Duff:1996rs} is manifested in the simple case of $B_2=\mathbb{F}_0$ as the interchange of $K3$ fibrations in the F-theory dual. 

With this review in hand, there are several important generalizations of this idea in $6$-dimensions that can be studied. First, it can be readily seen by considering more general CY $3$-folds, $Y_3$, that forms of the heterotic/heterotic duality of \cite{Duff:1996rs} can be realized for geometries much more general than the Weierstrass form illustrated above in \eref{stand_weier}. For example, geometries with multiple sections to the elliptic fibration will still obey the principles described above. For example this complete intersection:
\beq\label{K3CY3}
Y_3=\left[\begin{array}{c|cc}
\IP^{1} & 1 & 1 \\
\IP^{2} & 1 & 2   \\\hdashline
\IP^1 &    1& 1 \\
\IP^1 & 1 & 1 
\end{array}\right] \ 
\eeq 
is a smooth manifold $\pi_f:Y_3 \to \mathbb{F}_0$ with $h^{1,1}(Y_3)=4$, $h^{2,1}(Y_3)=50$. In this case there are \emph{two} rational sections to the fibration ({i.e.} ${\rm rk}\, MW =1$, see \cite{us_to_appear} for details), leading to a dual heterotic $K3$ surface with an elliptic fibration that also admits two sections. Once again, the interchange of the two $K3$ fibers leads to the same perturbative/non-perturbative correspondence described above for non-Abelian gauge fields. However, the Abelian ({i.e.} $U(1)$) gauge fields share an origin from reducible bundles in both the smooth and the singular cases. This generalization can also involve the presence of Green-Schwarz massive $U(1)$ symmetries in the $6$-dimensional theory \cite{Lukas:1999nh,Louis:2011hp} and has been the subject of several recent explorations in the context of heterotic/F-theory duality \cite{Cvetic:2015uwu,Anderson:2015cqy,Cvetic:2016ner}.

\subsubsection{Example of non-perturbative heterotic/heterotic duality and its F-theory dual}\label{new6dim}

In view of the above results, it is natural to go further and ask whether or not there are more general forms of multiple $K3$ fibrations in F-theory (with different bases to the elliptic fibration, other than $\mathbb{F}_0$) and whether they can lead to any generalizations of the heterotic/heterotic duality. Beginning with $K3$ fibrations which share an elliptic fibration (Case 1 in Table~\ref{tb:three_cases}) it is clear that we require a base $B_2$ which admits more than one $\mathbb{P}^1$ fibration (as in \eref{multp1_first}). For $6$-dimensional heterotic/F-theory dual pairs that are \emph{purely perturbative in the heterotic theory}\footnote{I.e., a large volume, smooth $K3$ surface and only one tensor multiplet in the $6$-dimensional effective theory.}  the only possible bases, $B_2$, are the Hirzebruch surfaces, $\mathbb{F}_n$ \cite{Morrison:1996na}. Considering these first then, the only Hirzebruch surface that admits more than one rational fibration is $\mathbb{F}_0=\mathbb{P}^1 \times \mathbb{P}^1$, which we have already studied.

However, for more general rationally fibered surfaces, $B_2$, it is easy to see that there can be more possibilities -- corresponding to non-perturbative $6$-dimensional theories \cite{Seiberg:1996vs,Candelas:1996ht}. In terms of F-theory geometry, the difference between perturbative/non-perturbative heterotic dual theories is visible in the structure of the base $B_2$. The Hirzebruch surfaces are $\mathbb{P}^1$-bundles over $\mathbb{P}^1$ ({i.e.}, with nowhere degenerate fiber). More general rationally fibered surfaces fall under the category of $\mathbb{P}^1$ fibrations rather than bundles (see for example, ``conic bundles" \cite{opac-b1102947}), in which the $\mathbb{P}^1$ fiber can degenerate over points in the base. Unlike elliptic fibrations, the possible degenerations of $\mathbb{P}^1$-fibrations are quite simple and generally consist only of a single $\mathbb{P}^1$ becoming multiple $\mathbb{P}^1$s over points in the base. One simple example of this is the del Pezzo surfaces \cite{opac-b1102947} which are rationally fibered for $n>0$ (as are the ``generalized del Pezzo surfaces" \cite{gen_del_pezzo}). For example, in addition to viewing $dP_2$ as $\mathbb{P}^2$ blown-up at two points, the surface can also be viewed as a single blow-up of $dP_1=\mathbb{F}_1$. 

The heterotic duals to F-theory on a blown-up Hirzebruch surface are well-understood to include additional tensor multiplets in the $6$-dimensional theory \cite{Morrison:1996pp,Candelas:1996ht,Aldazabal:1996fm}. That is, the effective theories include non-perturbative effects such as $NS5$ branes in the heterotic theory ({i.e.}, $M5$-branes along the $S^1/\mathbb{Z}_2$ interval direction in the language of heterotic M-theory \cite{Horava:1995qa}). Anomaly cancellation in the heterotic theory is generalized in this case to
\beq
c_2(V_1) + c_2(V_2)+n_T=25
\eeq
where $n_T$ is the number of tensor multiplets. It is clear that as additional tensors are included in the theory, the net instanton number of the two vector bundles over $K3$ must decrease. The extremal situation (for a smooth $K3$) consists of $24$ instantons (either point-like \cite{Aspinwall:1998he,Louis:2011aa} or in the interval).

Let us consider such non-perturbative vacua in the context of heterotic/F-theory duality. It is clear that F-theory on rationally fibered bases can generically involve more than one choice of $\mathbb{P}^1$ fibration. As an illustration, consider the following complete intersection threefold with $h^{1,1}(Y_3)=5$ and $h^{2,1}(Y_3)=43$, described by:
\beq\label{non_pert_eg}
Y_3=\left[\begin{array}{c|cc:cc}
\IP^{1} & 0& 0 & 1 & 1\\
\IP^{2} & 0 & 0& 1 & 2  \\\hdashline
\IP^2 &   1 & 1  &  0 & 1 \\
\IP^1 & 1& 0 & 1 & 0 \\
\IP^1 & 0& 1 & 1 & 0  
\end{array}\right], \ \quad
B_2=dP_2=\left[\begin{array}{c|cc}
\IP^2 &  1 & 1  \\
\IP^1 &  1  & 0 \\
\IP^1 & 0 & 1  
\end{array}\right].
\eeq  
This manifold is a smooth fibration, $\pi_f: Y_3 \to dP_2$. The elliptic fiber, described as a degree-$(1,1)$ hypersurface in $dP_1=\left[\begin{array}{c|c}
\IP^1 &  1  \\
\IP^2 &  2  \\
\end{array}\right]$, admits two rational sections (a zero section given by the unique global section $\cO(-1,1,0,1,1)$ and a second rational section associated to $\cO(2,-1,1,3,3)$) \cite{us_to_appear}. The divisors in $Y_3$ are divided according to those pulled back from the base and sections (in this case there are no other additional vertical divisors). Since $h^{1,1}(B_2)=3$, and there are two rational sections: $h^{1,1}(Y_3)=5=h^{1,1}(B_2)+1+{\rm rk}\,MW$ as expected from the Shioda-Tate-Wazir theorem \cite{shioda, shioda2,COM:213767}.

For this complete intersection description of $dP_2$, there are clearly two manifest ways to view $dP_2$ as a $\mathbb{P}^1$-fibration over $\mathbb{P}^1$. The first is
\beq\label{dp2base}
B_2=dP_2=\left[\begin{array}{c|cc}
\IP^2 &  1 & 1  \\
\IP^1 &  1 & 0 \\ \hdashline
\IP^1 & 0 & 1  
\end{array}\right] \ , 
\eeq
where the $\mathbb{P}^1$ fiber is described as 
\beq
\mathbb{P}^1=\left[\begin{array}{c|cc}
\IP^2 &  1 & 1  \\
\IP^1 &  1 & 0 
\end{array}\right] \simeq
\left[\begin{array}{c|c}
\IP^1 &  1  \\
\IP^1 &  1
\end{array}\right] \ , 
\eeq 
and the second is given by the equivalent configuration matrix with the last two $\mathbb{P}^1$ rows interchanged in \eref{dp2base}. As a result the CY threefold in \eref{non_pert_eg} once again inherits \emph{two distinct $K3$-fibrations}.

In this case, the heterotic/heterotic duality will exhibit a structure in some ways similar to that explored in the previous Subsection. In general, once again the ``weakly coupled" heterotic limits of the $K3$-fibration will be different and can involve a shift between a perturbative and non-perturbative origin of gauge fields. However, some features will differ. For instance, the instanton numbers of the two bundles on the $K3$ surface need no longer be equal ({i.e.}, an asymmetric instanton embedding into $E_8 \times E_8$). In addition, the presence of a 5-branes in the interval direction indicates that  the number of tensor multiplets will be $n_T>1$. The duality will act non-trivially not only on the hypermultiplets, but also on the two tensor multiplets in the example above.

The link between the heterotic coupling and the possible tensor multiplets in F-theory can be more complicated than the map \eref{coupling_match}, derived in the case of $n_T=1$. However, a logical conjecture is that as in the perturbative theory, the heterotic coupling is fixed by the F-theory geometry to be
\beq\label{first_coupl}
e^{-2\phi}\sim \frac{vol(\mathbb{P}^1_b)}{vol(\mathbb{P}^1_f)} \ . 
\eeq
Applying this to the concrete geometry given in \eref{non_pert_eg}, for the first $K3$-fibration we have
\beq
e^{-2\phi_1}\sim \frac{t^3}{t^1 + t^2} \ , 
\eeq
where $t^1, t^2, t^3$ are the K\"ahler parameters of each $\mathbb{P}^{n_i}$-hyperplane in the CICY base description of $B_2=dP_2$ given above. Similarly, for the second fibration we find
\beq
e^{-2\phi_2}\sim \frac{t^2}{t^1 + t^3} \ . 
\eeq
Thus, the action of ``interchanging" the two $K3$-fibrations takes $t^2 \leftrightarrow t^3$ and is not simply an inversion of the coupling in this case (though it will clearly generically involve a strong/weak coupling correspondence). The validity of the formulae above can be tested by considering the blown-down limit in which $dP_2 \to \mathbb{F}_1$. In this case, this is realized by 
\beq
dP_2=\left[\begin{array}{c|cc}
\IP^2 &  1 & 1  \\
\IP^1 &  1 & 0 \\ \hdashline
\IP^1 & 0 & 1  
\end{array}\right] \longrightarrow \mathbb{F}_1=\left[\begin{array}{c|c}
\IP^2 &  1  \\
\IP^1 &  1
\end{array}\right] \ , 
\eeq
which can be obtained by sending the volume of \emph{either} the first or second ambient $\mathbb{P}^1$ to zero. In the case of the first $K3$ fiber then, we see that the limit back to the perturbative heterotic theory can be taken by considering $t^2 \to 0$ which leads to exactly the form expected by \eref{coupling_match}. Explicitly this sends \eref{first_coupl} to
\beq
\frac{t^3}{t^1}=\frac{S+E}{E}= \frac{vol(\mathbb{P}^1_b)}{vol(\mathbb{P}^1_f)}
\eeq
in the usual notation of $\mathbb{F}_1$ with $S^2=-1$, $S\cdot E=1$ and $E^2=0$ (with $S$ and $S+E$ both corresponding to sections and $E$ to the fiber class of $\mathbb{F}_1$). Likewise the second $K3$ fibration, (obtained by interchanging $t^2$ and $t^3$) also limits correctly to the case of $\mathbb{F}_1$ realized by sending $t^3 \to 0$. 

From the above it can be seen that for some geometries with $n_T \gg 1$, the F-theory compactification can correspond to a Calabi-Yau threefold with dozens of $K3$-fibrations, corresponding to various non-perturbative dualities of the form described above. It would be intriguing to characterize these more systematically in the future.

\subsubsection{Mirror symmetry and Heterotic/F-theory duality}
It is worth observing briefly that the dualities described above  -- including the interchange of perturbative/non-perturbative sectors under heterotic/heterotic duality -- share many similarities with features of mirror symmetry \cite{Hori:2003ic} as it appears in $6$-dimensional heterotic/F-theory dual pairs (or equivalently $4$-dimensional heterotic/Type IIA pairs). 
It has been conjectured (see e.g. \cite{Berglund:1998ej,Louis:2011aa,Alexandrov:2014jua}) that \emph{if a CY $3$-fold is $K3$-fibered, its mirror must also admit a $K3$-fibration} (see \cite{Louis:2011aa} for some evidence of this fact in heterotic/TypeIIA duality) and one can construct the following links:
\begin{equation}
\begin{array}{lllll}
& {\rm Heterotic} && {\rm F}\text-{\rm theory}& \\
&~~~~K3&{\Leftrightarrow}&~~~~Y_3&\\
&~~~~\,\,\Big\updownarrow&&~~~\,~\Big\updownarrow~{\rm mirror}& \\
&~~~~{K3}' &{\Leftrightarrow}&~~~~{Y}_3'&
\end{array}
\label{nested_fib2}
\end{equation}

This has lead to interesting observations regarding the expectation for the heterotic theories connected by mirror symmetry of a CY 3-fold \cite{Berglund:1998ej}: the dual heterotic theories to the mirror pair of threefolds are expected to exchange a gauge background consisting of a smooth vector bundle and $K3$ surface with one on a singular $K3$ surface involving non-perturbative gauge fields. This is reminiscent of the correspondences seen in the previous Sections. 

Moreover, mirror symmetry and T-duality once again appear in the context of ${\cal N}=2$ Type IIA compactifications where it has been observed that given a mirror pair of CY $3$-folds $(M_3, W_3)$ will lead to the same effective theory upon reducing on an $S^1$ to three-dimensions -- the same type of relationship seen in the previous Sections.

If it is true that the mirror of a $K3$-fibered CY 3-fold is also $K3$-fibered this raises an interesting possibility concerning \emph{multiple fibrations and mirror symmetry} (see also \cite{yau_mirror}). If a generic CY 3-fold admits multiple $K3$ fibrations (see \cite{usscanning,Rohsiepe:2005qg} for explorations of $K3$-fibrations in known datasets of CY $n$-folds) we propose the following conjecture:

\vspace{0.3cm}

\noindent \emph{Conjecture: Calabi-Yau three-folds in a mirror pair admit the same number of K3 fibrations.}

\vspace{0.3cm} 
 
\noindent Such a geometric conjecture would lead to an important topic of future exploration and could yield further insight into the dataset of Calabi-Yau three-folds as well as string dualities.

\subsection{Multiple fibrations and $4$-dimensional theories}
In what follows, we turn to heterotic/F-theory duality in $4$-dimensions and highlight ways that multiple fibrations can provide new windows into the structure of the ${\cal N}=1$ effective theories in $4$-dimensions. 

Heterotic/F-theory duality in $4$-dimensions is best understood in the case that both the CY $3$- and CY $4$-fold geometries are in minimal (Weierstrass) form with a single holomorphic section. We will restrict ourselves to the case of a heterotic theory without $5$-branes and in which the holomorphic vector bundles are irreducible, and will briefly review the geometric correspondences below. By the required fibration structure of the CY $4$-fold \eref{nested_fib}, the base ${\cal B}_3$ must be $\mathbb{P}^1$-fibered. As in the case of the Hirzebruch surfaces in the $6$-dimensional theory, the simplest class of geometries will correspond to bases that are $\mathbb{P}^1$ bundles over $B_2$. As in \cite{Friedman:1997yq}, such a bundle can be defined as the projectivization of two line bundles,
\beq\label{p1bund}
{\cal B}_3=\mathbb{P}(\cO \oplus {\cal L}) \ , 
\eeq
where $\cO$ is the trivial bundle and ${\cal L}$ is a general line bundle on $B_2$. In this case the topology of ${\cal B}_3$ is completely fixed by the choice of line bundle ${\cal L}$. More precisely, it is fixed by the so-called ``twist"of the rational fibration which is determined by a $(1,1)$-form, $T$, on $B_2$ -- corresponding to $c_1({\cal L})$. It is the twist which allows for a geometric matching of the degrees of freedom in the $4$-dimensional heterotic/F-theory dual pairs.

In the $E_8 \times E_8$ heterotic theory, the topology of the bundles $V_i$ ($i=1,2$) can be decomposed as
\beq\label{c2twist}
c_2(V_i)=\eta_i \wedge \omega_{\hat 0} + \zeta_i \ , 
\eeq
where $\eta_i$ and $\zeta_i$ are respectively $(1,1)$- and $(2,2)$-forms pulled back from $B_2$, and $\omega_{\hat 0}$ is the $(1,1)$-form dual to the zero section of $\pi: X_3 \stackrel{\mathbb{E}}{\longrightarrow} B_2$. For any CY $3$-fold in Weierstrass form as described above, $c_2(TX_3)=12c_1(B_2)\wedge \omega_{\hat 0} +(c_2(B_2) +11c_1(B_2)^2)$ \cite{Friedman:1997yq}. Anomaly cancellation then requires
\beq\label{twisty}
\eta_{1,2}=6c_1(B_2) \pm T
\eeq
where $T$ is a $(1,1)$-form on $B_2$. This choice of labels is not an accident and is indeed exactly the ``twist" introduced in the geometry of the F-theory base, ${\cal B}_3$. The correspondence between $\eta_i$ and $T$ was explicitly made in the stable degeneration limit in \cite{Friedman:1997yq} and generalized in \cite{Grimm:2012yq}. Note that the form of the geometry assumed -- exactly one (holomorphic) section and Weierstrass form for both $X_3$ and $Y_4$ -- is restrictive and natural extensions of this geometric correspondence (including higher rank Mordell-Weil groups, multisections, etc.) would be interesting to explore in future work. For now, we stay with the standard correspondence and explore the consequences of multiple fibrations.

\subsubsection{Adiabatic $4$-dimensional realizations of higher dimensional dualities}
To begin, we should note that all the dualities given above will have $4$-dimensional realizations obtained by fibering the higher-dimensional correspondences over a shared base manifold. Briefly, this structure includes the following possibilities:
\begin{itemize}
\item As seen in previous Sections the ``Case 2" correspondence -- Calabi-Yau $4$-folds with multiple $K3$ fibrations and distinct elliptic fibrations -- will lead to a collection of $4$-dimensional heterotic F-theory dual pairs with very different effective theories. However, as in Section \ref{m_f_duals} this network of $4$-dimensional theories will all lead to identical $3$-dimensional theories upon reduction on a circle (and going to the Coulomb branch). This is more than a shared branch to the paired theories. Since they are all described by the complex structure moduli of a single CY $3$-fold, these very distinct $4$-dimensional theories have a shared (infinitesimal) moduli space.
\item The ``Case 3" correspondence of a single $K3$ fibration with multiple elliptic fibrations can once again play a role in $SO(32)$ and $E_8 \times E_8$ dual theories in $4$-dimensions. In $4$-dimesnsions it has been established \cite{Grimm:2012yq} that the twist
\beq
T=2c_1(B_2)
\eeq
gives rise to the paired heterotic theories (the analog of the $\mathbb{F}_4$ base in F-theory in $6$-dimensions). Once again, the generic symmetry of this CY $4$-fold is $SO(8)$.
\item Lastly, the ``Case 1" geometries -- multiple $K3$ fibrations with a shared elliptic fibration -- are once again a rich playground for $4$-dimensional heterotic/heterotic correspondence. 

Within the standard set-up of heterotic/F-theory duality described above, we will consider first the case that each of the $K3$ and elliptic fibrations of $Y_4$ admits a section and that the two fibrations are compatible ({i.e.}, the $K3$ fiber is itself elliptically fibered with section, etc.). This structure is summarized by the following diagram in $4$-dimensions:
\begin{equation}
\begin{array}{lllll}
&~~~~~Y_4&\xrightarrow{~~\mathbb{E}~~}&{\cal B}_3&\\
& K3~\Big\downarrow&&\,\Big\downarrow~\mathbb{P}^1& \\
&~~~~~B_2 &\xleftrightarrow{~~=~~}&B_2&
\end{array}
\label{first_def}
\end{equation}
with the requirement of compatibility of the $K3$/elliptic fibrations and sections for each. In the case of multiple fibrations in ``Case 1", the base, ${\cal B}_3$, must admit more than one rational ({i.e.} $\mathbb{P}^1$) fibration.
\beq\label{basep1_which_wayup}
 \xymatrix{
& {\cal B}_3 \ar[ld]^{\rho}_{\mathbb{P}^1} \ar[rd]^{\mathbb{P}^1}_{\rho'} &\\
B_{2}& & B'_{2}} 
\eeq
Examples of such bases are easy to observe in all known constructions of CY $4$-folds.

As in the previous Subsection, it is useful to begin with the heterotic theory in the absence of $5$-branes. This class of bases will correspond to the Hirzebruch surfaces in $6$-dimensions and the multiple fibration structure in this case will be the analog of the heterotic/heterotic duality of Duff, Minasian and Witten \cite{Duff:1996rs}.

A class of $3$-(complex) dimensional bases that are of this form are the ``generalized Hirzebruch" manifolds, sometimes denoted $\mathbb{F}_{nmk}$ in the literature (see \cite{Berglund:1998ej}). These are $\mathbb{P}^1$ bundles of the form described in \eref{p1bund} where the subscript denotes a twist of $T=nS+mE$ over the Hirzebruch surface $\mathbb{F}_k$ (with $S^2=-k$, $S \cdot E=1$ and $E^2=0$). 

In this case, the analog of \cite{Duff:1996rs} and the base $\mathbb{F}_0=\mathbb{P}^1 \times \mathbb{P}^1$ explored in Section \ref{dmw_recap} is the threefold defined by the $(m,0)$ twist over $\mathbb{F}_0$ or equivalently, the zero-twist over $\mathbb{F}_m$. This $3$-fold can be denoted:
\beq
\mathbb{F}_{0,0,m} \simeq \mathbb{F}_{0,m,0} \simeq \mathbb{F}_{m,0,0}
\eeq
An inspection of the toric description of such a manifold yields immediately that
\beq\label{baseagain}
 \xymatrix{
& {\cal B}_3 \ar[ld]^{\rho}_{\mathbb{P}^1} \ar[rd]^{\mathbb{P}^1}_{\rho'} &\\
\mathbb{F}_{0}& & \mathbb{F}_{m}}
\eeq
both with section. As pointed out in \cite{Berglund:1998ej}, it is clear that as in the $6$-dimensional correspondence, these pairs will generically correspond to a perturbative/non-perturbative interchange of gauge fields in a $4$-dimensional heterotic/heterotic duality. The heterotic CY $3$-fold Weierstrass models over $\mathbb{F}_0$ are generically smooth, while those over $\mathbb{F}_m$ for $m>2$ are generically singular \cite{Morrison:1996pp}. Thus, a symmetry group realized by a smooth bundle $V \to X_3$ and CY $3$-fold, $\pi_h: X_3 \to \mathbb{F}_0$ for one heterotic theory ({i.e.}, one $K3$-fibration) will correspond to gauge fields associated to singularities in $X'_3 \to \mathbb{F}_m$ in the other ({i.e.}, the second $K3$-fibration). As in Section \ref{new6dim}, the inclusion of $5$-branes in the heterotic theory will vastly extend the possible geometry of the threefold bases ${\cal B}_3$ in \eref{first_def} (see for example \cite{Diaconescu:1999it}).
\end{itemize}

\subsubsection{Multiple elliptic fibrations in $4$-dimensional heterotic compactifications}\label{double_het}
In ${\cal N}=1$ compactifications to $4$-dimensions, not only can we adiabatically fiber the Heteterotic/F-theory dualities from previous sections over shared base manifolds, we also have a quantitatively new set of string dualities at our disposal. Perhaps the most interesting of these involves the study of multiply-fibered Calabi-Yau threefolds as the base manifolds of $4$-dimensional heterotic compactifications. In particular, suppose that $X$ is a CY 3-fold with the following form
\beq\label{het_which_wayup}
 \xymatrix{
& X_3 \ar[ld]^{\pi}_{\mathbb{E}} \ar[rd]^{\mathbb{E}}_{\pi'} &\\
B_{2}& & B'_{2}}
\eeq
That is, it is genus-one fibered over two different base manifolds, $\pi: X_3 \to B_2$ and $\pi': X_3 \to B'_2$. Several important observations must be made about the heterotic theory in this case. First, \emph{the structure of the heterotic theory is completely independent of the `orientation' of the heterotic fibration}. That is, the massless states of the theory, the gauge symmetry and couplings, a priori do not depend on the fibration structure of $X_3$!

However, it is clear that under the heterotic/F-theory dictionary \cite{Friedman:1997yq} described above, the F-theory duals of the two possibillites in \eref{het_which_wayup} can look very different. That is, in the case that both fibrations in \eref{het_which_wayup} admit a section, we have in principle two well-understood F-theory duals, consisting of CY 4-folds with $K3$ fibers over the distinct base manifolds $B_2$ and $B'_2$
\beq
\tau: Y_4 \stackrel{K3}{\longrightarrow} B_2 \ ,\quad\quad \tau': Y'_4 \stackrel{K3}{\longrightarrow} B'_2
\eeq
These 4-folds will be potentially topologically distinct, with different G-flux backgrounds. However, by the observation above, they must give rise to truly \emph{identical $4$-dimensional dual theories}. This then, is a new example of string duality appearing in $4$-dimensions and it will be our goal here to briefly sketch some of the possible structure of $Y_4$ and $Y'_4$. It should be noted as well that for known datasets of CY $3$-folds, the number of such inequivalent genus-one fibrations can number in the dozens -- implying that there can exist vast networks of dual CY $4$-folds whose effective theories can be linked. We will illustrate this phenomenon for a pair of fibrations below.

In order to apply the standard maps of heterotic/F-theory duality \cite{Friedman:1997yq}, we will focus here on the case where $X_3$ is smooth and has no fibral divisors (i.e no blow-ups in the fiber). In this case, $h^{1,1}(X_3)=h^{1,1}(B) + {\rm rk} \, MW+1$ for each of its fibrations. Having chosen a given fibration, say, $\pi$ in \eref{het_which_wayup}, we can divide the divisors into ``horizontal" and ``vertical" types with respect to this choice:
\bea\label{divisors_in_x}
\sigma_m~&: ~~\text{sections to fibration}~ \pi: X \to B~~~~(m=0 \ldots {\rm rk}\, MW) \ , \\ 
D^{\rm b}_{\alpha}~&: ~~\text{divisors pulled back from base}~B,~{\text{as}}~\pi^*(\hat{D}^{\rm b}_\alpha)~~~~(\alpha=1 \ldots h^{1,1}(B)) \ . 
\eea
The fact that a single threefold $X_3$ is multiply elliptically fibered as in \eref{het_which_wayup} means that its topology can be expanded in a basis of forms dual to \eref{divisors_in_x} in \emph{distinct ways}. Focusing, in particular, on cases with ${\rm rk}\, MW = 0$, the second Chern class of a bundle in the $E_8 \times E_8$ theory, as described in \eref{c2twist}, takes the form:
\beq\label{2waysc2}
c_2(V)=\eta \wedge \omega_{\hat 0} + \zeta = \eta'\wedge \omega_{\hat 0} + \zeta' \ . 
\eeq
This leads naturally to the twists over $B_2$ and $B'_2$,
\beq
\eta=6c_1(B_2) +T \ ~~\text{and}~~ \eta'=6c_1(B'_2)+ T' \ , 
\eeq
and a natural question arises as to \emph{how $T$ and $T'$ are related}. The divisors (equivalently $(1,1)$-forms) on $X_3$ are fixed and must be expanded as in \eref{divisors_in_x} in two different ways, related by an integral basis change. This schematically takes the form
\beq\label{basischange}
\left(\begin{array}{c}
\sigma'_{m'}  \\
D'_{\a'}
\end{array}\right)=M_{\Lambda\Sigma}\left(\begin{array}{c}
\sigma_m  \\
D_{\a}
\end{array}\right)
\eeq
where $M_{\Lambda\Sigma}$ is an invertible matrix and $\Lambda, \Sigma$ run over the full range of $h^{1,1}(X_3)$. Note that although the length of the vectors is fixed by $h^{1,1}(X_3)$ the fiber/base division in \eref{divisors_in_x} can be very different in $B_2$ and $B'_2$.

For a given example, this correspondence can be determined explicitly. Moreover, in general the basis change $M_{\Lambda\Sigma}$ will be constrained by the topology of $X_3$ itself. For simply connected manifolds, Wall's theorem \cite{wall} says that the diffeomorphism class of $X_3$ is determined by the collection of numbers, 
\beq
\{d_{\Lambda\Sigma\Psi}\,,~ c_1(X_3)\,,~ c_2(X_3)\,,~ c_3(X_3)\,,~ h^{1,1}(X_3)\,,~ h^{2,1}(X_3)\} \ , 
\eeq
where $d_{\Lambda\Sigma\Psi}$ for $\Lambda,\Sigma,\Psi =1,\ldots h^{1,1}(X)$ are the triple intersection numbers of $X_3$. But for an elliptically fibered manifold with section (and no fibral divisors), the Chern class and triple intersection numbers take a constrained form. See Appendix \ref{topology} for a collection of such useful results. In general these can be used to determine the relationship between $T$ and $T'$ for any given example.

To illustrate this, for simplicity, we will consider an example in which there is a single section $\sigma_0$ to \emph{each} fibration. In this case, $h^{1,1}(X_3)=1+h^{1,1}(B)$ (with $h^{1,1}(B)=h^{1,1}(B_2)=h^{1,1}(B'_2)$). Furthermore, the triple intersection numbers and links between $c_2(TX)$ and $c_1(B_2)$ are highly constrained (see Appendix \ref{topology}). It is clear however that some bases could satisfy these conditions -- including the Hirzebruch surfaces with $h^{1,1}(\mathbb{F}_n)=2$ and $\chi(\mathbb{F}_n)=4$ for all $n \geq 0$. Simplifying still further, the heterotic/F-theory duality map is most easily defined in the case that both fibrations are in fact \emph{holomorphic} (rather than merely rational). As we can see from the examples of previous sections, requiring both fibrations to have exactly one holomorphic section and no fibral divisors is by no means generic, however even this case can demonstrate some interesting possibilities.

In the constrained case of exactly one holomorphic section for each fibration, the triple intersection numbers obey \cite{Friedman:1997yq}:
\begin{align}\label{intersecs_hir}
&d_{000}=\eta_{\alpha\beta}K^{\alpha}K^{\beta} & d_{00\a}=\eta_{\alpha\beta}K^{\beta} \\
&d_{0\a\beta}=\eta_{\a \beta} & d_{\alpha \beta \gamma}=0 
\end{align}
where $K^{\alpha}$ are the coefficients in the expansion of the base canonical class, $K=-\left[c_1(B_2) \right]=K^{\alpha}\hat D^{\rm b}_{\alpha}$, and $\eta_{\alpha\beta}=\hat D^{\rm b}_{\alpha} \cdot \hat D^{\rm b}_{\beta}$ are the double intersection numbers for the divisors on $B_2$.

For illustration, let us choose the two bases to be Hirzebruch surfaces, $B_2=\mathbb{F}_m$ and $B'_2=\mathbb{F}_k$. To avoid singular Weierstrass models that must be resolved (and added fibral divisors), we will restrict ourselves to the case when $m,k \leq 2$. In this case, it can readily be observed that the equivalence of the intersection numbers in \eref{intersecs_hir} for integral basis changes fixes the choice of integers to $k=2$ and $m=0$. Then, many changes of basis in \eref{basischange} are possible. These include the following map from the description of $X_3 \to \mathbb{F}_0$ to that of $X_3 \to \mathbb{F}_2$:
\beq
\left(\begin{array}{c}
\sigma'_{0}  \\
h' \\
f'
\end{array}\right)=\left(\begin{array}{ccc}
1 & 0 & 0 \\
0& 1 & -1 \\
0 & 0 & 1
\end{array}\right)
\left(\begin{array}{c}
\sigma_0  \\
h \\
f \end{array}\right) \ , 
\eeq
with $h^2=f^2=0$, $h\cdot f=1$, $h'^2=-2$, $h'\cdot f'=1$, and $f'^2=0$. Note that this solution is purely a basis change within the Hirzebruch base itself, a fact that is not surprising since $\mathbb{F}_0$ and $\mathbb{F}_2$ are infinitesimally close in moduli space \cite{Hubsch:1992nu}. For this choice, the twists (and hence the F-theory geometry) are related as
\beq
T'_{\mathbb{F}_2}=ah'+bf'~~~\Leftrightarrow~~~T_{\mathbb{F}_0}=ah+(b-a)f \ . 
\eeq

More novel possibilities that actually mix fiber/base descriptions include for example:
\beq
\left(\begin{array}{c}
\sigma'_{0}  \\
h' \\
f'
\end{array}\right)=\left(\begin{array}{ccc}
1 & 2 & 2 \\
0& -1 & 1 \\
0 & 0 & 1
\end{array}\right)
\left(\begin{array}{c}
\sigma_0  \\
h \\
f \end{array}\right)
\eeq
which yields (by choosing to look at the $\eta=6c_1(B)+T$ case)
\beq
T'_{\mathbb{F}_2}=ah'+bf'~~~\Leftrightarrow~~~T_{\mathbb{F}_0}=-(a+24)h+(b+24)f \ . 
\eeq 

It would be an intriguing prospect to construct such a duality explicitly for specific bundles on a realistic CY $3$-fold and construct the F-theory duals, including G-flux. It has been observed  \cite{Anderson:2013rka,Anderson:2014gla,Anderson:2016kuf} that in the context of heterotic/F-theory duality novel solutions for G-flux (such as T-branes) depend on the decompositions such as \eref{2waysc2}. We hope to explore such examples in future work.

\subsubsection{Exploring the moduli space of stable bundles}\label{eta_bounds}
It has recently been observed \cite{Rajesh:1998ik,Berglund:1999qk,Anderson:2014gla} that heterotic/F-theory duality in $4$-dimensions can provide a useful window into the allowed geometry of both heterotic and F-theory compactifictions. In particular, it was noted that simple criteria in F-theory could help to constrain how possible structure groups are linked to the topology of slope-stable vector bundles over Calabi-Yau $3$-folds. In the context of heterotic model building, the following question is a notoriously difficult one, with few mathematical tools available to address it: \\

\noindent \emph{Given a CY 3-fold, $X_3$, does there exist a stable bundle, $V \to X_3$, with given rank, $rk(V)$, structure group, $H \subset E_8$, and total Chern class, $c(V)$?} \\

\noindent In general, very little is known about the structure of the moduli space ${\cal M}_{X_3}(c(V))$ of semi-stable sheaves on $X_3$ with fixed rank and total Chern class. 

In \cite{Anderson:2014gla}, the singularity structure of $Y_4$ was used to link the form of $\eta$ in \eref{c2twist} to the structure group of $V \to X_3$. In particular, the existence of generic (non-Higgsable) symmetries on $Y_4$ provides a simple criteria for the triviality/vanishing of ${\cal M}_{X_3}(c(V))$. Briefly, if a generic symmetry $G$ (arising from singular fibers in $Y_4$) cannot be Higgsed in the $4$-dimensional effective theory, it follows that for the given topology of the bundle ($\eta$ in \eref{c2twist}) there cannot exist a bundle with a structure group larger than $H$, the commutant of $G \subset E_8$ ({i.e.}, in order to build a bundle with structure group $H$ over the elliptically fibered CY threefold, there is a minimum ``size" for $\eta$). Table \ref{eta_boundy} gives a sample of such constraints on $\eta$ in the case of a dual $4$-fold (without G-flux).
\begin{table}
\begin{center}
\begin{tabular}{|c ||c |}
\hline
$H$ & $a$~~in~~$\eta \geq a  c_1(B_2)$\\
\hline
  $SU(n) $&$ n~~(n \geq 2) $\\
$SO(7) $ &$ 4  $\\
 $SO(m) $ &$ \frac{m}{2}~~(m\geq 8)$\\
 $Sp(k)$ &$ 2k~~(k \geq 2) $\\
 $F_4 $ &$ 4 $\\
 $ G_2 $ & $3$\\
 $E_6$ & $\frac{9}{2}$ \\
 $E_7$ & $\frac{14}{3}$ \\
 $E_8$ & $\frac{24}{5}$ \\
\hline
\end{tabular}\end{center}
\caption{Constraints on the ``size" of $\eta$ required for certain structure groups $H$ of heterotic bundles derived in \cite{Rajesh:1998ik,Berglund:1999qk,Anderson:2014gla}. Here, $\eta \geq a c_1(B_2)$ indicates that $\eta - ac_1(B_2)$ is an effective divisor on $B_2$.} 
\label{eta_boundy}
\end{table}

Importantly, these bounds involve only a part of the second Chern class of the bundle, $V$ ({i.e.}, the twist $T$, or equivalently, $\eta$ in \eref{twisty}). In this context then, multiple fibrations to the heterotic threefold play a remarkable role in that each fibration constrains a further component of $c_2(V)$. A systematic exploration of the bounds on all possible $\eta,\eta'$s in the context of \eref{het_which_wayup} could provide information on the full $c_2(V)$ and a more complete view of the moduli space of heterotic bundles. Information linking structure group (and further, the zero-mode spectrum \cite{Anderson:2014gla}) to the topology of a bundle would make the large scale systematic searches \cite{Anderson:2013xka,Anderson:2012yf,Anderson:2011ns} for phenomenologically relevant heterotic vacua dramatically simpler.

\section{Conclusions and Other Dualities in Diverse Dimensions} \label{conc}
In this work we have taken some first steps in an exploration of some of the rich duality structures that can be brought to light by multiple fibrations in CY geometries. Our primary results include the following
\begin{itemize}
\item Within the M-/F-theory correspondence, the use of M-theory to determine the structure of the F-theory effective field theory has been a powerful tool which has lead to considerable recent progress (see e.g \cite{Bonetti:2011mw,Grimm:2013oga,Grimm:2015wda,Bonetti:2013cza}). In this work, we have seen that the multiple fibration structures explored in Section \ref{m_f_duals} provide numerous examples of related EFTs -- e.g. in Section \ref{9waysup} for example, 9 different elliptic fibrations arise from a single CY 3-fold. The $6$-dimensional F-theory compactifications are all distinct theories -- with different numbers of vector, tensor and hypermultiplets. However, the correspondence of a shared M-theory limit indicates that in some way, the collected F-theory vacua \emph{share a broader moduli space}. Since the complex structure moduli of one CY $n$-fold controls the deformations of all of the F-theory limits, any deformation of the geometry will deform the EFTs linked by different fibrations in a correlated and anomaly-cancelling way. This suggests that essential properties (and deformations) of these theories can be connected in deep and previously unseen ways. It will be intriguing to further explore the consequences of such multiple CY fibrations on the effective theories in more detail in future work.

One area of particular interest includes the linking of non-flat fibrations (known to give rise to superconformal theories \cite{Morrison:1996na,Morrison:1996pp,Seiberg:1996vs,Candelas:2000nc,Braun:2013nqa}) to ordinary, flat fibrations through the shared M-theory limits seen in Section \ref{m_f_duals}.  As expected, we find that the construction of the Weierstrass models for CY geometries with non-flat fibers give rise to discriminant loci with vanishing orders $(f,g,\Delta) \geq (4,6,12)$, as expected. Interestingly, the occurrence of such superconformal points in discriminant loci has primarily been systematically studied to date via blowing-up points in the base \cite{Morrison:1996na,Morrison:1996pp}. This work highlights the fact that by considering more exotic, non-flat fibers (at higher co-dimension) there are other choices of resolution leading to smooth CY geometries. These possibilities are all clearly linked to such superconformal theories and they may have a further role to play in string dualities.

\item Within heterotic/F-theory duality (and heterotic/Type IIA dualities), we find that multiple fibrations are present in most known correspondences -- including heterotic/heterotic duality in $6$-dimensions, T-duality and Mirror symmetry. 

Perhaps most novel in this collection of results is the observation that the choice of an elliptic fibration does not effect the form of a ${\cal N}=1$ heterotic compactification on a (multiply) elliptically fibered CY 3-fold in $4$-dimensions. However, the form of its $4$-dimensional F-theory dual can appear very different -- with topologically distinct CY $4$-folds and G-flux being linked by these common heterotic duals. It seems that these observations could shed light on the notoriously difficult problem of classifying (and explicitly constructing) G-flux in $4$-dimensional compactifications.

\item Finally, the tools developed here and in \cite{us_to_appear} make it possible to systematically scan for fibration structures and the corresponding string dualities in known datasets of CY $n$-folds. Already in this work we have found examples of CY geometries with high rank Mordell-Weil groups, intriguing non-flat fibration structures and novel examples of dualities. As observed in Section \ref{sec:geometry} and in \cite{us_to_appear} the vast majority of known CY $3$- and $4$-folds admit multiple elliptic and $K3$-fibrations and it has been conjectured that in fact, \emph{all CY $n$-folds with large enough Hodge numbers may be elliptically fibered} \cite{gross_finite}. As a result, we hope in future work to use these tools to complete a ``duality cartography"  \cite{usscanning} and survey fibrations and dualities in CY datasets beginning with the CICY $3$- and $4$-folds.
\end{itemize}

To conclude we briefly outline a number of other interesting areas in which multiple fibrations in CY geometries may play a role in new string dualities and other correspondences.

\subsection{Further directions and applications}
\subsubsection*{$(0,2)$ Target space duality and Heterotic/F-theory duality}
It was observed in \cite{Distler:1995bc} that apparently distinct $(0,2)$ Gauged Linear Sigma Models (GLSMs) \cite{Witten:1993yc} can be related to one another by a shared non-geometric phase. The nature of this correspondence and extensions of it which allowed for topologically distinct geometric limits (with CY $3$-folds with different Hodge numbers and vector bundles over them) were explored and extended in \cite{Blumenhagen:1997vt, Blumenhagen:1997cn}. Despite numerous examples of target space dual GLSMs, thus far, a concrete linking of the underlying $(0,2)$ sigma models has remained elusive. It is an open question as to whether this target space duality corresponds to two distinct $(0,2)$ sigma models with a shared locus in their moduli space (akin to a conifold transition between CY 3-folds in Type II theories \cite{Candelas:1989ug}) or is instead a true duality of the underlying theories. Some recent evidence in favor of the latter possibility was assembled in a landscape scan \cite{Blumenhagen:2011sq,Rahn:2011jw} which systematically explored the zero mode spectrum of thousands of models and in \cite{Anderson:2016byt} which explored the ${\cal N}=1$ potential ({i.e.}, D-terms and F-terms) and the vacuum structure of the paired theories. In both cases the target space ``dual" theories agreed to a high level of detail.

One simple class of target space dual theories involves Calabi-Yau threefolds that are related by simple conifold-type transitions \cite{Blumenhagen:2011sq}. In this case, if the CY 3-fold associated to the starting GLSM is elliptically fibered, in many cases the dual theory will be as well. In this case, it is expected that both theories will admit F-theory duals defined by CY $4$-folds. It was conjectured in \cite{Blumenhagen:2011sq} that perhaps the F-theory dual geometries of the two heterotic target space theories \emph{could be the same CY $4$-fold}. This could be possible if the single geometry $Y_4$, was of the ``Case 1" type and admitted two $K3$- fibrations of the form discussed in Section \ref{section:hetf} and equation \eref{multp1_first} -- that is, a single $4$-fold, $Y_4$, which admits \emph{two distinct $K3$-fibrations with a shared elliptic fibration}. It would be intriguing to put this conjecture to the test using the tools developed here and in \cite{us_to_appear}.

Finally, it should be noted that understanding the target space duality (and the corresponding relationships in F-theory vacua) could play an important role in highlighting key structures and removing ``redundancy" from systematic/algorithmic searches for phenomenologically relevant $4$-dimensional, ${\cal N}=1$ vacua such as the scans undertaken in \cite{Anderson:2011cza,Anderson:2007nc,Anderson:2008uw,Anderson:2009mh,Anderson:2011ns,Anderson:2012yf,Anderson:2013xka,Anderson:2014hia,Buchbinder:2016jqr}.

\subsubsection*{Superconformal theories in $6$-dimensions and multiple fibrations}
Multiple elliptic fibrations and T-duality have been recently observed to play a role in Little String Theories (LSTs) \cite{Bhardwaj:2015oru,Hohenegger:2015btj,Morrison:2016djb}. Both LSTs and superconformal theories (SCFTs) in $6$-dimensions have been systematically studied as geometric phases of F-theory. Recent results \cite{Heckman:2013pva,DelZotto:2014hpa,Heckman:2015bfa,Heckman:2015ola} include a demonstration that all $6$-dimensional superconformal theories can be classified according to an ``atomic classification" of blow-ups of the base $B_2$ of an F-theory elliptic fibration, and that in addition, all $6$-dimensional SCFTs naturally embed into an LST. These results also shed light on SCFTs in $4$-dimensional compactifications \cite{Morrison:2016nrt}. Importantly, in the context of LST examples have been found \cite{Bhardwaj:2015oru,Morrison:2016djb} of geometries admitting multiple elliptic fibrations. As in the examples presented here, distinct $6$-dimensional vacua were linked through further dimensional reduction on a circle -- with T-duality of the circle radius mapping distinct LSTs into one another. 

In view of the present study of multiple fibrations in CY geometries, an intriguing open question remains how the structure of non-flat fibers in $6$- and $4$-dimensions can be linked to the superconformal theories studied above? In addition, in geometries admitting SCFT or LST limits, there can be many more than two fibrations and it would be fruitful to see how this network of theories could be linked.
 
 \subsubsection*{$2$-dimensional $(0,2)$ theories and multiple fibrations in Calabi-Yau $5$-folds}
 Interesting recent results \cite{Schafer-Nameki:2016cfr,Apruzzi:2016iac} have demonstrated that F-theory compactifications on CY $5$-folds give rise to $(0,2)$ supersymmetric theories in $2$-dimensions. The form of the $2$-dimensional sigma models are directly linked to the structure of the elliptic fibration of $Y_5$ and its singularities. Many non-trivial features of the elliptically fibered CY geometry appear in the cancellation of anomalies, Chern-Simons terms and more. In addition, the resulting $2$-dimensional $(0,2)$ theories can be linked to, and interpreted as, heterotic worldsheet theories (and GLSMs).
 
 In this context, it is clear that CY $5$-folds admitting multiple fibrations could play a significant role in dualities linking $2$-dimensional $(0,2)$ theories. In particular, as seen in Section \ref{section:hetf}, \emph{nested fibration structures} in which $Y_5$ admits any combination of the following chain of possible sub-fibrations 
 \begin{align}
 &\pi_1: Y_5 \stackrel{\mathbb{E}}{\longrightarrow} B_4~~~,~~~\pi_2: Y_5 \stackrel{K3}{\longrightarrow} B_3 \\
  &\pi_3: Y_5 \stackrel{CY_3}{\longrightarrow} B_2~~~,~~~\pi_4: Y_5 \stackrel{CY_4}{\longrightarrow} \mathbb{P}^1
 \end{align}
 with potentially nested or distinct sub-fibers
 \beq
 \mathbb{E} \subset K3 \subset CY_3 \subset CY_4 \subset Y_5
 \eeq
 and rationally fibered bases, $B_n$, could have consequences for the $(0,2)$ GLSMs introduced in \cite{Schafer-Nameki:2016cfr,Apruzzi:2016iac}. 
 
\vspace{15pt}

In summary, we view the tools developed here and in \cite{us_to_appear, usscanning} as an important first step towards understanding how the geometry of CY fibrations sheds light on fundamental string dualities. We hope the questions above and many others will be explored in future work.

\section*{Acknowledgements}
The authors would like to thank A. Grassi, Y. L\"u, T. Pantev and W. Taylor for useful discussions. The work of LA (and XG in part) is supported by NSF grant PHY-1417337. The work of JG (and SJL in part) is supported by NSF grant PHY-1417316. This project is part of the working group activities of the 4-VA initiative ``A Synthesis of Two Approaches to String Phenomenology".

\appendix

\section{Counting of Nodes in a Complex Curve} \label{sec:plane}
\begin{figure}[t!]
\centering{
\includegraphics[width=0.25\textwidth]{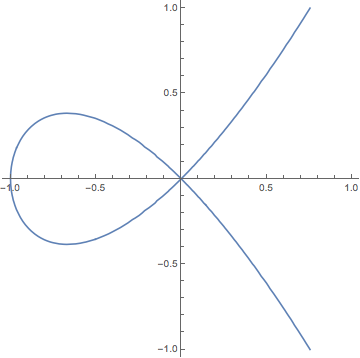}
\quad\quad 
\includegraphics[width=0.25\textwidth]{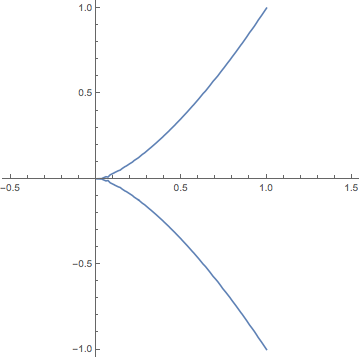}
\quad\quad   
\includegraphics[width=0.25\textwidth]{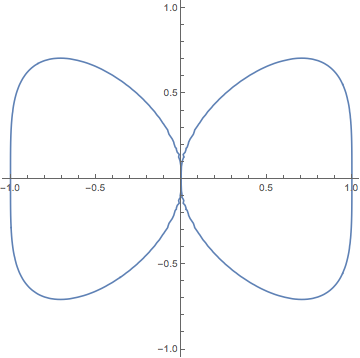}
\caption{Examples of double point singularities from left to right: node, cusp, and tacnode. See \cite{Kunz} for details of multiple point singularities.}
\label{fig:double_points}}
\end{figure}

Here, we give a brief review of how the nodes of a complex curve are counted. In general, for complex curves the singularities can include nodes, cusps, tacnodes (See Figure \ref{fig:double_points}) and multiple point singularities. In the context of the main text of this paper, the algebraic curve $\mathcal C$ in question is defined as a component (i.e. the $I_1$ component) of the  discriminant locus, $\{\Delta=0\}$, in a two-fold surface $B$, which may itself be defined as a complete intersection of $N$ polynomials, $P_{j=1, \cdots, N}$ in $\mathcal A_B$. Defining ${\cal C}$ via the zero locus of a polynomial $\{F=0\}$ we have
\beq
\mathcal C \overset{F}\hookrightarrow B \overset{P_j}\hookrightarrow \mathcal A_B \ .
\eeq
We denote by $x_i \in \IC$ for $i=1, \cdots, N+2$, the affine coordinates of the $(N+2)$-dimensional ambient space, $\mathcal A_B$.

\subsection{A hypersurface curve}
When $N=0$, the algebraic curve $\mathcal C$ is a single hypersurface, $\mathcal C = \{ F(x_1, x_2) = 0 \} \subset B$, and the singularities are associated with the following ideal,
\beq
\mathcal I=\left<F, {\rm d} F \right> =\left<F, F_{x_1}, F_{x_2}\right> \ , 
\eeq
where, by an abuse of notation, we denote the polynomial coefficients of the expansion of a differential form in a basis of forms of an appropriate degree, by the form expression itself. Here, $F_{x_i}$ denote the first derivatives of $F$ with respect to $x_i$. 
Next, we form the ideal,
\beq\label{hideal}
\mathcal H = \left<h\right> \ ,
\eeq
generated by the determinant, $h={\rm Det}(H)$, of the Hessian matrix,
\beq\label{hessian}
H=\left(\begin{array}{cc}
F_{x_1 x_1} & F_{x_1 x_2} \\
F_{x_2 x_1} & F_{x_2 x_2} \\
\end{array}\right) \ .
\eeq
The ideal associated with the nodes can then can be obtained via the commutative algebra procedure of ``saturation" as follows: 
\beq
{\cal I}_{\rm node}=({\mathcal I}: \mathcal{H}^{\infty}) \ . 
\eeq
Intuitively, this corresponds to (the algebraic closure of) the part of the ideal ${\mathcal I}$ that remains when the generators of ${\mathcal  H}$ do \emph{not} vanish. Then we arrive at last at 
\beq\label{node_total}
\#({\rm Nodes})= \textnormal{dim}_V({\cal I}_{\rm node}) \ , 
\eeq
where by $ \textnormal{dim}_V({\cal I}_{\rm node})$ we mean the vector space dimension of the zero-dimensional ideal ${\cal I}_{\rm node}$. This counting can be done practically with the tools in \cite{DGPS,Gray:2008zs,Gray:2006gn}.

In general, there may arise double point singularities that are not nodes but either cusps or tacnodes. Tacnode singularities never appear in any of the examples of flat fibrations analyzed in this paper. Nevertheless, they do appear in non-flat cases and we believe that they are relevant to the physics of non-flat fibrations. To complete the story, let us also describe the relevant ideals that can be used to count both cusps  and tacnodes.  
We first define a combined ideal
\beq
{\mathcal I}_{{\rm non-node}}= \langle F, F_{x_1},F_{x_2}, h \rangle
\eeq
and then form the ideal associated with cusps as a saturation
\beq
{\mathcal I}_{{\rm cusp}}=(I_{{\rm non-node}} : {{\mathcal H}_{\nabla}}^{\infty})
\eeq
where ${\mathcal H}_{\nabla}$ is defined by taking the derivative of the determinant of the Hessian (itself as defined in \eref{hessian}), and forming the following ideal,
\beq
{\mathcal H}_{\nabla}= \langle h_{x_1}, h_{x_2} \rangle \ , 
\eeq
which yields a count
\beq
\#({\rm Cusps})=  \textnormal{dim}_V({\cal I}_{\rm cusp}) \ , 
\eeq
Finally, to count the number of tacnodes, it is necessary to quantify the possible multiple point singularities, 
\beq\label{mult_cond}
{\mathcal I}_{{\rm mult}}=\left<F, F_{x_1},F_{x_2},F_{x_1x_1},F_{x_1x_2},F_{x_2x_2}\right>  \ , 
\eeq
where $F_{x_i}$ and $F_{x_i x_j}$ denote the first and the second derivatives of $F$ with respect to the respective variables. With these definitions in hand, we can then describe the following ideal
\beq
{\mathcal I}_{\rm non-node-non-cusp}= \langle F, F_{x_1},F_{x_2},h, h_{x_1}, h_{x_2} \rangle \ . 
\eeq
The tacnodes are then simply counted by removing the multiple point singularities from the dimension of ${\mathcal I}_{\rm non-node-non-cusp}$ as
\beq
\#({\rm Tacnodes})= \textnormal{dim}_V({\mathcal I}_{\rm non-node-non-cusp})- \textnormal{dim}_V({\mathcal I}_{{\rm mult}})
\eeq
Again, we note that tacnodes (and multiple point singularities) only appear in the context of the non-flat examples considered in this work. See Section \ref{9waysup}. 

\subsection{A complete intersection curve}
The node-counting algorithm for a complete intersection curve, with $N>0$, is given as a straightforward generalization of that for the hypersurface case. 
The formula~\eref{node_total} still applies once the relevant ideals, $\mathcal I$ and $\mathcal H$, are defined in an appropriately generalized manner as follows.
Firstly, the ideal describing all of the singular points of the curve is given by
\beq
\mathcal I=\left<F, P_1, \cdots, P_{N}, (\wedge_{j=1}^N {\rm d} P_j)\wedge {\rm d} F \right> \ .
\eeq
The ideal $\mathcal H$ is then defined as in \eref{hideal} where the Hessian matrix $H$, in the presence of base defining equations, $P_1, \cdots, P_N$, is constructed as \cite{neudecker}
\beq\label{}
H=\left(\begin{array}{cccccc}
0 & \cdots & 0 & \frac{\partial P_1}{\partial x_1} & \cdots & \frac{\partial P_1}{\partial x_{N+2}} \\
\vdots &\ddots & \vdots & \vdots &\ddots &\vdots \\
0 & \cdots & 0 & \frac{\partial P_N}{\partial x_1} & \cdots & \frac{\partial P_N}{\partial x_{N+2}} \\
\frac{\partial P_1}{\partial x_1}  & \cdots & \frac{\partial P_N}{\partial x_1} & F_{x_1  x_1} & \cdots & F_{x_{1}  x_{N+2}}  \\
\vdots  & \ddots & \vdots  & \vdots & \ddots& \vdots  \\
\frac{\partial P_1}{\partial x_{N+2}}  & \cdots & \frac{\partial P_N}{\partial x_{N+2}} &  F_{x_{N+2}  x_1} & \cdots & F_{x_{N+2}  x_{N+2}} \\
\end{array}\right) \ .
\eeq
As in \eref{node_total} this allows us to obtain a count of the number of nodes (with analogous formulae for cusps, tacnodes, etc) for a more general description of the base manifold, $B$.

\section{The Topology of an Elliptically Fibered Calabi-Yau $3$-fold}\label{topology}
In this Section, a brief review is provided of a collection of useful results
regarding the geometry and topology of elliptically fibered Calabi-Yau
$3$-folds (see \cite{Friedman:1997yq,Klemm:1996ts,Grimm:2013oga} for a more complete
treatment). 

In the case of a flat fibration, $\pi:X_3\to B_2$, the Shioda-Tate-Wazir theorem \cite{shioda,shioda2,COM:213767} guarantees that we can decompose the divisors $D_\Lambda$ of $X_3$, for $\Lambda=1, \ldots, h^{1,1}(X_3)$, as
\begin{align}
&D^{\rm b}_{\a}=\pi^{*}(D^{\rm b}_{\a})&: &~~ \text{divisors pulled back from the base} \\
&D_I &:  &~~ \text{fibral divisors} \\ 
&D_{\hat{0}}&:  &~~ \text{the zero section} \\
&D_m &: &~~ \text{rational sections, elements of the Mordell-Weil group of}~X \ ,
\end{align}
where the indices of different types run over the following ranges:
\beq
\a,\beta, \cdots =1, \ldots, h^{1,1}(B_2)\, ;\quad\quad I,J,\cdots = 1, \ldots, {\rm rk}\, G\, ;\quad\quad m, p, \cdots = 1, \ldots,  {\rm rk}\, MW \ . 
\eeq
Here, $G$ denotes the non-Abelian fiber symmetry associated to the resolved singular fibers of $X$, whose rank counts the fibral divisors. In particular, we have
\beq
h^{1,1}(X_3)=h^{1,1}(B_2)+ \#\text{(fibral divisors)}+1 + {\rm rk}\, MW \ . 
\eeq
We
will denote the basis of $(1,1)$-forms dual to the different types of divisors above as
\beq
\omega_{\Lambda}=\{\omega_{\a},\,\omega_I,\,\omega_{\hat 0},\,\omega_m\} \ , 
\eeq
with the index $\Lambda$ running over the full range of $h^{1,1}(X_3)$.

We begin by exploring the triple intersection numbers of $X_3$:
\begin{equation}
d_{\Lambda\Sigma\Gamma}=D_{\Lambda}\cdot D_\Sigma \cdot D_\Gamma=\int_{X_3} \omega_{\Lambda} \wedge \omega_{\Sigma} \wedge \omega_{\Gamma} \ . 
\end{equation}
First, since the base is a $2$-fold it is clear that
\begin{equation}
D^{\rm b}_{\a} \cdot D^{\rm b}_{\beta} \cdot D^{\rm b}_{\gamma}=0 \ . 
\end{equation}
Moreover, from the very definition of what it means for $D_{\hat{0}}$ and $D_m$ to be
rational sections (rather than multisections), it is guaranteed that for any
four-form $\hat \zeta$ on $B_2$, the following relation holds:
\beq
\int_{X_3} \omega_{\hat{0}} \wedge \zeta=\int_{B_2} \hat \zeta \ , 
\eeq
where $\zeta = \pi^*(\hat \zeta)$, and likewise for $\omega_a$. That is, a section intersects the generic elliptic fiber precisely once. It then follows that for either a holomorphic or rational section, 
\begin{equation}
D_m \cdot D^{\rm b}_{\alpha} \cdot D^{\rm b}_{\beta} = \eta_{\a \beta} \ , 
\end{equation}
where $\eta_{\alpha\beta}= D^{\rm b}_{\a} \cdot  D^{\rm b}_{\beta}$. 

The observations above are enough to derive the following important double intersection formula which holds for \emph{holomorphic} sections, $D_{hol}$,
\begin{equation}\label{basetop}
D_{hol} \cdot D_{hol} =K_{B_2} \cdot D_{hol} \ , 
\end{equation}
where $K_{B_2}$ is the canonical class of the base. In the case that a section $D_{rat}$ is merely rational, the (slightly weaker) triple intersection formula holds (frequently used to help identify rational sections in $X_3$ \cite{us_to_appear}),
\beq
D_{rat} \cdot D_{rat} \cdot D^{\rm b}_{\a} =K_{B_2} \cdot D_{rat} \cdot D^{\rm b}_{\a} \ , ~~~\text{for}~ \a=1, \ldots, h^{1,1}(B_2) \ , 
\eeq
which in particular must hold for $D_{\hat 0}$ and $D_m$. 

Following \cite{Grimm:2013oga,Grimm:2015wda} it is possible to define a simple, shifted version of the zero section which obeys convenient ``orthogonality" properties under the Shioda map. These can be found in general in \cite{Grimm:2013oga,Grimm:2015wda}. For the zero section $D_{\hat{0}}$, a shifted version, $D_0$, can be obtained from
\beq
D_0 =D_{\hat{0}} -\frac{1}{2}(D_{\hat{0}} \cdot D_{\hat{0}} \cdot D^{{\rm b}, \a} )D^{\rm b}_{\a} \ , 
\eeq
where the indices $\a,\beta,\cdots$ are raised and lowered using the $\eta_{\a\beta}=D^{\rm b}_{\a} \cdot D^{\rm b}_{\beta}$.
Such a shift guarantees that $D_0 \cdot D_0 \cdot D^{\rm b}_{\a}=0$, for $\a=1, \ldots, h^{1,1}(B_2)$. Then, with respect to this basis (with the shifted zero section), whether the zero section is holomorphic or not, the triple intersection numbers are given as
\begin{align}
&d_{\alpha\beta\gamma}=0 && d_{0\a\beta}=\eta_{\a\beta} && d_{00\alpha}=0 \\
&d_{\a\beta I}=0 && d_{\a 0 I}=0 && d_{\a I J}=- C_{IJ}(\cS^{\rm b} \cdot D_{\a}^{\rm b}) \\
&d_{\a\beta m}=0 && d_{\a I m}=0 && d_{0 \alpha m}=0 \\
&d_{\a m n}=\pi(D_m \cdot D_n)_{\a}   && & & 
\end{align}
Here, $\cS^{\rm b}$ is a divisor in $B_2$ over which the elliptic fiber develops singularities and $C_{IJ}$ is the co-root matrix. See \cite{Grimm:2013oga,Grimm:2015wda} for further details.

For the analysis in Section \ref{double_het} we will consider the form of the intersection numbers in the \emph{blown-down} limit in which all fibral divisors $D_I$ go to zero volume. Furthermore, in the case that the zero section is holomorphic we have that
\beq
D_{\hat{0}} \cdot D_m=0 \ , 
\eeq
and it follows that the remaining intersection numbers take the simple form \cite{Grimm:2013oga}:
\begin{align}\label{intersecs2}
&d_{000}=\frac{1}{4}\eta_{\alpha\beta}K^{\alpha}K^{\beta}& &d_{0mn}=-\frac{1}{2}\pi(D_m \cdot D_n)_{\a}K^{\a} \\
&d_{00m}=0 \ , & &  \nonumber
\end{align}
where $K^\a$ are the coefficients in the expansion, $K_{B_2}=-\left[c_1(B_2)\right]=K^{\a}D^{\rm b}_{\a}$, of the base canonical class.
Once again the equalities in \eref{intersecs2} only hold in the case of a holomorphic zero section.

The fibration structure guarantees if $X_3$ has a holomorphic zero section and no fibral divisors, then the second Chern class of $X_3$ can be written as \cite{Friedman:1997yq}
\begin{equation}\label{c2tangent}
c_2(TX_3)=12c_1(B_2)\wedge \omega_{\hat 0} + c_2(B_2)+11c_1(B_2)^2 \ , 
\end{equation}
where in addition the topology of $B_2$ satisfies
\begin{equation}
\chi(B_2)=\int_{B_2} c_2(B_2)=2+h^{1,1}(B_2)~~~,~~~\int_{B_2} c^{2}_{1}(B_2)=K^{\a}K^{\beta}\eta_{\a\beta}=10-h^{1,1}(B_2) \ . 
\end{equation}

In general, the second Chern class of an elliptically fibered CY $3$-fold obeys \cite{Friedman:1997yq}
\beq
\int_{X_3} \omega_{\a} \wedge c_2(X_3)=-12 K_{\a} \ . 
\eeq
Finally, in the case of a single section obeying \eref{basetop} the second Chern class of any bundle, $V$, on $X_3$ can be written:
\begin{equation}\label{c2special}
c_2(V)=\eta \wedge \omega_{\hat 0} + \zeta \ , 
\end{equation}
where $\eta$ and $\zeta$ are pullbacks through $\pi$ of $(1,1)$- and $(2,2)$-forms on $B_2$, respectively.

\nocite{*}
\bibliography{F-theoryDuality}
\bibliographystyle{utphys}

\end{document}